\documentclass[a4-paper,12pt]{article}
\usepackage{amssymb, amsmath,amsthm}
\renewcommand{\Bbb}{\mathbb}
\date{}

\makeatletter

\@addtoreset{equation}{section}
\makeatother

\newtheorem{dfn}{Definition}[section]
\newtheorem{thm}[dfn]{Theorem}%[section]
\newtheorem{prp}[dfn]{Proposition}%[section]
\newtheorem{lemma}[dfn]{Lemma}%[section]
\newtheorem{cor}[dfn]{Corollary}

\numberwithin{equation}{section}

\begin{document}
%%%%%%%%%%%%%%%%%%%%%%%%%%%%%%%%%%

\title{Improved Sobolev Embedding Theorems 
           for Vector-valued Functions\\
{\it\normalsize  To the memory of our friend Rentaro AGEMI}
}
\author{Takashi Ichinose* and Yoshimi Sait$\hbox{{\= o}}${**}
%\thanks{}
\\\normalsize{
*Department of Mathematics, Faculty of Science, Kanazawa University,} 
\\\normalsize{Kanazawa, 920-1192, Japan; 
E-mail: ichinose@kenroku.kanazawa-u.ac.jp}
\\\normalsize{ 
**Department of Mathematics, University of Alabama at Birmingham, }
\\\normalsize{Birmingham, AL 35294, USA; E-mail: ysaito@math.uab.edu}
}

\maketitle

\abstract
{The aim of this paper is to give an extension of the improved Sobolev
embedding theorem for single-valued functions to the case of vector-valued
functions which is involved with the three-dimensional massless Dirac operator 
together with the three- or two-dimensional Weyl--Dirac (or Pauli) operator, 
the Cauchy--Riemann operator and also the four-dimensional Euclidian Dirac 
operator. }

\medskip\noindent
{\bf Mathematics Subject Classification (2010)}: 46E35, 46E40, 35Q40.

\noindent
{\bf Keywords}: Sobolev inequality; Gagliardo--Nirenberg inequality;
improved Sobolev embedding theorem; Dirac-Sobolev inequality; 
(Sobolev inequality for) vector-valued functions; Dirac operator.

%%%%%%%%%%%%%%%%%%%%%%%%%%%%%%%%%%%%%%%%%%%%%%%%%%

\section{Introduction and Results}%sect.1

The improved Sobolev embedding theorem is the following
inequality: For $1\leq p<q<\infty$, there exists a positive constant 
$C$ only depending on $p$ and $q$ (and $n$) such that
\begin{equation}
 \|\psi \|_{q} \leq C\|\nabla \psi\|_{p}^{p/q}
  \|\psi \|_{B^{p/(p-q)}_{\infty,\infty}}^{1-(p/q)}
\end{equation}%\tag1.1
for every ${\Bbb C}$-valued function $\psi$ on ${\Bbb R}^n$ which satisfies
$\nabla \psi \in L^p({\Bbb R}^n)$ and  belongs to the Banach space 
$B^{p/(p-q)}_{\infty,\infty}({\Bbb R}^n)$,
where $\nabla = (\partial_1, \dots, \partial_n),\, 
\partial_j = \partial/\partial x_j,\, j=1,2, \dots, n$. 
Here with $a <0$, $B^{a}_{\infty,\infty}({\Bbb R}^n)$ stands for
the homogeneous Besov space  of indices $(a,\infty,\infty)$ 
with norm
\begin{equation}
 \|\psi \|_{B^{a}_{\infty,\infty}} 
:= \sup_{t>0} t^{-a/2}\|e^{t\Delta}\psi\|_{\infty}
\end{equation}%\tag1.2
(e.g. [T, Sect.2.5.2, pp.190--192]). Here $e^{t\Delta}$ stands for 
the heat semigroup acting on the ${\Bbb C}$-valued functions $\psi$ 
on ${\Bbb R}^n$,
where $\Delta$ is the Laplacian in ${\Bbb R}^n$, 
and
$\|e^{t\Delta}\psi\|_{\infty} := \sup_{x}|(e^{t\Delta} \psi)(x)|$.
This was shown by Cohen et al. [CDPX] 
(cf. Cohen et al. [CMO]) and Ledoux [Le].
In fact,  (1.1) is a very general inequality which covers not only
the classical Sobolev inequalities
$\|\psi\|_q \leq C \|\nabla \psi\|_p$ with 
$\frac1{q}=\frac1{p}-\frac1{n}, \, 1\leq p< n$, 
for every function $\psi$ vanishing at infinity in a certain mild sense, 
but also the Gagliardo--Nirenberg inequalities
\begin{equation}%
 \|\psi\|_q \leq C \|\nabla \psi\|_p^{p/q}\|\psi\|_r^{1-(p/q)},
\quad \frac1{q}=\frac1{p}-\frac{r}{qn}.
\end{equation}%\tag1.3

In all the inequalities the functions $\psi$ 
are supposed to be {\it single-valued} functions. 
In this work we will show an inequality 
like (1.1) for the case where the $\psi$ are {\it vector-valued} functions. 
Of course, inequality (1.1) holds also if one replaces 
single-valued functions $\psi$ by vector-valued functions $f$, understanding 
their semi-norm $\|\nabla f\|_p$ on the right-hand side of (1.1)
in the sense of (1.11) as below. But what we want to have 
is an inequality in the situation where the semi-norm concerned with the 
first-order derivatives is related to the massless Dirac operator 
\begin{equation}%
\alpha\cdot \operatorname{p} 
    = \alpha_1\operatorname{p}_1 
      +\alpha_2\operatorname{p}_2 +\alpha_3\operatorname{p}_3
    = \alpha\cdot(-i\nabla)
    = -i(\alpha_1\partial_1 +\alpha_2\partial_2 +\alpha_3\partial_3),
\end{equation}%\tag1.4
therefore, acting on ${\Bbb C}^4$-valued 
functions $f(x) = {}^t(f_1(x), f_2(x),f_3(x),f_4(x))$
defined in special 3-dimensional space ${\Bbb R}^3$, 
though not in general ${\Bbb R}^n$.
In (1.4), $\alpha := (\alpha_1,\alpha_2,\alpha_3)$ is the triple of 
the $4\times 4$ Dirac matrices which satisfy the anti-commutation relation
$\alpha \alpha_k + \alpha_k \alpha_j = 2\delta_{jk} I_{4}\,\, j,k = 1,2,3$,
where $I_4$ is the $4\times 4$-identity matrix. 
We are concerned mainly with what are usually called ``Dirac matrices":
\begin{equation}%
   \alpha_j =\left(
             \begin{array}{cc}
               0_2& \sigma_j\\
             \sigma_j & 0_2
             \end{array}\right)     \qquad (j= 1, 2, 3)
\end{equation}%\tag1.5
with the $2\times 2$ zero matrix $0_2$ and the triple of $2\times 2$ Pauli
matrices
\begin{equation}%
 \sigma_1 = \left(\begin{array}{cc}
              0& 1 \\ 1& 0 \end{array}\right)\,, \quad 
 \sigma_2 = \left(\begin{array}{cc}
              0& -i \\ i& 0 \end{array}\right)\,, \quad 
 \sigma_3 = \left(\begin{array}{cc}
              1& 0 \\ 0& -1 \end{array}\right)\,.
\end{equation}%\tag1.6

In the beginning let us confirm the notations to be used
about norms for vectors and functions. 
First of all, the absolute value of a number 
$c:= a+ib \in {\Bbb C}$ is denoted, as usual, by $|c| := \sqrt{a^2+b^2}$.
Next, we shall use the standard notations of 
the $\ell^p$ and $\ell^{\infty}$ norm
 for an $m$-vector $a = {}^t (a_1,a_2, \dots,a_m) \in {\Bbb C}^m$ :
\begin{eqnarray}%$$\aligned
 &&|a|_{\ell^p} := (\sum_{k=1}^m |a_k|^p)^{1/p} 
   = (|a_1|^p +|a_2|^p + \dots +|a_m|^p)^{1/p}, 
\quad 1\leq p <\infty, \nonumber\\ 
 &&|a|_{\ell^{\infty}} := \vee_{k=1}^m |a_k| 
= |a_1| \vee |a_2|\vee \cdots \vee|a_m|,
\end{eqnarray}%\endaligned\tag1.7$$
where $b_1 \vee b_2\vee \cdots \vee b_m$ denotes $\max\{b_1,b_2, \dots,b_m\}$. 
The $L^p$ and $L^{\infty}$ norms for a 
${\Bbb C}^m$-valued function $f(x) = {}^t(f_1(x), f_2(x), \dots,f_m(x))$
are given, respectively, by 
\begin{equation}%
 \|f\|_p = \Big(\int |f(x)|_{\ell^p}^p dx\Big)^{1/p}\,, \quad 1\leq p < \infty
; \quad\|f\|_{\infty} = \sup_x |f(x)|_{\ell^{\infty}}, \quad p=\infty.
\end{equation}%\tag1.8

In [IS] we considered the case $m=4$ and introduced the semi-norm
\begin{eqnarray}%$$\aligned
&&\|(\alpha\cdot \operatorname{p})f\|_{p}
= 
\Big(\int|(\alpha\cdot \operatorname{p})f(x)|_{\ell^p}^p dx\Big)^{1/p},
  \qquad 1\leq p <\infty, \nonumber\\
&&|(\alpha\cdot \operatorname{p})f(x)|_{\ell^p}^p
= |\sum_{j=1}^3 \alpha_j \operatorname{p}_jf(x)|_{\ell^p}^p
= \sum_{k=1}^4|(\sum_{j=1}^3 \alpha_j\operatorname{p}_jf)_k(x)|^p
= \sum_{k=1}^4|(\sum_{j=1}^3 \alpha_j\partial_jf)_k(x)|^p\,.\nonumber\\
\end{eqnarray}%\endaligned\tag1.9$$
for $f(x) = {}^t(f_1(x), f_2(x), f_3(x), f_4(x))$ defined on
${\Bbb R}^3$.
The Banach spaces obtained as completion in the norm 
$\|f\|_{\alpha\cdot \operatorname{p},1,p} 
:= (\|f\|_p^p +\|(\alpha\cdot \operatorname{p})f\|_p^p)^{1/p}$
of the linear space $C_0^{\infty}({\Bbb R}^3;{\Bbb C}^4)$ and 
the linear space 
$\{f \in C^{\infty}({\Bbb R}^3;{\Bbb C}^4)\,;\, f,\,
 (\alpha\cdot \operatorname{p})f \in  L^p({\Bbb R}^3;{\Bbb C}^4), j=1,2,3 \}$
 were denoted in [IS] by ${\Bbb H}_0^{1,p}({\Bbb R}^3)$ and 
${\Bbb H}^{1,p}({\Bbb R}^3)$, respectively. However, in the present
paper we denote them by 
$H_{\alpha\cdot \operatorname{p},0}^{1,p}({\Bbb R}^3;{\Bbb C}^4)$ and 
$H_{\alpha\cdot \operatorname{p}}^{1,p}({\Bbb R}^3;{\Bbb C}^4)$, respectively. 

Note that %by H\"older's inequality we have
\begin{equation}%
\|(\alpha\cdot \operatorname{p})f\|_{p} \leq 3^{1-(1/p)}\|\nabla f\|_{p}\,,
\end{equation}%\tag1.10
where 
\begin{equation}
\|\nabla f\|_{p} \equiv \Big(\int|\nabla f(x)|_{\ell^p}^p dx\Big)^{1/p}, \,\,\,
 |\nabla f(x)|_{\ell^p}^p 
 := \sum_{j=1}^3 |\partial_jf(x)|_{\ell^p}^p
 = \sum_{j=1}^3\sum_{k=1}^4|\partial_j f_k(x)|^p\,. 
\end{equation}%\tag1.11
A proof of (1.10) only uses that 
$\|\sum_{j=1}^m \psi_j\|_p \leq m^{1-(1/p)}(\sum_{j=1}^m \|\psi_j\|_p^p)^{1/p}$
for single-valued functions $\psi_j,\, j=1,2,\dots,m$, 
an inequality following from H\"older's inequality.

As is the case for the Sobolev spaces of single-valued functions, 
so does coincidence hold for our Dirac--Sobolev spaces of vector-valued 
functions:
$H_{{\alpha\cdot \operatorname{p}},0}^{1,p}({\Bbb R}^3;{\Bbb C}^4)
= H_{\alpha\cdot \operatorname{p}}^{1,p}({\Bbb R}^3;{\Bbb C}^4)
= W_{\alpha\cdot \operatorname{p}}^{1,p}({\Bbb R}^3;{\Bbb C}^4)$,
where the last space is the Banach space of all 
$f \in L^p({\Bbb R}^3;{\Bbb C}^4)$ such that 
$(\alpha\cdot \operatorname{p})f$ belongs to $L^p({\Bbb R}^3;{\Bbb C}^4)$.
It is shown in [IS] that, for $1<p<\infty$, 
$H_{{\alpha\cdot \operatorname{p}},0}^{1,p}({\Bbb R}^3;{\Bbb C}^4)$
 coincides with 
$H_{0}^{1,p}({\Bbb R}^3;{\Bbb C}^4)$, the completion of 
$C_0^{\infty}({\Bbb R}^3; {\Bbb C}^4)$ in the norm 
$\|f\|_{1,p} := (\|f\|_p^p+ \|\nabla f\|_p^p)^{1/p}$,
while for $p=1$ the latter is a proper subspace of the former. 

With $a<0$, let $B^{a}_{\infty,\infty}({\Bbb R}^n;{\Bbb C}^4)$ be
the homogeneous Besov space  
for ${\Bbb C}^4$-valued functions $f(x)$ on ${\Bbb R}^n$
of indices $(a,\infty,\infty)$ with norm
\begin{equation}%
 \|f\|_{B^{a}_{\infty,\infty}} 
:= \sup_{t>0} t^{-a/2}\|P_tf\|_{\infty}.
\end{equation}%\tag1.12
Here $P_t := e^{t\Delta I_4} = e^{t\Delta}I_4$ 
($I_4: 4\times 4$-identity matrix) stands for 
the heat semigroup acting on the ${\Bbb C}^4$-valued functions 
$f$ on ${\Bbb R}^n$, where $\Delta$ is the Laplacian in ${\Bbb R}^n$, 
$e^{t\Delta}$ being the heat semigroup acting on the ${\Bbb C}$-valued 
functions on ${\Bbb R}^n$,
and
$\|P_t f\|_{\infty} := \sup_{x}|P_t f(x)|_{\ell^{\infty}}
 =  \sup_{x} \vee_{k=1}^4  |e^{t\Delta} f_k(x)|$.

With the notations above concerning vector-valued functions, 
it is easy to see the following {\it trivial} version of (1.1) for
${\Bbb C}^4$-valued functions $f$ holding : For $1\leq p<q<\infty$,
there exists a positive constant $C$ such that
\begin{equation}%
 \|f\|_{q} \leq C\|\nabla f\|_{p}^{p/q}
    \|f\|_{B^{p/(p-q)}_{\infty,\infty}}^{1-(p/q)}
\end{equation}%\tag1.13
for every ${\Bbb C}^4$-valued function $f \in$ 
$B^{a}_{\infty,\infty}({\Bbb R}^n;{\Bbb C}^4)$ which satisfies 
$\|\nabla f\|_p <\infty$, therefore, in particular, for every $f$ 
in the Sobolev space $H_{0}^{1,p}({\Bbb R}^n;{\Bbb C}^n) 
= H({\Bbb R}^n;{\Bbb C}^4) = W^{1,p}({\Bbb R}^n;{\Bbb C}^n)$
as well as in $B^{a}_{\infty,\infty}({\Bbb R}^n;{\Bbb C}^n)$.

%%%%%%%%%%%%%%%%%%%%%%%%%%%
\medskip
Then the first attempt to get a version of (1.1) for vector-valued functions
in our sense was done in the paper [BES] where the authors
showed, replacing the $L^q$ norm of $f$
on its left-hand side by the weak $L^q$ norm of $f$, the following inequality, 
which they called {\it Dirac--Sobolev inequality}
 : For $1\leq p<q<\infty$,  there exists a constant $C>0$ such that  
\begin{equation}%
\|f\|_{q,\infty} \leq C\|(\alpha\cdot \operatorname{p})f\|_{p}^{p/q}
 \|f\|^{1-(p/q)}_{{\Bbb B}^{p/(p-q)}_{\infty,\infty}}
\end{equation}%\tag1.14
for every $f \in  B^{p/(p-q)}_{\infty,\infty}({\Bbb R}^3;{\Bbb C}^4)$
which satisfies
$(\alpha\cdot \operatorname{p})f \in L^p({\Bbb R}^3;{\Bbb C}^4)$,
therefore, in particular, for every 
$\in H_{{\alpha\cdot \operatorname{p}},0}^{1,p}({\Bbb R}^3;{\Bbb C}^4)
\cap B^{p/(p-q)}_{\infty,\infty}({\Bbb R}^3;{\Bbb C}^4)$.
As a result, this $f$ belongs to the weak $L^q$ space 
with the weak $L^q$ norm defined by
\begin{equation}%
 \|f\|_{q,\infty} 
:= \big[\sup_{u>0}\,\, u^q \big|\{|f|_{\ell^{\infty}}\geq u\}\big|\big]^{1/q},
\end{equation}%\tag1.15
where 
$\big|\{|f|_{\ell^{\infty}} \geq u\}\big|
= \int \chi_{\{|f|_{\ell^{\infty}} \geq u\}}(x)\,dx$ 
is the measure of the set 
$\{|f|_{\ell^{\infty}} \geq u\}$ on which 
$u \leq |f(x)|_{\ell^{\infty}} := \vee_{k=1}^4 |f_k(x)|$,
$dx$ being the Lebesgue measure on ${\Bbb R}^3$, and $\chi_E(x)$ stands
for the characteristic function of a subset $E$ of ${\Bbb R}^3$.

Now one may ask oneself whether or not, for any $1\leq p<q<\infty$, 
inequality (1.14) can hold valid, if 
replacing the weak $L^q$ norm of $f$ on the left-hand side by its strong 
$L^q$ one as in the vector-valued version (1.13) of the original (1.1) 
but eqipping on the right-hand side with 
{\it either} the first-order-derivative semi-norm 
$\|(\alpha\cdot \operatorname{p})f\|_{p}$ as in (1.14) {\it or} some 
other one related to the massless Dirac operator 
$\alpha\cdot \operatorname{p}\,$. In particular, we ask whether or not 
there exists a positive constant $C$ such that
\begin{equation}%
\|f\|_{q} \leq C\|(\alpha\cdot \operatorname{p})f\|_{p}^{p/q}
  \|f\|^{1-(p/q)}_{B^{p/(p-q)}_{\infty,\infty}}
\end{equation}%\tag1.16
for every $f \in B^{p/(p-q)}_{\infty,\infty}({\Bbb R}^3;{\Bbb C}^4)$
which satisfies $(\alpha\cdot \operatorname{p})f \in 
L^p({\Bbb R}^3;{\Bbb C}^4)$. 
However, this replacement does not work so well; 
indeed (1.16) cannot hold for $p=1$, although
it holds for $1< p<q<\infty$. A counterexample for this is essentially
found in Balinsky--Evans--Umeda [BEU], 
which we will refer to in Section 2 below.
This suggest us that in order to get an inequality like (1.16) with
the strong $L^q$ norm of $f$ kept on the left-hand side, 
we have to replace the semi-norm $\|(\alpha\cdot \operatorname{p})f\|_p$ 
on the right-hand side by a somewhat stronger one. 
This leads us to introduce a third semi-norm 
$M_{\alpha\cdot \operatorname{p}; p}(f)$ concerned with $L^p$-norm of 
the first-order derivatives of functions $f = {}^t(f_1,f_2,f_3,f_4)$
in the space $C_0^{\infty}({\Bbb R}^3;{\Bbb C}^4)$. Noting that the 
massless Dirac operator (1.4) can be rewritten, 
based on the representations (1.5) of the Dirac matrices 
$\alpha_j,\, j=1,2,3$, as 
\begin{equation}%
\alpha\cdot \operatorname{p}
= \left(\begin{array}{cccc}
0&0& \operatorname{p}_3& \operatorname{p}_1-i\operatorname{p}_2 \\ 
0&0& \operatorname{p}_1+i\operatorname{p}_2 &-\operatorname{p}_3\\
\operatorname{p}_3&\operatorname{p}_1-i\operatorname{p}_2 & 0 &0 \\ 
\operatorname{p}_1+i\operatorname{p}_2 &-\operatorname{p}_3& 0 &0
\end{array}\right)\,,
\end{equation}%\tag1.17
 decompose it into the sum of its two parts:
\begin{eqnarray}%$$\aligned
\alpha\cdot \operatorname{p} 
&=& (\alpha\cdot \operatorname{p})P_{13} 
            + (\alpha\cdot \operatorname{p})P_{24}\nonumber\\
&=& \left(\begin{array}{cccc}
           0&0&\operatorname{p}_3&0 \\ 
           0&0&\operatorname{p}_1+i\operatorname{p}_2&0 \\ 
           \operatorname{p}_3&0&0&0 \\ 
           \operatorname{p}_1+i\operatorname{p}_2&0&0&0 
  \end{array}\right)
+ \left(\begin{array}{cccc}
           0&0&0&\operatorname{p}_1-i\operatorname{p}_2 \\ 
           0&0&0&-\operatorname{p}_3 \\ 
           0&\operatorname{p}_1-i\operatorname{p}_2&0&0 \\ 
           0&-\operatorname{p}_3&0&0 
  \end{array}\right) \,,
\end{eqnarray}%\endaligned\tag1.18$$  
where $P_{13} : = \hbox{\rm diag} (1, 0, 1, 0)$ 
and $P_{24} := \hbox{\rm diag} (0, 1, 0, 1)$ 
are two projection matrices acting on the space ${\Bbb C}^4$
of four-vectors, which satisfies that $P_{13}+P_{24} = I_4$, and define
\begin{equation}%
M_{\alpha\cdot \operatorname{p}; p}(f)
:=\, \big[\,\|(\alpha\cdot \operatorname{p})P_{13}f\|_p^p 
      +\|(\alpha\cdot \operatorname{p})P_{24}f\|_p^p \,\big]^{1/p}.
\end{equation}%\tag1.19
{\it At first sight, this introduction of the semi-norm 
$M_{\alpha\cdot \operatorname{p}; p}(f)$ here may appear
to be artificial but we shall see soon that the semi-norm
turns out to be rather intrinsic.}

Let us see how this semi-norm 
$M_{\alpha\cdot \operatorname{p}; p}(f)$ in (1.19) 
is related to the other
semi-norms, $\|(\alpha\cdot \operatorname{p})f\|_p$ and $\|\nabla f\|_p$.
We have from (1.17) 
$$
(\alpha\cdot \operatorname{p})f
= \left(\begin{array}{cccc}
\operatorname{p}_3f_3+ (\operatorname{p}_1-i\operatorname{p}_2)f_4 \\ 
(\operatorname{p}_1+i\operatorname{p}_2)f_3-\operatorname{p}_3f_4\\
\operatorname{p}_3f_1 +(\operatorname{p}_1-i\operatorname{p}_2)f_2\\ 
(\operatorname{p}_1+i\operatorname{p}_2)f_1-\operatorname{p}_3f_2
\end{array}\right),
$$ 
so that, recalling the definition (1.9) of the $\ell^p$ norm,  we have 
\begin{eqnarray*}%$$\aligned
|(\alpha\cdot \operatorname{p})f|_{\ell^p}^p
&=&|\operatorname{p}_3f_{3}+(\operatorname{p}_1-i\operatorname{p}_2)f_{4}|^p
  +|(\operatorname{p}_1+i\operatorname{p}_2)f_{3}-\operatorname{p}_3f_{4}|^p \\
&&\qquad
+ |\operatorname{p}_3f_{1}+ (\operatorname{p}_1-i\operatorname{p}_2)f_{2}|^p
+|(\operatorname{p}_1+i\operatorname{p}_2)f_{1}-\operatorname{p}_3f_{2}|^p\\
&=& |(\operatorname{p}_1+i\operatorname{p}_2)f_{1}-\operatorname{p}_3f_{2}|^p
+ |(\operatorname{p}_1-i\operatorname{p}_2)f_{2}+\operatorname{p}_3f_{1}|^p\\
 &&\qquad
+|(\operatorname{p}_1+i\operatorname{p}_2)f_{3}-\operatorname{p}_3f_{4}|^p
 +|(\operatorname{p}_1-i\operatorname{p}_2)f_{4}+\operatorname{p}_3f_{3}|^p,
\end{eqnarray*}%\endaligned$$
where we have rearranged the four terms, when passing through the second 
equality. Hence
\begin{eqnarray}%$$\aligned
\|(\alpha\cdot \operatorname{p})f\|_p^p
&=& \|(\partial_1+i\partial_2)f_{1}-\partial_3f_{2}\|_p^p
  + \|(\partial_1-i\partial_2)f_{2}+\partial_3f_{1}\|_p^p \nonumber\\
 &&+\|(\partial_1+i\partial_2)f_{3}-\partial_3f_{4}\|_p^p
  +\|(\partial_1-i\partial_2)f_{4}+\partial_3f_{3}\|_p^p\,. 
\end{eqnarray}%\endaligned\tag1.20$$
Then one can calculate the right-hand side of (1.19) to get
\begin{eqnarray}%$$\aligned
M_{\alpha\cdot \operatorname{p}; p}(f)^p 
\!&\!=\!&\!
 \big[\big(\|\partial_3 f_3\|_p^p+\|(\partial_1+i\partial_2) f_3\|_ p^p\big) 
 +\big(\|\partial_3 f_1\|_p^p+\|(\partial_1+i\partial_2) f_1\|_ p^p\big)\big]
   \nonumber\\
&&+\big[\big(\|(\partial_1-i\partial_2) f_4\|_p^p 
                         +\|\partial_3 f_3\|_ p^p\big)
 + \big(\|(\partial_1-i\partial_2) f_2\|_p^p 
                         +\|\partial_3 f_2\|_ p^p\big)\big] \nonumber\\
\!&\!=\!&\! 
  \big(\|(\partial_1+i\partial_2) f_1\|_p^p+\|\partial_3 f_1\|_ p^p\big) 
  +\big(\|(\partial_1-i\partial_2) f_2\|_p^p+\|\partial_3 f_2\|_ p^p\big)
   \nonumber\\
 &&+\big(\|(\partial_1+i\partial_2) f_3\|_p^p +\|\partial_3 f_3\|_ p^p\big)
   +\big(\|(\partial_1-i\partial_2) f_4\|_p^p +\|\partial_3 f_4\|_ p^p\big).
\end{eqnarray}%\endaligned\tag1.21$$

%%%%%%%%%%%%
We can compare (1.20) and (1.21) and recall (1.10) to show with aid of 
H\"older's inequality that for $1\leq p < \infty$,
\begin{equation}%
2^{-(1-(1/p))} \|(\alpha\cdot \operatorname{p})f\|_{p}
\leq M_{\alpha\cdot \operatorname{p}; p}(f) \,
\leq\, 2^{1-(1/p)} \|\nabla f\|_{p}\,, 
\end{equation}%\tag1.22
so that the semi-norm 
$M_{\alpha\cdot \operatorname{p}; p}(f)$ is an intermediate one
in strength lying between the other two first-order-derivative semi-norms 
$\|(\alpha\cdot \operatorname{p})f\|_{p}$ and $\|\nabla f\|_{p}$.
We shall denote by 
$H_{M_{\alpha\cdot \operatorname{p}},0}^{1,p}({\Bbb R}^3;{\Bbb C}^4)$
the Banach space obtained as completion in the norm 
$\|f\|_{M_{\alpha\cdot \operatorname{p}},1,p} 
:= (\|f\|_p^p +M_{\alpha\cdot \operatorname{p}; p}(f)^p)^{1/p}$
of the space $C_0^{\infty}({\Bbb R}^3;{\Bbb C}^4)$. 
>From (1.22) we see the following inclusion relation among the three
Banach spaces:
\begin{equation}%
 H_{0}^{1,p}({\Bbb R}^3;{\Bbb C}^4) \quad \subseteq\quad  
 H_{M_{\alpha\cdot \operatorname{p}},0}^{1,p}({\Bbb R}^3;{\Bbb C}^4)
 \quad\subseteq\quad 
 H_{({\alpha\cdot \operatorname{p}}),0}^{1,p}({\Bbb R}^3;{\Bbb C}^4).
\end{equation}%\tag1.23

Now we are going to see a significant character of 
the semi-norm $M_{\alpha\cdot \operatorname{p}; p}(f)$ introduced in (1.19),
by considering the other decompositions of the Dirac opearator 
$\alpha\cdot \operatorname{p}$ in (1.17) than  the one (1.18). In fact, 
there are a few other decompositions:

\medskip\noindent
$\underline{M_{\alpha}^{(1)}}$
\begin{eqnarray}%$$\aligned
\alpha\cdot \operatorname{p}
\!&\!=\!&\! 
(\alpha\cdot \operatorname{p})P_{14} +(\alpha\cdot \operatorname{p})P_{23}
  \nonumber\\
\!&\!\equiv\!&\! \left(\begin{array}{cccc}
0&0&0& \operatorname{p}_1-i\operatorname{p}_2 \\ 
0&0&0&-\operatorname{p}_3\\
\operatorname{p}_3&0&0&0 \\ 
\operatorname{p}_1+i\operatorname{p}_2 &0&0&0
\end{array}\right)
+ \left(\begin{array}{cccc}
0&0& \operatorname{p}_3&0\\ 
0&0& \operatorname{p}_1+i\operatorname{p}_2 &0\\
0&\operatorname{p}_1-i\operatorname{p}_2 &0&0\\ 
0&-\operatorname{p}_3&0&0
\end{array}\right),\,\,
\end{eqnarray}%\endaligned\tag1.24$$ 
where $P_{14} : = \hbox{\rm diag} (1, 0, 0, 1)$ 
and $P_{23} := \hbox{\rm diag} (0, 1, 1, 0)$ 
are two projection matrices acting on the space ${\Bbb C}^4$
of four-vectors, so that $P_{14}+P_{23} = I_4$. Note that both the 
operators $(\alpha\cdot \operatorname{p})P_{14}$ and 
$(\alpha\cdot \operatorname{p})P_{23}$ on the right are selfadjoint,
i.e. 
$((\alpha\cdot \operatorname{p})P_{14})^* 
      = (\alpha\cdot \operatorname{p})P_{14},\,\,
 ((\alpha\cdot \operatorname{p})P_{23})^*
      = (\alpha\cdot \operatorname{p})P_{23}$.

\medskip\noindent
$\underline{M_{\alpha}^{(2)}}$
\begin{eqnarray}%$$\aligned
\alpha\cdot \operatorname{p}
&=\,& \left(\begin{array}{cc}
             0_2&\sigma_1\operatorname{p}_1+\sigma_2\operatorname{p}_2\\
            \sigma_3\operatorname{p}_3&0_2
   \end{array}\right)
  +\left(\begin{array}{cc} 
            0_2&\sigma_3\operatorname{p}_3\\
            \sigma_1\operatorname{p}_1+\sigma_2\operatorname{p}_2&0_2
   \end{array}\right)\nonumber\\
&\equiv\,& \left(\begin{array}{cccc}
0&0&0&\operatorname{p}_1-i\operatorname{p}_2 \\ 
0&0&\operatorname{p}_1+i\operatorname{p}_2 &0\\
\operatorname{p}_3&0 & 0 &0 \\ 
0&-\operatorname{p}_3& 0 &0\end{array}\right) 
+ \left(\begin{array}{cccc}
0&0&\operatorname{p}_3&0 \\ 
0&0&0&-\operatorname{p}_3\\
0&\operatorname{p}_1-i\operatorname{p}_2 & 0 &0 \\ 
\operatorname{p}_1+i\operatorname{p}_2 &0& 0 &0\end{array}\right) 
     \nonumber\\
&=:& (\alpha\cdot \operatorname{p})_1 + (\alpha\cdot \operatorname{p})_2,
\end{eqnarray}%\endaligned\tag1.25$$ 
where note that $(\alpha\cdot \operatorname{p})_2$ is the adjoint of
$(\alpha\cdot \operatorname{p})_1$ as operators, say,  
in $L^2({\bold R}^3;{\Bbb C}^4)$, i.e. 
$(\alpha\cdot \operatorname{p})_2 = {(\alpha\cdot \operatorname{p})_1}^*$.
 
\medskip\noindent
$\underline{M_{\alpha}^{(3)}}$  
\begin{eqnarray}%$$\aligned
\alpha\cdot \operatorname{p}
&=\,&\left(\begin{array}{cc}
0&\sigma_1\operatorname{p}_1+\sigma_2\operatorname{p}_2 \\ 
\sigma_1\operatorname{p}_1+\sigma_2\operatorname{p}_2&0
\end{array}\right) 
+ \left(\begin{array}{cccc} 
0&\sigma_3\operatorname{p}_3\\ 
\sigma_3\operatorname{p}_3&0
\end{array}\right) \nonumber\\
&\equiv\,& \left(\begin{array}{cccc}
0&0&0&\operatorname{p}_1-i\operatorname{p}_2 \\ 
0&0&\operatorname{p}_1+i\operatorname{p}_2 &0\\
0&\operatorname{p}_1-i\operatorname{p}_2 & 0 &0 \\ 
\operatorname{p}_1+i\operatorname{p}_2 &0& 0 &0
\end{array}\right) 
+ \left(\begin{array}{cccc}
0&0&\operatorname{p}_3&0\\ 
0&0&0&-\operatorname{p}_3\\
\operatorname{p}_3&0&0& \\ 
0&-\operatorname{p}_3& &0
\end{array}\right) \nonumber\\
&=:& (\alpha\cdot \operatorname{p})_3 +(\alpha\cdot \operatorname{p})_4\,,
\end{eqnarray}%\endaligned\tag1.26$$ 
where note that both the operators 
$(\alpha\cdot \operatorname{p})_3$ and 
$(\alpha\cdot \operatorname{p})_4$ on the right are selfadjoint.

%%%%%%%%%%%%%%%%%%
\medskip
Then we can see in the following proposition that
the semi-norm $M_{\alpha\cdot \operatorname{p}; p}(f)$ of $f 
\in C_0^{\infty}({\Bbb R}^3;{\Bbb C}^4)$ defined  by (1.19), 
though with the rather artificial decomposition (1.18) dependent
on the pair $(P_{13}, P_{24})$ of projection matrices, 
turns out to be meaningful enough to have some universal character.

%%%%
\bigskip\noindent
{\bf Proposition 1.0} %Proposition 1.0
{\it The semi-norm $M_{\alpha\cdot \operatorname{p}; p}(f)$ 
in $\hbox{\rm (1.19)}$ coincides with the ones to be defined with 
the decompositions  \hbox{\rm (1.24)}, \hbox{\rm  (1.25)} 
and \hbox{\rm (1.26)}:
\begin{subequations}
\begin{align}
M^{(1)}_{\alpha\cdot \operatorname{p}; p}(f)
&:=  [\|(\alpha\cdot \operatorname{p})P_{14} f\|_p^p 
    + \|(\alpha\cdot \operatorname{p})P_{23} f\|^p]^{1/p}; \\%\tag1.27a
M^{(2)}_{\alpha\cdot \operatorname{p}; p}(f)
&:= [\|(\alpha\cdot \operatorname{p})_1 f\|_p^p 
    + \|(\alpha\cdot \operatorname{p})_2 f\|^p]^{1/p}
=  [\|(\alpha\cdot \operatorname{p})_1 f\|_p^p 
    + \|{(\alpha\cdot \operatorname{p})_1}^* f\|^p]^{1/p}; \\%\tag1.27b 
M^{(3)}_{\alpha\cdot \operatorname{p}; p}(f)
&:= [\|(\alpha\cdot \operatorname{p})_3 f\|_p^p 
    + \|(\alpha\cdot \operatorname{p})_4 f\|^p]^{1/p}. %\tag1.27c
\end{align}
\end{subequations}
More generally, in fact, every decomposition of 
$\alpha\cdot \operatorname{p}$ into its two parts,  
$\alpha\cdot \operatorname{p} 
= (\alpha\cdot \operatorname{p})_5+ (\alpha\cdot \operatorname{p})_6$,
such that each row of both the matrices $(\alpha\cdot \operatorname{p})_5$
and $(\alpha\cdot \operatorname{p})_6$ contains only one nonzero entry,
defines the semi-norm $M_{\alpha\cdot \operatorname{p}; p}(f)$ which has
the expression \hbox{\rm (1.21)}. 
}

\bigskip
{\it Proof}.  In fact, direct calculation of the right-hand sides of
(1.27a), (1.27b) and (1.27c)
in view of (1.24), (1.25) and (1.26) yields
nothing but a rearrangement of the last member of the expression (1.21)
of $M_{\alpha\cdot \operatorname{p}; p}(f)$. The assertion for the more 
general case is evident.
\qed 

\medskip
We note that (1.27a) says that our semi-norm
$M_{\alpha\cdot \operatorname{p}; p}(f)$, which is defined in (1.19) 
by using the pair $(P_{13}, P_{24})$ of projection matrices 
can be defined by using another pair $(P_{14}, P_{23})$.
However, among all three possible pairs of  projection matrices,
$(P_{13}, P_{24})$, $(P_{14}, P_{23})$, $(P_{12}, P_{34})$, whose sum 
becomes  the identity matrix $I_4$, the decomposition 
$\alpha\cdot \operatorname{p} 
= (\alpha\cdot \operatorname{p})P_{12} 
    +(\alpha\cdot \operatorname{p})P_{34}
$
to be defined with the remaining last pair consisting 
of $P_{12} = \hbox{\rm diag} (1, 1, 0, 0)$ and 
$P_{34}  =\hbox{\rm diag} (0, 0, 1, 1)$, 
is not fit for our semi-norm $M_{\alpha\cdot \operatorname{p}; p}(f)$, 
since this decomposition does not satisfy the condition for the more general 
case in Proposition 1.0. 
In Section 6, we shall come back to this decomposition to discuss 
the issue.

%%%%%%%%%%%%%
\medskip
The main result of this work is the following theorem.

\begin{thm}%{Theorem 1.1}
{\rm (with 3-dimensional massless Dirac operator)}
\hbox{\rm (i)} 
For $1\leq p<q<\infty$, a ${\Bbb C}^4$-valued function 
$f = {}^t(f_1,f_2,f_3,f_4)$ belongs to $L^q({\Bbb R}^3;{\Bbb C}^4)$, 
if
 $f$ belongs to $B^{p/(p-q)}_{\infty,\infty}({\Bbb R}^3;{\Bbb C}^4)$
and satisfies $M_{\alpha\cdot \operatorname{p}; p}(f) < \infty$, 
and further, there exists a positive constant $C$ such that
\begin{equation}%
 \|f\|_{q} \leq C M_{\alpha\cdot \operatorname{p}; p}(f)^{p/q}
  \|f\|_{B^{p/(p-q)}_{\infty,\infty}}^{1-(p/q)}\,.
\end{equation}%\tag1.28
Therefore this holds, in particular, for every $f \in 
H_{M_{\alpha\cdot \operatorname{p}},0}^{1,p}({\Bbb R}^3;
{\Bbb C}^4)
\cap B^{p/(p-q)}_{\infty,\infty}({\Bbb R}^3;{\Bbb C}^4)$.

\medskip\noindent
\hbox{\rm (ii)} 
For $\infty >p>1$, the three semi-norms 
$\|(\alpha\cdot \operatorname{p})f\|_{p}$, 
$M_{\alpha\cdot \operatorname{p}; p}(f)$ and
$\|\nabla f\|_p$ are equivalent, so that the corresponding three
Banach spaces in \hbox{\rm (1.23)} coincide with one another:
\begin{equation}%
 H_0^{1,p}({\Bbb R}^3;{\Bbb C}^4) \quad =\quad 
  H_{M_{\alpha\cdot \operatorname{p}},0}^{1,p}({\Bbb R}^3;{\Bbb C}^4)
 \quad=\quad  
H_{({\alpha\cdot \operatorname{p}}),0}^{1,p}({\Bbb R}^3;{\Bbb C}^4).
\end{equation}%\tag1.29
Therefore assertion \hbox{\rm (i)} turns out: For $1<p<q<\infty$, 
there exists a positive constant $C$ such that
\begin{equation}%
 \|f\|_{q} \leq C \|(\alpha\cdot \operatorname{p})f\|_{p}^{p/q}
  \|f\|_{B^{p/(p-q)}_{\infty,\infty}}^{1-(p/q)},
\end{equation}%\tag1.30
for every $f \in B^{p/(p-q)}_{\infty,\infty}({\Bbb R}^3;{\Bbb C}^4)$
whose semi-norm $\|(\alpha\cdot \operatorname{p})f\|_{p}$, 
$M_{\alpha\cdot \operatorname{p}; p}(f)$ or
$\|\nabla f\|_p$ is finite. Therefore this holds, in particular,
for every $f$ in the above space \hbox{\rm (1.29)} which belongs to
$B^{p/(p-q)}_{\infty,\infty}({\Bbb R}^3;{\Bbb C}^4)$. The inequality 
\hbox{\rm (1.30)} is equivalent to the vector-valued version  \hbox{\rm (1.13)} 
of \hbox{\rm (1.1)} with $n=3$. 
\end{thm}

%%%%%%
\medskip
Similarly we can also show the following five results in related
different situations.

First, replacing the Dirac operator $\alpha\cdot \operatorname{p}$ 
in Theorem 1.1 by the 3-dimensional Weyl--Dirac (or Pauli) operator 
\begin{equation}%
\sigma\cdot\operatorname{p}
:= \sigma_1 \operatorname{p}_1 + \sigma_2 \operatorname{p}_2 
   +\sigma_3 \operatorname{p}_3
= \left(\begin{array}{cc}
  \operatorname{p}_3& \operatorname{p}_1 -i\operatorname{p}_2\\
  \operatorname{p}_1 +i\operatorname{p}_2& -\operatorname{p}_3 
\end{array}\right)
\end{equation}%\tag 1.31 
acting on ${\Bbb C}^2$-valued $C^{\infty}$ function $h := {}^t (h_1,h_2)$
on ${\Bbb R}^3$, where the $\sigma_j,\, j=1,2,3$, are the Pauli 
matrices in (1.6), we have exactly the same result.
For  $h := {}^t (h_1,h_2)$ whose four first-order derivatives
$(\partial_1+i\partial_2)h_1$,  $\partial_3 h_1$, 
$(\partial_1-i\partial_2)h_2$ and $\partial_3 h_2$  are $p$-th power
integrable in ${\Bbb R}^3$, consider the semi-norm 
\begin{eqnarray}%$$\aligned
M_{\sigma\cdot\operatorname{p}; p}(h)  
\!\!&:=&\!\! \big[\|(\sigma\cdot\operatorname{p})P_{1}h\|_p^p 
      +\|(\sigma\cdot\operatorname{p})P_{2}h\|_p^p \big]^{1/p} \nonumber\\ 
\!\!&\,=&\!\! \big[\|(\partial_1+i\partial_2)h_1\|_p^p+\|\partial_3 h_1\|_p^p
    +\|(\partial_1-i\partial_2)h_2\|_p^p+\|\partial_3 h_2\|_p^p\big]^{1/p}, \qquad
\end{eqnarray}%\endaligned\tag1.32$$
decomposing $\sigma\cdot\operatorname{p}$ into the sum of its two parts:
$$
\sigma\cdot\operatorname{p}
= (\sigma\cdot\operatorname{p})P_1 +(\sigma\cdot\operatorname{p})P_2
= \left(\begin{array}{cc} 
   \operatorname{p}_3&0\\
   \operatorname{p}_1 +i\operatorname{p}_2&0 \end{array}\right)
 +\left(\begin{array}{cc} 
               0& \operatorname{p}_1 -i\operatorname{p}_2\\
               0& -\operatorname{p}_3 \end{array}\right).
$$
Here  $P_{1}= \left(\begin{array}{cc} 1&0\\ 0&0 \end{array}\right)$ and
$P_{2}= \left(\begin{array}{cccc} 0&0\\ 0&1 \end{array}\right)$ are
two projection matrices acting on the ${\Bbb C}^2$ of two-vectors
and note that $P_{1}+P_{2} = I_2$ ($: 2\times 2$-identity matrix).  
By the same argument as before around Proposition 1.0 for 
$\alpha\cdot \operatorname{p}$, it is also seen that this semi-norm 
$M_{\sigma\cdot\operatorname{p}; p}(h)$ defined by (1.32) with the 
decomposition (1.31) of $\sigma\cdot\operatorname{p}$ coincides with 
the one to be defined with another decomposition: 
$$\sigma\cdot\operatorname{p}
= \left(\begin{array}{cc} 0&\operatorname{p}_1 -i\operatorname{p}_2\\
  \operatorname{p}_1 +i\operatorname{p}_2&0 \end{array}\right)
+ \left(\begin{array}{cc} \operatorname{p}_3&0\\
  0&-\operatorname{p}_3 \end{array}\right)
=: (\sigma\cdot\operatorname{p})_1 +(\sigma\cdot\operatorname{p})_2,
$$
i.e. 
$M_{\sigma\cdot\operatorname{p}; p}(h)  
=\,\big[\|(\sigma\cdot\operatorname{p})_1h\|_p^p 
      +\|(\sigma\cdot\operatorname{p})_2h\|_p^p \big]^{1/p}$.

The Banach spaces obtained as completions 
of $C_0^{\infty}({\Bbb R}^3; {\Bbb C}^2)$ 
by the norms
$\|h\|_{M_{\sigma\cdot \operatorname{p}},1,p} 
:= (\|h\|_p^p + M_{\sigma\cdot \operatorname{p}; p}(h)^p)^{1/p}$ and
$\|h\|_{{\sigma\cdot \operatorname{p}},1,p} 
:= (\|h\|_p^p + \|(\sigma\cdot \operatorname{p})h\|_p^p)^{1/p}$
are denoted by  
$H_{M_{\sigma\cdot\operatorname{p}},0}^{1,p}({\Bbb R}^3;{\Bbb C}^2)$,
$H_{(\sigma\cdot\operatorname{p}),0}^{1,p}({\Bbb R}^3;{\Bbb C}^2)$, 
respectively.

\begin{cor} %{Corollary 1.3} 
{\rm (with 3-dimensional Weyl--Dirac operator) 
(i)} For $1\leq p<q<\infty$,  a ${\Bbb C}^2$-valued functions 
$h = {}^t(h_1,h_2)$ belongs to $L^q({\Bbb R}^3;{\Bbb C}^2)$, 
if $h$ belongs to $B^{p/(p-q)}_{\infty,\infty}({\Bbb R}^3;{\Bbb C}^2)$
and satisfies $M_{\sigma\cdot\operatorname{p}; p}(h) < \infty$,
and further, there exists a positive constant $C$ such that
\begin{equation}%
 \|h\|_{q} \leq C M_{\sigma\cdot\operatorname{p}; p}(h) ^{p/q}
  \|h\|_{B^{p/(p-q)}_{\infty,\infty}}^{1-(p/q)}\,.
\end{equation}%\tag1.33
Therefore this holds, in particular, for $h \in
H_{M_{\sigma\cdot\operatorname{p}},0}^{1,p}({\Bbb R}^3;{\Bbb C}^2)
\cap B^{p/(p-q)}_{\infty,\infty}({\Bbb R}^3;{\Bbb C}^2)$.

\medskip\noindent
\hbox{\rm (ii)}  For $\infty>p>1$, the three semi-norms 
$\|(\sigma\cdot\operatorname{p})h\|_p$, 
$M_{\sigma\cdot\operatorname{p}; p}(h)$ and
$\|\nabla h\|_p$ are equivalent, so that the corresponding three
Banach spaces coincide with one another:
\begin{equation}%
 H_0^{1,p}({\Bbb R}^3;{\Bbb C}^2) \quad =\quad 
  H_{M_{\sigma\cdot\operatorname{p}},0}^{1,p}({\Bbb R}^3;{\Bbb C}^2)
 \quad=\quad  
H_{\sigma\cdot\operatorname{p},0}^{1,p}({\Bbb R}^3;{\Bbb C}^2).
\end{equation}%\tag1.34
Therefore assertion \hbox{\rm (i)} turns out: For $1<p<q<\infty$, 
there exists a positive constant $C$ such that
\begin{equation}%
 \|h\|_{q} \leq C \|(\sigma\cdot\operatorname{p})h\|_{p}^{p/q}
  \|h\|_{B^{p/(p-q)}_{\infty,\infty}}^{1-(p/q)},
\end{equation}%\tag1.35
for every $h \in B^{p/(p-q)}_{\infty,\infty}({\Bbb R}^3;{\Bbb C}^2)$
whose semi-norm $\|(\sigma\cdot\operatorname{p})h\|_{p}$, 
$M_{\sigma\cdot\operatorname{p}; p}(h)$ or
$\|\nabla h\|_p$ is finite.
Therefore this holds, in particular, for every $f$ in the space 
\hbox{\rm (1.34)} which belongs to 
$B^{p/(p-q)}_{\infty,\infty}({\Bbb R}^3;{\Bbb C}^2)$. The inequality 
\hbox{\rm (1.35)} is equivalent to the vector-valued version  
\hbox{\rm (1.13)} of \hbox{\rm (1.1)} with $n=3$. 
\end{cor}

Second, for ${\Bbb C}$-valued $C^{\infty}$ functions $\psi$
whose two first-order derivatives 
$(\partial_1-i\partial_2) \psi$ and $\partial_3 \psi$
are $p$-th power integrable in ${\Bbb R}^3$, consider the semi-norm 
\begin{equation}%
M_{(\partial_1-i\partial_2)\vee \partial_3; p}(\psi)  
  := \big[\|(\partial_1-i\partial_2) \psi\|_p^p 
                  +\|\partial_3 \psi\|_p^p\big]^{1/p}. 
\end{equation}%\tag1.36
The Banach space obtained as completion of 
$C_0^{\infty}({\Bbb R}^3)$ by the norm 
$\|\psi\|_{M_{(\partial_1-i\partial_2)\vee \partial_3},1,p} 
:= (\|\psi\|_p^p
  + M_{((\partial_1-i\partial_2)\vee \partial_3);p}(\psi)^p)^{1/p}$
is denoted by
$H_{M_{(\partial_1-i\partial_2)\vee \partial_3},0}^{1,p}({\Bbb R}^3)$.

\bigskip\noindent
\begin{cor}%{Corollary 1.3} 
{\rm (i)} For $1\leq p<q<\infty$, a function $\psi$ belongs to 
$L^q({\Bbb R}^3)$, if $\psi$ belongs to 
$B^{p/(p-q)}_{\infty,\infty}({\Bbb R}^3)$
and satisfies 
$M_{(\partial_1-i\partial_2)\vee \partial_3; p}(\psi)< \infty$,
and further, there exists a positive constant $C$ such that
\begin{equation}%
\|\psi\|_{q} \leq 
   C\,M_{(\partial_1-i\partial_2)\vee \partial_3; p}(\psi)^{p/q}
  \|\psi\|_{B^{p/(p-q)}_{\infty,\infty}}^{1-(p/q)}\,.
\end{equation}%\tag1.37
Therefore, in particular, for every $\psi \in
H_{M_{(\partial_1-i\partial_2)\vee \partial_3},0}^{1,p}({\Bbb R}^3)
\cap B^{p/(p-q)}_{\infty,\infty}({\Bbb R}^3)$. 

\medskip\noindent
\hbox{\rm (ii)}  For $\infty>p>1$, the two semi-norms
$M_{(\partial_1-i\partial_2)\vee \partial_3; p}(\psi)$ and
$\|\nabla f\|_p$ are equivalent, so that the corresponding two
Banach spaces coincide with each other:
\begin{equation}%
 H_0^{1,p}({\Bbb R}^3;{\Bbb C}^2) \quad =\quad 
  H_{M_{(\partial_1-i\partial_2)\vee \partial_3},0}^{1,p}
  ({\Bbb R}^3;{\Bbb C}^2).
\end{equation}%\tag1.38
Therefore assertion \hbox{\rm (i)} turns out: For $1<p<q<\infty$, 
there exists a positive constant $C$ such that
\begin{equation}%
 \|\psi\|_{q} \leq 
C M_{(\partial_1-i\partial_2)\vee \partial_3; p}(\psi)^{p/q}
  \|\psi\|_{B^{p/(p-q)}_{\infty,\infty}}^{1-(p/q)},
\end{equation}%\tag1.39
for every $f \in B^{p/(p-q)}_{\infty,\infty}({\Bbb R}^3)$
whose semi-norm $M_{(\partial_1-i\partial_2)\vee \partial_3; p}(\psi)$ 
or $\|\nabla f\|_p$ is finite. Therefore this holds, in particular,
for every $f$ in the space \hbox{\rm (1.38)} which belongs to  
$B^{p/(p-q)}_{\infty,\infty}({\Bbb R}^3)$. The inequality 
\hbox{\rm (1.39)} is equivalent to the vector-valued version
 \hbox{\rm (1.13)} of \hbox{\rm (1.1)} with $n=2$. 
\end{cor}

%%%%%%%%%%%%%%%%%
Third, we shall consider the {\it two}-dimensional 
Weyl--Dirac (or Pauli) operators made from two of the three Pauli matrices 
(1.6). There are the following three:
\begin{subequations}
\begin{align}
&\!\!\!\!\!\!\!(\sigma\cdot\operatorname{p})^{(a)}f 
:= (\sigma_1 \operatorname{p}_1 + \sigma_2 \operatorname{p}_2)f 
= \left(\begin{array}{cc} 
  0 \!&\! \operatorname{p}_1-i\operatorname{p}_2\\ 
  \operatorname{p}_1+i\operatorname{p}_2 \!&\! 0\end{array}\right)
 \!\left(\begin{array}{c} f_1 \\ f_2\end{array}\right),      \\%\tag1.40a
%%%
&\!\!\!\!\!\!\!(\sigma\cdot\operatorname{p})^{(b)}f
:=(\sigma_3 \operatorname{p}_1 + \sigma_1 \operatorname{p}_2)f 
= \left(\begin{array}{cc} 
   \operatorname{p}_1 \!&\! \operatorname{p}_2 \\ 
   \operatorname{p}_2 \!&\! -\operatorname{p}_1\end{array}\right)
  \!\left(\begin{array}{c} f_1 \\ f_2\end{array}\right),      \\%\tag1.40b
%%%
&\!\!\!\!\!\!\!(\sigma\cdot\operatorname{p})^{(c)}f 
:= (\sigma_3 \operatorname{p}_1 + \sigma_2 \operatorname{p}_2)f 
= \left(\begin{array}{cc} \operatorname{p}_1 \!&\! -i\operatorname{p}_2 
\\ i\operatorname{p} \!&\! -\operatorname{p}_1\end{array}\right)
  \!\left(\begin{array}{c} f_1 \\ f_2\end{array}\right),   %\tag1.40c
\end{align}
\end{subequations}
for $f := {}^t(f_1,f_2)$. 
As we shall see later in Lemma 5.1, these three operators
$(\sigma\cdot\operatorname{p})^{(a)}, \, 
(\sigma\cdot\operatorname{p})^{(b)},\, 
(\sigma\cdot\operatorname{p})^{(c)}$ are unitarily equivalent,
so that the three semi-norms 
$\|(\sigma\cdot\operatorname{p})^{(a)}f\|_p$, 
$\|(\sigma\cdot\operatorname{p})^{(b)}f\|_p$, 
$\|(\sigma\cdot\operatorname{p})^{(c)}f\|_p$
are equivalent. Therefore we write any of these three operators
as $(\sigma\cdot\operatorname{p})^{(2)}$ so as to distinguish it 
from the {\it three}-dimensional Weyl--Dirac (or Pauli) operator 
$\sigma\cdot\operatorname{p}$ in (1.31), and
any of these semi-norms as 
$\|(\sigma\cdot\operatorname{p})^{(2)}f\|_p$ to consider
the norm $\|f\|_{(\sigma\cdot\operatorname{p})^{(2)},1,p} 
:= (\|f\|_p^p+ \|(\sigma\cdot\operatorname{p})^{(2)}f\|_p^p)^{1/p}$. 
What can be shown just in the same way as in [IS] is that the Banach space 
$H_{(\sigma\cdot\operatorname{p})^{(2)},0}^{1,p}({\Bbb R}^2; {\Bbb C}^2)$
obtained as completion of 
$C_0^{\infty}({\Bbb R}^2; {\Bbb C}^2)$ in this norm 
coincides for $1<p<\infty$ with the Sobolev spaces 
$H_0^{1,p}({\Bbb R}^2; {\Bbb C}^2)$ 
= $H^{1,p}({\Bbb R}^2; {\Bbb C}^2)$, but is for $p=1$ strictly larger.
Differing from Corollary 1.2 for 3-dimensional case, 
the following theorem for 2-dimensional case gives a {\it true} extension 
of inequality (1.1) for single-valued functions to the case for   
vector-valued functions. 

\begin{thm}%{Theorem 1.4} 
{\rm (with 2-dimensional Weyl--Dirac (or Pauli) operator)} 
For $1\leq p<q<\infty$ there exists a positive 
constant $C$ such that
\begin{equation}%
 \|f \|_{q} \leq C \|(\sigma\cdot\operatorname{p})^{(2)}f\|_p^{p/q}
  \|f \|_{B^{p/(p-q)}_{\infty,\infty}}^{1-(p/q)}
\end{equation}%\tag1.41
for every $f \in B^{p/(p-q)}_{\infty,\infty}({\Bbb R}^2; {\Bbb C}^2)$
which satisfies $\|(\sigma\cdot\operatorname{p})^{(2)}f\|_p < \infty$.
Therefore this holds, in particular, for every 
$f \in H_{(\sigma\cdot\operatorname{p})^{(2)},0}^{1,p}({\Bbb R}^2; {\Bbb C}^2)
\cap B^{p/(p-q)}_{\infty,\infty}({\Bbb R}^2; {\Bbb C}^2)$.
\end{thm}

Forth, from Corollary 1.3 or Theorem 1.4 we can get the following
inequality involved with the Cauchy--Riemman operator
$\frac12 (\partial_1+ i\partial_2)$ in ${\Bbb R}^2$. 

\begin{cor}%{Corollary 1.5} 
{\rm (with Cauchy--Riemann operator)}
For $1\leq p < q <\infty$, there exists
a positive constant $C$ such that
\begin{equation}%
 \|\psi\|_q \leq C \|(\partial_1 + i\partial_2)\psi\|_p^{p/q}
  \|\psi\|^{1-(p/q)}_{B_{\infty,\infty}^{p/(p-q)}}
\end{equation}%\tag1.42
for every $\psi \in B_{\infty,\infty}^{p/(p-q)}({\Bbb R}^2)$
which satisfies $\|(\partial_1 + i\partial_2)\psi\|_p < \infty$.
\end{cor}

\bigskip
Finally, we are going to consider the 
four-dimensional Euclidian Dirac operator
\begin{equation}%
\beta\cdot \operatorname{p}
    = \sum_{k=1}^4\beta_k \cdot \operatorname{p}_k 
    = -i\sum_{k=1}^4 \beta_k \partial_k,
\end{equation}%\tag1.43
with $ \operatorname{p}
= (\operatorname{p}_1,\operatorname{p}_2,\operatorname{p}_3,
\operatorname{p}_4),\,\,\operatorname{p}_k =-i\partial_k,\,\, k=1,2,3,4$,
which acts on ${\Bbb C}^4$-valued 
functions $f(x) = {}^t(f_1(x), f_2(x),f_3(x),f_4(x))$
defined in 4-dimensional Euclidian space-time ${\Bbb R}^4$.
Here we are using the symbol $\beta$ for a quadruple 
$\beta := (\beta_1,\beta_2,\beta_3, \beta_4)$ 
of the Dirac matrices which are $4\times 4$ Hermitian matrices
 satisfying the anti-commutation relation
$\beta_j\beta_k + \beta_k\beta_j =2\delta_{jk}I_4,\,\, j,k = 1,2,3,4$. 
As the first three of it, we take here, with the same triple of Pauli matrices 
as in (1.6), 
\begin{equation}%
   \beta_j := \alpha_j =\left(\begin{array}{cc} 0_2& \sigma_j\\ \sigma_j & 0_2
               \end{array}\right)     \qquad (j= 1, 2, 3),
\end{equation}%\tag1.44
and, as the fourth $\beta_4$, we adopt
\begin{equation}%
  \beta_4 := \alpha_5 = \left(\begin{array}{cc} 
                        0_2& -iI_2\\ iI_2 & 0_2  \end{array}\right),
\end{equation}%\tag1.45
but not the usual $\alpha_4$ given by 
$$
 \qquad \alpha_4 = \left(\begin{array}{cc} 
                   I_2& 0_2\\ 0_2 & -I_2 \end{array}\right).
$$
The $\alpha_4$ is often written as ``$\beta$", but of course,  
different from our $\beta$ on the left-hand side of (1.43) above 
(e.g. [BeSa, p.48]). 
For this, see e.g. [W] where $\alpha_5$ is given as in (1.45) and read in 
[ItZ, p.693] as 
$\alpha_5 := i\gamma^5\gamma^0 = \alpha_1\alpha_2\alpha_3\alpha_4$ 
(see also [G]). Note that as the five $\alpha_k,\, k=1,2,3,4,5$,  
are mutually anti-commuting, Hermitian matrices satisfying 
$\alpha_j\alpha_k + \alpha_k\alpha_j =2\delta_{jk}I_4, \,\,j, k=1,2,3,4$, 
so are the four $\beta_k,\, k=1,2,3,4$. (Here $\delta_{jk}$ is 
the usual Kronecker delta, one when the indices are the same, othewise zero.) 
Therefore 
$\beta\cdot \operatorname{p} 
= \sum_{k=1}^4\beta_k\cdot \operatorname{p}_k$ is a
 selfadjoint operator in $L^2({\bold R}^4; {\Bbb C}^4)$ as well as 
$\sum_{k=1}^4 \alpha_k \operatorname{p}_k$.

Then similarly to the 3-dimensional case before (see around  (1.18)),
we consider the semi-norm $M_{\beta\cdot \operatorname{p}; p}(f)$
as well as the semi-norm $\|(\beta\cdot \operatorname{p})f\|$ concerning the  
first-order derivatives of functions of functions $f = {}^t(f_1,f_2,f_3,f_4)$
in the space $C_0^{\infty}({\Bbb R}^4;{\Bbb C}^4)$. 
To define $M_{\beta\cdot \operatorname{p}; p}(f)$, note first that 
the 4-dimensional Euclidian Dirac operator (1.43) can be rewritten, 
based on the representation (1.44) together with (1.45) for the matrices 
$\beta_k$, $k=1,2,3,4$, as  
\begin{equation}%
\beta\cdot \operatorname{p}
= \left(\begin{array}{cccc}
0&0&\operatorname{p}_3-i\operatorname{p}_4
&\operatorname{p}_1-i\operatorname{p}_2 \\ 
0&0&\operatorname{p}_1+i\operatorname{p}_2 
&-(\operatorname{p}_3+i\operatorname{p}_4)\\
\operatorname{p}_3+i\operatorname{p}_4
&\operatorname{p}_1-i\operatorname{p}_2&0&0\\ 
\operatorname{p}_1+i\operatorname{p}_2 
&-(\operatorname{p}_3-i\operatorname{p}_4)
& 0 &0\end{array}\right).
\end{equation}%\tag1.46
Then decompose it into the sum of its two parts:
\begin{eqnarray}%$$\aligned
\beta\cdot \operatorname{p}
&=& (\beta\cdot \operatorname{p})P_{13} +(\beta\cdot \operatorname{p})P_{24}
     \nonumber\\
&=& \left(\begin{array}{cccc} 
      0&0&\operatorname{p}_3-i\operatorname{p}_4&0 \\ 
      0&0&\operatorname{p}_1+i\operatorname{p}_2 &0\\
      \operatorname{p}_3+i\operatorname{p}_4&0 & 0 &0 \\ 
      \operatorname{p}_1+i\operatorname{p}_2&0 & 0 &0
  \end{array}\right)
 +\left(\begin{array}{cccc}
   0&0&0&\operatorname{p}_1-i\operatorname{p}_2 \\ 
   0&0&0&-(\operatorname{p}_3+i\operatorname{p}_4)\\
   0&\operatorname{p}_1-i\operatorname{p}_2 & 0 &0 \\ 
   0&-(\operatorname{p}_3-i\operatorname{p}_4)& 0 &0
  \end{array}\right),\nonumber\\
\end{eqnarray}%\endaligned\tag1.47$$
where $P_{13} : = \hbox{\rm diag} (1, 0, 1, 0)$ 
  and $P_{24} := \hbox{\rm diag} (0, 1, 0, 1)$ 
are the same two projection matrices acting on the space ${\Bbb C}^4$ 
of four-vectors as before, and define 
\begin{equation}%
M_{\beta\cdot \operatorname{p}; p}(f)
:=\, \big[\,\|(\beta\cdot \operatorname{p})P_{13}f\|_p^p
        +\|(\beta\cdot \operatorname{p})P_{24}f\|_p^p \,\big]^{1/p}.
\end{equation}%\tag1.48

Let us see how this semi-norm 
$M_{\beta\cdot \operatorname{p}; p}(f)$ in (1.48) 
is related to the other
semi-norms $\|(\beta\cdot \operatorname{p})f\|_p$ and $\|\nabla f\|_p$.
However, we should note here that 
the latter $\|\nabla f\|_p$ differs from (1.11),
since in the present case we have the 4-dimensional gradient 
$\nabla = (\partial_1,\partial_2, \partial_2, \partial_2)$, so that
$|\nabla f(x)|_{\ell^p}^p 
 := \sum_{j=1}^4 |\partial_jf(x)|_{\ell^p}^p
 = \sum_{j=1}^4\sum_{k=1}^4|\partial_j f_k(x)|^p$.

Then 
$$
(\beta\cdot \operatorname{p})f
= \left(\begin{array}{cccc}
(\operatorname{p}_3-i\operatorname{p}_4)f_3
 + (\operatorname{p}_1-i\operatorname{p}_2)f_4 \\ 
(\operatorname{p}_1+i\operatorname{p}_2)f_3
-(\operatorname{p}_3+i\operatorname{p}_4)f_4\\
(\operatorname{p}_3+i\operatorname{p}_4)f_1 
+\operatorname{p}_1-i\operatorname{p}_2)f_2\\ 
(\operatorname{p}_1+i\operatorname{p}_2)f_1
-(\operatorname{p}_3-i\operatorname{p}_4)f_2
\end{array}\right),  
$$
so that, recalling the definition of the $\ell^p$ norm in (1.9), 
we have 
\begin{eqnarray*}%$$\aligned
|(\beta\cdot \operatorname{p})f|_{\ell^p}^p
&=&|(\operatorname{p}_3-i\operatorname{p}_4)f_{3}
    +(\operatorname{p}_1-i\operatorname{p}_2)f_{4}|^p
 +|(\operatorname{p}_1+i\operatorname{p}_2)f_{3}
    -(\operatorname{p}_3+i\operatorname{p}_4)f_{4}|^p\\
 &&\qquad
  +|(\operatorname{p}_3+i\operatorname{p}_4)f_{1}
    +(\operatorname{p}_1-i\operatorname{p}_2)f_{2}|^p
  +|(\operatorname{p}_1+i\operatorname{p}_2)f_{1}
    -(\operatorname{p}_3-i\operatorname{p}_4)f_{2}|^p\\
&=& |(\operatorname{p}_1+i\operatorname{p}_2)f_{1}
   -(\operatorname{p}_3-i\operatorname{p}_4)f_{2}|^p
  + |(\operatorname{p}_1-i\operatorname{p}_2)f_{2}
  +(\operatorname{p}_3+i\operatorname{p}_4)f_{1}|^p\\
 &&\qquad+|(\operatorname{p}_1+i\operatorname{p}_2)f_{3}
  -(\operatorname{p}_3+i\operatorname{p}_4)f_{4}|^p
 +|(\operatorname{p}_1-i\operatorname{p}_2)f_{4}
 +(\operatorname{p}_3-i\operatorname{p}_4)f_{3}|^p,
\end{eqnarray*}%\endaligned$$
where we have rearranged the four terms, when passing through the second 
equality. Hence
\begin{eqnarray}%$$\aligned
\|(\beta\cdot \operatorname{p})f\|_p^p
\!\!&\!=\!&\!\! 
  \|(\partial_1+i\partial_2)f_{1}-(\partial_3-i\partial_4)f_{2}\|_p^p
  + \|(\partial_1-i\partial_2)f_{2}+(\partial_3+i\partial_4)f_{1}\|_p^p
    \nonumber\\
 &&+\!\! \|(\partial_1+i\partial_2)f_{3}-(\partial_3+i\partial_4)f_{4}\|_p^p
  +\|(\partial_1-i\partial_2)f_{4}+(\partial_3-i\partial_4)f_{3}\|_p^p. 
   \nonumber\\
\end{eqnarray}%\endaligned\tag1.49$$
Then one can calculate the right-hand side of (1.48) to get
\begin{eqnarray}%$$\aligned
&&M_{\beta\cdot \operatorname{p}; p}(f) ^p \nonumber\\ 
&=& \|(\beta\cdot \operatorname{p})P_{13}f\|_p^p 
   + \|(\beta\cdot \operatorname{p})P_{24}f\|_p^p \nonumber\\ 
&=& \big(\|(\partial_1+i\partial_2) f_1\|_p^p
            +\|(\partial_3+i\partial_4) f_1\|_ p^p\big)
  +\big(\|(\partial_1-i\partial_2) f_2\|_p^p
            +\|(\partial_3-i\partial_4) f_2\|_ p^p\big) \nonumber\\
 &&+\big(\|(\partial_1+i\partial_2) f_3\|_p^p 
            +\|(\partial_3-i\partial_4) f_3\|_ p^p\big)
   +\big(\|(\partial_1-i\partial_2) f_4\|_p^p 
            +\|(\partial_3+i\partial_4) f_4\|_ p^p\big).\nonumber\\
\end{eqnarray}%\endaligned\tag1.50$$

%%%%%%%%%%%%
Similarly to the 3-dimensional case before (see (1.22), (1.23)), 
for the semi-norms (1.49) and (1.48)/(1.50) we have with $1\leq p < \infty$,
\begin{equation}%
2^{-(1-(1/p))} \|(\beta\cdot \operatorname{p})f\|_{p}
\leq M_{\beta\cdot \operatorname{p}; p}(f) \,
\leq\, 2^{1-(1/p)} \|\nabla f\|_{p}\,. 
\end{equation}%\tag1.51
The Banach space $H_{({\beta\cdot \operatorname{p}}),0}^{1,p}
({\Bbb R}^4;{\Bbb C}^4)$
/$H_{M_{\beta\cdot \operatorname{p}},0}^{1,p}
({\Bbb R}^4;{\Bbb C}^4)$
is defined as completion of the space $C_0^{\infty}({\Bbb R}^4;{\Bbb C}^4)$
in the norm 
$\|f\|_{(\beta\cdot \operatorname{p}),1,p} 
:= (\|f\|_p^p +\|(\beta\cdot \operatorname{p})f\|_p^p)^{1/p}$
/$\|f\|_{M_{\beta\cdot \operatorname{p}},1,p} 
:= (\|f\|_p^p +M_{\beta\cdot \operatorname{p}; p}(f)^p)^{1/p}$.
>From (1.51) we see the following inclusion relation :
\begin{equation}%
 H_{0}^{1,p}({\Bbb R}^4;{\Bbb C}^4) \quad \subseteq\quad  
 H_{M_{\beta\cdot \operatorname{p}},0}^{1,p}({\Bbb R}^4;{\Bbb C}^4)
 \quad\subseteq\quad 
 H_{({\beta\cdot \operatorname{p}}),0}^{1,p}({\Bbb R}^4;{\Bbb C}^4).
\end{equation}%\tag1.52

Now we note the semi-norm 
$M_{\beta\cdot \operatorname{p}; p}(f)$ has a significant character as that of 
$M_{\alpha\cdot \operatorname{p}; p}(f)$ in Proposition 1.0, 
by considering other decompositions 
of the Euclidian Dirac operator $\beta\cdot \operatorname{p}$ in (1.46),
than (1.47), into the sum of its two parts: 

\medskip\noindent
$\underline{M_{\beta}^{(1)}}$
\begin{eqnarray}%$$\aligned
\beta\cdot \operatorname{p}
&=& (\beta\cdot \operatorname{p})P_{14} +(\beta\cdot \operatorname{p})P_{23}
    \nonumber\\
&\equiv& \left(\begin{array}{cccc}
0&0&0& \operatorname{p}_1-i\operatorname{p}_2 \\ 
0&0&0&-(\operatorname{p}_3+i\operatorname{p}_4)\\
\operatorname{p}_3+i\operatorname{p}_4&0&0&0 \\ 
\operatorname{p}_1+i\operatorname{p}_2 &0&0&0
\end{array}\right)\,
+ \left(\begin{array}{cccc}
0&0& \operatorname{p}_3-i\operatorname{p}_4&0\\ 
0&0& \operatorname{p}_1+i\operatorname{p}_2 &0\\
0&\operatorname{p}_1-i\operatorname{p}_2 &0&0\\ 
0&-(\operatorname{p}_3-i\operatorname{p}_4)&0&0
\end{array}\right), \nonumber\\
\end{eqnarray}%\endaligned\tag1.53$$ 
where $P_{14} : = \hbox{\rm diag} (1, 0, 0, 1)$ 
and $P_{23} := \hbox{\rm diag} (0, 1, 1, 0)$ 
are the same two projection matrices acting on the space ${\Bbb C}^4$
of four-vectors as before, and note that both the 
operators $(\beta\cdot \operatorname{p})P_{14}$ and 
$(\beta\cdot \operatorname{p})P_{23}$ on the right are selfadjoint,
i.e. 
$((\beta\cdot \operatorname{p})P_{14})^* 
      = (\beta\cdot \operatorname{p})P_{14},\,\,
 ((\beta\cdot \operatorname{p})P_{23})^*
      = (\beta\cdot \operatorname{p})P_{23}$.

\medskip\noindent
$\underline{M_{\beta}^{(2)}}$
\begin{eqnarray}%$$\aligned
\beta\cdot \operatorname{p}
&=\,& \left(\begin{array}{cc} 
            0_2&\sigma_1\operatorname{p}_1+\sigma_2\operatorname{p}_2\\
            \sigma_3\operatorname{p}_3+iI_2\operatorname{p}_4&0_2
   \end{array}\right)
  +\left(\begin{array}{cccc}
            0_2&\sigma_3\operatorname{p}_3-iI_2\operatorname{p}_4\\
            \sigma_1\operatorname{p}_1+\sigma_2\operatorname{p}_2&0_2
   \end{array}\right)\nonumber\\
&\equiv\,& \left(\begin{array}{cccc}
0&0&0&\operatorname{p}_1-i\operatorname{p}_2 \\ 
0&0&\operatorname{p}_1+i\operatorname{p}_2 &0\\
\operatorname{p}_3-i\operatorname{p}_4&0 & 0 &0 \\ 
0&-(\operatorname{p}_3-i\operatorname{p}_4)& 0 &0\end{array}\right) 
     \nonumber\\
&&\qquad\qquad+ \left(\begin{array}{cccc}
0&0&\operatorname{p}_3-i\operatorname{p}_4&0 \\ 
0&0&0&-(\operatorname{p}_3+i\operatorname{p}_4)\\
0&\operatorname{p}_1-i\operatorname{p}_2 & 0 &0 \\ 
\operatorname{p}_1+i\operatorname{p}_2 &0& 0 &0\end{array}\right) 
      \nonumber\\
&=:& (\beta\cdot \operatorname{p})_1 + (\beta\cdot \operatorname{p})_2,
\end{eqnarray}%\endaligned\tag1.54$$ 
where note that $(\beta\cdot \operatorname{p})_2$ is the adjoint of
$(\beta\cdot \operatorname{p})_1$ as operators, say,  
in $L^2({\bold R}^3;{\Bbb C}^4)$, i.e. 
$(\beta\cdot \operatorname{p})_2 = {(\beta\cdot \operatorname{p})_1}^*$.

\medskip\noindent
$\underline{M_{\beta}^{(3)}}$  
\begin{eqnarray}%$$\aligned
\beta\cdot \operatorname{p}
&=\,&\left(\begin{array}{cc}
0&\sigma_1\operatorname{p}_1+\sigma_2\operatorname{p}_2 \\ 
\sigma_1\operatorname{p}_1+\sigma_2\operatorname{p}_2&0
\end{array}\right) 
+ \left(\begin{array}{cccc} 
0&\sigma_3\operatorname{p}_3-iI_2\operatorname{p}_4\\ 
\sigma_3\operatorname{p}_3+iI_2\operatorname{p}_4&0
\end{array}\right) \nonumber\\
&\equiv\,& \left(\begin{array}{cccc}
0&0&0&\operatorname{p}_1-i\operatorname{p}_2 \\ 
0&0&\operatorname{p}_1+i\operatorname{p}_2 &0\\
0&\operatorname{p}_1-i\operatorname{p}_2 & 0 &0 \\ 
\operatorname{p}_1+i\operatorname{p}_2 &0& 0 &0
\end{array}\right) \nonumber\\
&&\qquad\qquad+ \left(\begin{array}{cccc}
0&0&\operatorname{p}_3-i\operatorname{p}_4&0\\ 
0&0&0&-(\operatorname{p}_3+i\operatorname{p}_4)\\
\operatorname{p}_3+i\operatorname{p}_4&0&0& \\ 
0&-(\operatorname{p}_3-i\operatorname{p}_4)& &0
\end{array}\right) \nonumber\\
&=:& (\beta\cdot \operatorname{p})_3 +(\beta\cdot \operatorname{p})_4\,,
\end{eqnarray}%\endaligned\tag1.55$$ 
where note that both the operators 
$(\beta\cdot \operatorname{p})_3$ and 
$(\beta\cdot \operatorname{p})_4$ on the right are selfadjoint.

\medskip
Then we can confirm, in the same way as in Proposition 1.0  for 
$M_{\alpha\cdot \operatorname{p}; p}(f)$ with $\alpha\cdot \operatorname{p}$, 
that the semi-norm $M_{\alpha\cdot \operatorname{p}; p}(f)$ of $f$ defined
 by (1.48) with the rather artificial decomposition (1.47)
turns out to be equal to the ones to be defined with the other decompositions
 (1.53), (1.54) and (1.55), taking account of the expression (1.50) for 
$M_{\beta\cdot \operatorname{p}; p}(f)$:
\begin{subequations}
\begin{align}
M^{(1)}_{\beta\cdot \operatorname{p}; p}(f)
&:=  [\|(\alpha\cdot \operatorname{p})P_{14} f\|_p^p 
    + \|(\alpha\cdot \operatorname{p})P_{23} f\|^p]^{1/p}; \\%\tag1.56a
M^{(2)}_{\beta\cdot \operatorname{p}; p}(f)
&:= [\|(\beta\cdot \operatorname{p})_1 f\|_p^p 
    + \|(\alpha\cdot \operatorname{p})_2 f\|^p]^{1/p}
=  [\|(\beta\cdot \operatorname{p})_1 f\|_p^p 
    + \|{(\beta\cdot \operatorname{p})_1}^* f\|^p]^{1/p}; \nonumber\\
           \\%\tag1.56b 
M^{(3)}_{\beta\cdot \operatorname{p}; p}(f)
&:= [\|(\beta\cdot \operatorname{p})_3 f\|_p^p 
    + \|(\beta\cdot \operatorname{p})_4 f\|^p]^{1/p}. %\tag1.56c
\end{align}
\end{subequations}
Further, more generally, every decomposition of 
$\beta\cdot \operatorname{p}$ into its two parts,  
$\beta\cdot \operatorname{p} 
= (\beta\cdot \operatorname{p})_5+ (\beta\cdot \operatorname{p})_6$,
such that each row of both the matrices $(\beta\cdot \operatorname{p})_5$
and $(\beta\cdot \operatorname{p})_6$ contains only one nonzero entry, 
defines the semi-norm 
$M_{\beta\cdot \operatorname{p}; p}(f)$ which has the expression (1.50).
However, as mentioned for the operator $\alpha\cdot \operatorname{p}$
after Proposition 1.0, the decomposition
$\beta\cdot \operatorname{p} 
= (\beta\cdot \operatorname{p})P_{12}+(\beta\cdot \operatorname{p})P_{34}$
is not fit for the semi-norm 
$M_{\beta\cdot \operatorname{p}; p}(f)$, to which we will come back 
in Section 6 to discuss the issue.

\begin{thm}%{Theorem 1.6}
{\rm  (with 4-dimensional Euclidian Dirac operator)}. 
\hbox{\rm (i)} {\it For $1\leq p<q<\infty$, a ${\Bbb C}^4$-valued function 
$f = {}^t(f_1,f_2,f_3,f_4)$ belongs to $L^q({\Bbb R}^4;{\Bbb C}^4)$, 
if  $f$ belongs to $B^{p/(p-q)}_{\infty,\infty}({\Bbb R}^4;{\Bbb C}^4)$
and satisfies $M_{\beta\cdot \operatorname{p}; p}(f) < \infty$, 
and further, there exists a positive constant $C$ such that
\begin{equation}%
 \|f\|_{q} \leq C\,M_{\beta\cdot \operatorname{p}; p}(f)^{p/q}
  \|f\|_{B^{p/(p-q)}_{\infty,\infty}}^{1-(p/q)}\,.
\end{equation}%\tag1.57
Therefore this holds, in particular, for every $f \in 
H_{M_{\beta\cdot \operatorname{p}},0}^{1,p}({\Bbb R}^4;
{\Bbb C}^4)
\cap B^{p/(p-q)}_{\infty,\infty}({\Bbb R}^4;{\Bbb C}^4)$.
}

\medskip
(ii) {\it For $\infty >p>1$, the three semi-norms 
$\|(\beta\cdot \operatorname{p})f\|_{p}$, 
$M_{\beta\cdot \operatorname{p}; p}(f)$ and
$\|\nabla f\|_p$ are equivalent, so that the corresponding three
Banach spaces \hbox{\rm (1.52)} coincide with one another:
\begin{equation}%
 H_0^{1,p}({\Bbb R}^4;{\Bbb C}^4) \quad =\quad 
 H_{M_{\beta\cdot \operatorname{p}; p},0}^{1,p}({\Bbb R}^4;{\Bbb C}^4)
 \quad=\quad  
H_{({\beta\cdot \operatorname{p}}),0}^{1,p}({\Bbb R}^4;{\Bbb C}^4). 
\end{equation}%\tag1.58
Therefore assertion \hbox{\rm (i)} turns out: For $1<p<q<\infty$, 
there exists a positive constant $C$ such that
\begin{equation}%
 \|f\|_{q} \leq C \|(\beta\cdot \operatorname{p})f\|_{p}^{p/q}
  \|f\|_{B^{p/(p-q)}_{\infty,\infty}}^{1-(p/q)},
\end{equation}%\tag1.59
for every $f \in B^{p/(p-q)}_{\infty,\infty}({\Bbb R}^4;{\Bbb C}^4)$
whose semi-norm $\|(\beta\cdot \operatorname{p})f\|_{p}$, 
$M_{\beta\cdot \operatorname{p}; p}(f)$ or
$\|\nabla f\|_p$ is finite. Therefore this holds, in particular,
for every $f$ in the above space \hbox{\rm (1.58)} which belongs to
$B^{p/(p-q)}_{\infty,\infty}({\Bbb R}^4;{\Bbb C}^4)$. 
The inequality \hbox{\rm (1.59)} 
is equivalent to the vector-valued version  \hbox{\rm (1.13)} 
of \hbox{\rm (1.1)} with $n=4$. 
}
\end{thm}

\medskip
We note here that the 4-dimensional Euclidian 
Dirac operator $\sum_{k=1}^4 \beta_k \operatorname{p}_k$ in (1.47)
turns, if $\beta_4\operatorname{p}_4 =-i\beta_4\partial_4$ is removed from it, 
the 3-dimensional massless Dirac operator 
$\sum_{j=1}^3 \alpha_j \operatorname{p}_j$ in (1.17),
which reduces Theorem 1.6  to Theorem 1.1.

Finally, as is the case for Sobolev spaces of single-valued functions,
it is seen for the two spaces of vector-valued functions
which we introduced in (1.23) and (1.52) that 
 each of them coincides with the following two spaces: 
\begin{eqnarray*}%$$\aligned
H_{M_{\alpha\cdot \operatorname{p}; p},0}^{1,p}
({\Bbb R}^3;{\Bbb C}^4)
&=& H_{M_{\alpha\cdot \operatorname{p}; p}}^{1,p}
({\Bbb R}^3;{\Bbb C}^4)\\
&=& \{f \in L^p({\Bbb R}^3;{\Bbb C}^4)\,;\,
(\alpha\cdot \operatorname{p})P_{13}f,\, 
(\alpha\cdot \operatorname{p})P_{24}f \, 
\in \, L^p({\Bbb R}^3;{\Bbb C}^4)\}\\
&=& \{f \in L^p({\Bbb R}^3;{\Bbb C}^4)\,;\,
(\alpha\cdot \operatorname{p})_1f,\, 
(\alpha\cdot \operatorname{p})_2f \, 
\in \, L^p({\Bbb R}^3;{\Bbb C}^4)\}\,;\\
%%%
H_{M_{\beta\cdot \operatorname{p}; p},0}^{1,p}
({\Bbb R}^4;{\Bbb C}^4)
&=& H_{M_{\beta\cdot \operatorname{p}; p}}^{1,p}
({\Bbb R}^4;{\Bbb C}^4)\\
&=& \{f \in L^p({\Bbb R}^4;{\Bbb C}^4)\,;\,
(\beta\cdot \operatorname{p})P_{13}f,\, 
(\beta\cdot \operatorname{p})P_{24}f \, 
\in \, L^p({\Bbb R}^4;{\Bbb C}^4)\}\\
&=& \{f \in L^p({\Bbb R}^4;{\Bbb C}^4)\,;\,
(\beta\cdot \operatorname{p})_1f,\, 
(\beta\cdot \operatorname{p})_2f \, 
\in \, L^p({\Bbb R}^4;{\Bbb C}^4)\}.
\end{eqnarray*}%\endaligned$$
In each of these two formulas, the second space is 
the Banach space obtained as completion with respect to the norm 
$\|f\|_{M_{\alpha\cdot \operatorname{p}},1,p}\,$
[\hbox{\sl resp.} $\,\|f\|_{M_{\beta\cdot \operatorname{p}},1,p}$]
of the linear space of all 
$f \in C^{\infty}({\Bbb R}^3;{\Bbb C}^4) \cap L^p({\Bbb R}^3;{\Bbb C}^4)\,$ 
[\hbox{\sl resp.} 
$\,C^{\infty}({\Bbb R}^4;{\Bbb C}^4) \cap L^p({\Bbb R}^4;{\Bbb C}^4)$]. 
In the third and fourth spaces 
the first-order derivatives are taken in the distribution sense.

%%%%%%%%%%%%%%%%%%%%%%%
\bigskip
The proof of the improved Sobolev inequality (1.1) for single-valued functions
in [CDPX] and [CMO] was based on wavelet analysis, while 
Ledoux [Le] made a different approach by a direct semigroup argument.
We do our proof, modifying the method used by Ledoux so as to be able 
to apply to vector-valued functions.

\medskip
The plan of this paper is as follows.
Section 2 collects remarks to the results, stated in Section 1,
for vector-valued functions to compare them with the 
improved Sobolev inequality (1.1) and the Dirac--Sobolev inequality (1.14)
obtained in [BES]. Section 3 gives examples where 
the simple-minded, vector-valued version (1.16) 
connected not only with the three-dimensional massless Dirac operator 
but also with the four-dimensional Euclidian Dirac operator 
fails to hold for $p=1$.
In Section 4, we give proof of Theorem 1.1, and in Section 5, 
proofs of all the other five Corollaries 1.2, 1.3, Theorem 1.4, 
Corollary 1.5, Theorem 1.6. 
In Section 6 we make concluding comments 
on the first-order-derivative 
semi-norm connected with the Dirac operators which we have introduced 
in Section 1. It is defined at first with a rather artificial decomposition
of the Dirac operator into two parts, but later turns out to be meaningful 
enough to have universal character.
The final Section 7 briefly summarizes all our results to exhibit 
their significance  and difference from the case of single-valued functions.

%%%%%%%%%%%%%%%%%%%%%%%%%%%%
\medskip
\section{Remarks}%sect.2

\medskip\noindent
1$^o$. Theorem 1.1 (i) (ii): 
We compare our inequality 
(1.28) with (1.16)/(1.30),  the trivial version (1.13) and the first 
vector-valued one (1.14) of inequality (1.1) shown in [BES].

To do so, first we collect the results of equivalence and non-equivalence 
among the three first-order-detrivative semi-norms 
$\|\nabla f\|_p$ in (1.11), $\|(\alpha\cdot \operatorname{p})f\|_p$ in (1.9)
and $M_{\alpha\cdot \operatorname{p}; p}(f)$ in (1.18), which are
under relation (1.22).
When $1<p<\infty$, these three are all equivalent, which we shall see 
in the proof of Theorem 1.1 (ii) in Section 3 below, but different when $p=1$. 
In this case  $p=1$, we showed 
non-equivalence between $\|\nabla f\|_1$ and 
$\|(\alpha\cdot \operatorname{p})f\|_1$ in [IS, Theorem 1.3 (iii)]. 
Non-equivalence between $\|(\alpha\cdot \operatorname{p})f\|_1$ and 
$M_{\alpha\cdot \operatorname{p}; 1}(f)$ can be seen in view 
of their respective explicit expressions (1.20) and (1.21), 
and that between $\|\nabla f\|_1$ and
$M_{\alpha\cdot \operatorname{p}; 1}(f)$ in view of their respective 
definition (1.11) and explicit expression (1.21), both from the fact 
that (2.2) below cannot hold. 
In particular, the two inclusions in (1.23) are strict.

Next we going to observe the difference and coincidence among 
inequalities (1.28), (1.16)/(1.30), (1.13) and (1.14). 
For $1<p<\infty$, the first three, i.e. (1.28), (1.16)/(1.30) and (1.13),
are equivalent, and strictly sharper than and hence an improvement 
of the last one, (1.14). The former is because of equivalence of the three 
first-order-derivative semi-norms concerned as just seen above, 
and the latter because the $L^q$ norm $\|f\|_{q}$ 
on the left of (1.28) is stronger 
than the weak $L^q$ norm $\|f\|_{q,\infty}$ on the left of (1.14).  
For $p=1$, (1.16)/(1.28) does not hold in general, and (1.28) is sharper
than (1.13), because the semi-norm 
$M_{\alpha\cdot \operatorname{p}; 1}(f)$ on the right 
of (1.28) is weaker than the semi-norm $\|\nabla f\|_1$ on the right 
of (1.13). In the case $p=1$, however, two inequalities (1.28) and (1.14)
 cannot be compared so as to say which of them is sharper, because 
$M_{\alpha\cdot \operatorname{p}; 1}(f)$ 
on the right of (1.28) is not weaker than 
$\|(\alpha\cdot \operatorname{p})f\|_1$
on the right of (1.14), though $\|f\|_q$ on the left 
of (1.28) is stronger than $\|f\|_{q,\infty}$ on the left of (1.14).
As a result, (1.28) for $p=1$ is a new inequality for vector-valued version
of (1.1).

\medskip\noindent
2$^o$. Corollary 1.2 (i) (ii): The same remark as 1$^o$
above applies to the case for the 3-dimensional Weyl--Dirac  (or Pauli) 
operator $\sigma\cdot\operatorname{p}$ 
in place of the Dirac operator $\alpha\cdot\operatorname{p}$.

%%%%%%%
\medskip\noindent
3$^o$. Corollary 1.3 (i) (ii): For $p=1$, the semi-norm 
$M_{(\partial-i\partial_2)\vee \partial_3); 1}(\psi)$ 
in (1.36) is bounded by the semi-norm $\|\nabla \psi\|_{1}$, i.e.
\begin{equation}%
 M_{(\partial-i\partial_2)\vee \partial_3; 1}(\psi) 
    \leq \|\nabla \psi\|_{1},
\end{equation}%\tag2.1
but not reversely (See [St, pp.59--60, III, Propositions 3, 4, and
p.48, 6.1] and [IS, Lemma 4.3]). 
Therefore the Banach space 
$H_{M_{(\partial-i\partial_2)\vee \partial_3},0}^{1,p}
                    ({\Bbb R}^3;{\Bbb C}^2)$
obtained as 
completion of $C_0^{\infty}({\Bbb R}^3)$ with respect to the norm 
$\|\psi\|_{M_{(\partial-i\partial_2)\vee \partial_3},1,p} 
:= \|\psi\|_1+ M_{(\partial-i\partial_2)\vee \partial_3;p}(\psi)$
is strictly larger than the space $H_0^{1,1}({\Bbb R}^3)$.  
Therefore for $p=1$, Corollary 1.3 (i) gives a slightly more general
result than (1.1) of Ledoux [Le] though only in the case $n=3$.  
However, for $1<p<\infty$, it is nothing but his result though our result only concerns the case $n=3$, since the semi-norm 
$M_{(\partial-i\partial_2)\vee \partial_3; p}(\psi)$ is equivalent 
to the semi-norm $\|\nabla \psi\|_{p}$.
In this sense, therefore our inequality (1.37) for ${\Bbb C}$-valued
functions $\psi$ is more general, though only  for $n=3$.
Here it should be noted that it holds that for $1<p<\infty$, 
\begin{equation}%
 \|\partial_1 \psi\|_p +\|\partial_2\psi\|_p 
  \leq C_p\|(\partial_1-i\partial_2)\psi\|_p, 
\end{equation}%\tag2.2
for all $\psi \in C_0^{\infty}({\Bbb R}^2)$
with a positive constant $C_p$, but cannot for $p=1$
(cf. [St, pp.59--60, III, Propositions 3, 4, and p.48, 6.1] 
and [IS, Lemma 4.3]).
Therefore (2.2) implies that for $1<p<\infty$,
\begin{eqnarray}%$$\aligned
\|\nabla \psi\|_p &\equiv& (\sum_{j=1}^3 \|\partial_j \psi\|_p^p)^{1/p}
 \leq (C_p^{p/(p-1)}+1)^{(p-1)/p}
[\|(\partial_1-i\partial_2)\psi\|_p^p+\|\partial_3 \psi\|_p^p]^{1/p}
        \nonumber\\
&\equiv& (C_p^{p/(p-1)}+1)^{(p-1)/p}
 M_{(\partial_1-i\partial_2)\vee \partial_3; p}(\psi),
\end{eqnarray}%\endaligned\tag2.3$$
so that the two semi-norms 
$M_{(\partial_1-i\partial_2)\vee \partial_3; p}(\psi)$ and 
$\|\nabla \psi\|_p$ are equivalent.

%%%%%%%%%%%%%%%%%%%%%%
\medskip\noindent
4$^o$. Corollary 1.5: 
By analogous discussion made in Remark 3$^0$ to Corollary 1.3, 
(1.42) is also more general than (1.1) with $n=2$ for $p=1$, 
but equivalent to it for $\infty >p>1$.

%%%%%%%%%%%%%%%%%%%%%%%
\medskip\noindent
5$^o$. Theorem 1.6 and again Theorem 1.1: It can be seen that these 
two theorems hold also for some different representations of 
the 3-dimensional massless Dirac operator 
and 4-dimensional Euclidian Dirac operator than (1.17) and (1.46).

In fact, consider first the 4-dimensional Euclidian Dirac operators.
Let $\beta' = (\beta'_1,\beta'_2,\beta'_3,\beta'_4)$
be another quadruple of anti-commuting, Hermitian $4\times 4$-matrices 
satisfying $\beta'_j\beta'_k + \beta'_k\beta'_j =2\delta_{jk}I_4,\, 
j,k=1,2,3,4$. Then Theorem 1.6 holds for the Euclidian Dirac operator 
$\beta'\cdot\operatorname{p}= \sum_{k=1}^4 \beta'_k\operatorname{p}_k$
with corresponding projections $P_{13}',\, P_{24}'$.
Indeed, by the `fundamental theorem' in [P, p.8] or [G, p.190], there exists
a non-singular $4\times 4$-matrix $S$ such that 
$\beta'_k = S\beta_kS^{-1}$ for $k=1,2,3,4$. So $S$ is 
a similarity transformation which maps ${\Bbb C}^4$ 
one-to-one onto ${\Bbb C}^4$, and in fact can be taken to be a unitary 
matrix, because the $\beta_k$ and $\beta'_k$ are Hermitian. 
Then
$$
 \beta\cdot\operatorname{p} = S^{-1}(\beta'\cdot\operatorname{p})S,
\quad
 (\beta\cdot\operatorname{p})P_{13} 
   = S^{-1}(\beta'\cdot\operatorname{p})P'_{13}S, 
\quad
 (\beta\cdot\operatorname{p})P_{24} 
   = S^{-1}(\beta'\cdot\operatorname{p})P'_{24}S,
$$
where $P'_{13} := SP_{13}S^{-1}$ and $P'_{24} := SP_{24}S^{-1}$ are 
projection matrices acting on ${\Bbb C}^4$ such that 
$P'_{13} + P'_{24} = I_4$.
It implies equivalence of the related semi-norms concerning 
$\beta'\cdot\operatorname{p}$ and $\beta\cdot\operatorname{p}$
in the following sense:
\begin{eqnarray*}%$$\aligned
 &&(\|S^{-1}\|_{\ell^p\rightarrow \ell^p})^{-1}\|
  (\beta\cdot\operatorname{p})f\|_p 
  \leq \|(\beta'\cdot\operatorname{p})(Sf)\|_p
  \leq \|S\|_{\ell^p\rightarrow \ell^p}
          \|(\beta\cdot\operatorname{p})f\|_p, \\ 
&&(\|S^{-1}\|_{\ell^p\rightarrow \ell^p})^{-1}\|
  (\beta\cdot\operatorname{p})P_{13}f\|_p 
  \leq \|(\beta'\cdot\operatorname{p})P'_{13}(Sf)\|_p
  \leq \|S\|_{\ell^p\rightarrow \ell^p}
          \|(\beta\cdot\operatorname{p})P_{13}f\|_p, \\ 
&&(\|S^{-1}\|_{\ell^p\rightarrow \ell^p})^{-1}\|
  (\beta\cdot\operatorname{p})P_{24}f\|_p 
  \leq \|(\beta'\cdot\operatorname{p})P'_{24}(Sf)\|_p
  \leq \|S\|_{\ell^p\rightarrow \ell^p}
          \|(\beta\cdot\operatorname{p})P_{24}f\|_p, 
\end{eqnarray*}%\endaligned$$
with $1\leq p < \infty$, where $f = {}^t(f_1,f_2,f_3,f_4)$,  
which yields equivalence of the semi-norms 
$M_{\beta'\cdot\operatorname{p};p}(Sf)$
and $M_{\beta\cdot\operatorname{p};p}(f)$:  
$$
 C_p^{-1} M_{\beta\cdot\operatorname{p};p}(f) 
 \leq M_{\beta'\cdot\operatorname{p};p}(Sf) 
 \leq C_p M_{\beta\cdot\operatorname{p};p}(f) 
$$
with a positive constant $C_p$ depending on $p$.
In particular, all this holds also for the 
4-dimensional Euclidian Dirac operator
$\sum_{k=1}^4 \alpha_k\operatorname{p}_j$.

Though above we have dealt only the case corresponding to decomposition (1.47)
of $\beta\cdot\operatorname{p}$, the same 
is true for the cases correstonding to the other decompositions 
(1.53), (1.54) or (1.55). 

Next, for Theorem 1.1, the same is valid, if one may consider,
for $\alpha' = (\alpha'_1,\alpha'_2,\alpha'_3)$ another triple
 of anti-commuting, Hermitian $4\times 4$-matrices satisfying
$\alpha'_j\alpha'_k + \alpha'_k\alpha'_j =2\delta_{jk}I_4,\, j,k=1,2,3$,
the Dirac operator
$\alpha'\cdot\operatorname{p}= \sum_{j=1}^3 \alpha'_j\operatorname{p}_j$
together with the corresponding projection matrices $P'_{13},\, P'_{24}$
to introduce the related semi-norms.

%%%%%%%%%%%%%%%%%%%%%%%
\medskip
\section{Counterexamples for $p=1$}%\sect.3

Inequalities of the type (1.16), i.e. (1.30) of Theorem 1.1 for the 
three-dimensional massless Dirac operator $\alpha\cdot\operatorname{p}$, 
(1.35) of Corollary 1.2 with 3-dimensional Weyl--Dirac (or Pauli) 
operator $\sigma\cdot \operatorname{p}$, (1.59) of Theorem 1.6
with 4-dimensional Euclidian Dirac operator
$\beta\cdot \operatorname{p}$, do not in general hold for $p=1$,
although they do for $1<p<\infty$. This is why, for $p=1$, 
we had to introduce 
the intermediate first-order-derivative semi-norms 
$M_{\alpha\cdot\operatorname{p};p}(f)$ in (1.19), 
$M_{\sigma\cdot\operatorname{p};p}(h)$ in (1.32), 
$M_{\beta\cdot\operatorname{p};p}(f)$ in (1.48).
Here, before going further, we keep Theorem 1.4 in mind that {\it nevertheless
it holds for all $\,\, 1\leq p < \infty$ with the 2-dimensional Weyl--Dirac (or Pauli) 
operator $(\sigma\cdot \operatorname{p})^{(2)}$}, i.e. (1.40abc).

In this section, following the idea in the  recent paper [BEU] 
for the 3-dimensional Weyl--Dirac (or Pauli) operator,
we construct counterexamples not only for (1.30) with 
$\alpha\cdot\operatorname{p}$ but also for (1.59) with 
$\beta\cdot\operatorname{p}$, 
though the construction for both is only slightly  different. 
To the latter, as a matter of fact, we will come back 
in Section 6 to make some important comments on the semi-norms 
concerned.  

In [BEU], they observed, for the 3-dimensional Weyl--Dirac (or Pauli) 
operator $\sigma\cdot \operatorname{p}$,  
that, for $1 < p < 3$ with $q= \frac{3p}{3-p}$, 
the following inequality: 
\begin{equation}%
 \|h\|_q \leq C(p) \|(\sigma\cdot \operatorname{p})h\|_p\,
\end{equation}%\tag3.1
holds for all $h \in C_0^{\infty}({\Bbb R}^3;{\Bbb C}^2)$ with a positive 
constant $C(p)$ depending on $p$. This is a consequence from 
the usual Sobolev inequality together with the fact that, 
for $1 < p < \infty$, 
the two semi-norms $\|(\sigma\cdot \operatorname{p})h\|_p$
and $\|\nabla h\|_p$ are equivalent
(cf. [IS] and Lemma 3.2 of the present paper where analogous
results are given for the Dirac operator $\alpha\cdot \operatorname{p}$
instead of Weyl--Dirac (or Pauli) $\sigma\cdot \operatorname{p}\,$). 
They showed also that (3.1) is untrue when $p=1$, by using a zero mode for an 
appropriate Weyl--Dirac (or Pauli) operator constructed by Loss--Yau [LoY] 
to make a sequence $\{h_n\} \subset C_0^{\infty}({\Bbb R}^3;{\Bbb C}^2)$
such that $\{\|(\sigma\cdot \operatorname{p}) h_n\|_1\}$ 
is uniformly bounded for all over $n$,
but that $\|h_n\|_{3/2} \geq 
(\hbox{\rm positive}\,\,\hbox{\rm constant})\cdot (\log n)^{2/3}$,
concluding invalitity of (3.1) for $p=1$. As a result, this sequence
will turn out to violate (1.35) in Corollary 1.2.

We will modify their argument so as to apply to our cases of Theorems 1.1 
and 1.6 to construct an example.
First we consider the case for three-dimensional massless Dirac operator 
$\alpha\cdot\operatorname{p}$ and next for 4-dimensional Euclidian Dirac 
operator $\beta\cdot \operatorname{p}$.

\bigskip\noindent
{\it An example for \hbox{\rm (1.30)} of Theorem \hbox{\rm 1.1} 
with $p=1$ to fail to hold.} 

So with $x \in {\Bbb R}^3$ and $|x| = (x_1^2+x_2^2+x_3^2)^{1/2}$, let
\begin{eqnarray}%$$\aligned
e(x) 
&:=& \frac1{(1+|x|^2)^{3/2}}(I_4+ i\alpha\cdot x)
         \left(\begin{array}{cccc} 1\\ 0\\ 0\\ 0 \end{array}\right)
                                                           \nonumber\\
&\,=& \frac1{(1+|x|^2)^{3/2}}
    \left(\begin{array}{cccc}
    1&0&ix_3&ix_1+x_2 \\ 
    0&1&ix_1-x_2&-ix_3\\
    ix_3&ix_1+x_2&1&0\\
    ix_1-x_2&-ix_3&0&1\end{array}\right)
    \left(\begin{array}{cccc} 1\\ 0\\ 0\\ 0 \end{array}\right) 
                                                            \nonumber\\
&\,=& \frac1{(1+|x|^2)^{3/2}}
    \left(\begin{array}{cccc} 1\\0\\ix_3\\ix_1-x_2 \end{array}\right)\,, 
\end{eqnarray}%\endaligned\tag3.2$$
where $I_4$ is the $4\times 4$-identity matrix. 
Then we can see $e(x)$ satisfies the following equation
\begin{equation}%
(\alpha\cdot \operatorname{p})e(x) = \frac3{1+|x|^2} e(x), 
\end{equation}%\tag3.3
and inequalities:
\begin{eqnarray}%$$\align
|e(x)|_{\ell^{\infty}} 
  &=& \frac{1 \vee |ix_3| \vee |ix_1-x_2|}{(1+|x|^2)^{3/2}}
  = \frac{1 \vee |x_3| \vee (x_1^2+x_2^2)^{1/2}}{(1+|x|^2)^{3/2}}
              \nonumber\\
&\leq& \frac{(1+x_1^2+x_2^2+x_3^2)^{1/2}}{(1+|x|^2)^{3/2}}
   = \frac{1}{1+|x|^2},                                   \\%\tag3.4\\ 
|e(x)|_{\ell^{q}}^q
  &=& \frac{1 +|ix_3|^{q} +|ix_1-x_2|^{q}}{(1+|x|^2)^{3q/2}}
   = \frac{1 +(x_3^2)^{q/2} +(x_1^2+x_2^2)^{q/2}}{(1+|x|^2)^{3q/2}}
              \nonumber\\
&\geq& \frac{(1+|x|^2)^{q/2}}{(1+|x|^2)^{3q/2}}
=\Big(\frac{1}{1+|x|^2}\Big)^q \quad (1\leq q \leq 2),   %\tag3.5\\
\end{eqnarray}%\endalign$$
where (3.5) is due to that $a^{q/2}+b^{q/2} \geq (a+b)^{q/2}$
for $a\geq 0,\, b\geq 0$ and $1\leq q \leq 2$.

For each positive integer $n$, put 
$f_n(x) = \rho_n(|x|) e(x)$, where $\rho_n(r)$ is a nonnegative
cutoff function in $ C_0^{\infty}({\Bbb R})$ such that 
 $\rho_n(r) = 1\, \, (r \leq n)\,\,; \, = 0 \,\,(r \geq n+2)$, and further 
$|\rho_n'(r)| \equiv |(d/dr) \rho_n(r)| \leq 1$  for all $r \geq 0$.
Then it is evident that $f_n$ belongs to 
$C_0^{\infty}({\Bbb R}^3;{\Bbb C}^4)$.

We are going to see that 
inequality (1.16)/(1.30) does not hold with any constant $C>0$ 
for $p=1$, $q=\frac32$ and hence $\frac{p}{q} =\frac23$. Indeed, 
there exists no constant $C$ such that, for all $n$,
\begin{equation}%
\|f_n\|_{3/2} \leq C \|(\alpha\cdot \operatorname{p})f_n\|_{1}^{2/3}
  \|f_n\|^{1/3}_{B^{-2}_{\infty,\infty}}\,.
\end{equation}%\tag3.6

First, we show that the sequence 
$\{(\alpha\cdot \operatorname{p})f_n\}_{n=1}^{\infty}$ 
is uniformly bounded in $L^1$.
Indeed,  since
\begin{eqnarray*}%$$\aligned
 (\alpha\cdot \operatorname{p})f_n(x)
&=& \rho_n(|x|) (\alpha\cdot \operatorname{p})e(x)
  +\big((\alpha\cdot \operatorname{p})\rho_n(|x|)\big)e(x)\\
&=& \rho_n(|x|) \,\frac{3}{1+|x|^2}e(x)
  -i\rho_n'(|x|)\,\frac{\alpha\cdot x}{|x|}e(x)\\
&=& \frac{3\rho_n(|x|)}{(1+|x|^2)^{5/2}}
    \left(\begin{array}{cccc} 1\\0\\ix_3\\ix_1-x_2\end{array}\right)
   +\frac{\rho_n'(|x|)}{|x|(1+|x|^2)^{3/2}}
    \left(\begin{array}{cccc} 
          |x|^2\\0\\-ix_3\\-ix_1+x_2\end{array}\right),
\end{eqnarray*}%\endaligned$$
we can estimate the $L^1$ norm of $(\alpha\cdot \operatorname{p})f_n$, 
noting $\rho_n'(|x|) = 0$ for $|x| \leq n$ and $|x| \geq n+2$ and using 
 polar coordinates, to get
\begin{eqnarray}%$$\aligned
 \|(\alpha\cdot \operatorname{p})f_n\|_1
&\leq& \int_{\{x\in {\Bbb R}^3; \,|x|\leq n+2\}}
           \frac{3(1+|ix_3|+|ix_1-x_2|)}{(1+|x|^2)^{5/2}}dx \nonumber\\  
&&\qquad+\int_{\{x\in {\Bbb R}^3; \,n \leq |x|\leq n+2\}}
    \frac{|x|^2+|-ix_3|+|-ix_1+x_2|}{|x|(1+|x|^2)^{3/2}} dx \nonumber\\
&\leq& \int_{|x|\leq n+2} \frac{3(\sqrt{2}|x|+1)}{(1+|x|^2)^{5/2}} dx
+\int_{n\leq |x|\leq n+2} \frac{|x|^2+\sqrt{2}|x|}{|x|(1+|x|^2)^{3/2}} dx 
                                                             \nonumber\\
&=& \int_0^{n+2} \frac{3(\sqrt{2}r+1)4\pi r^2 dr}{(1+r^2)^{5/2}}
 +\int_n^{n+2} \frac{(r+\sqrt{2})4\pi r^2 dr}{(1+r^2)^{3/2}} \nonumber\\
&\leq& 12\pi \int_0^{n+2} \frac{2}{1+r^2}dr +4\pi\int_n^{n+2} 2 dr 
                                                             \nonumber\\
&=& 24\pi\tan^{-1}(n+2) +16\pi
      \leq 24\pi\cdot\frac{\pi}2 + 16\pi\,,   
\end{eqnarray}%\endaligned\tag3.7$$
where we have used that
$(r+\sqrt{2})r^2 \leq 2(1+r^2)^{3/2}$
and $(\sqrt{2}r+1)r^2 \leq 2(1+r^2)^{3/2}$ for all $r\geq 0$.
Thus we have shown the sequence 
$\{\|(\alpha\cdot \operatorname{p})f_n\|_1\}$ is uniformly bounded.

Next, we study how  $\{f_n\}_{n=1}^{\infty}$ behaves in the norm of 
$B_{\infty,\infty}^{-2}({\Bbb R}^3; {\Bbb C}^4)$ for large $n$. In fact,
we shall show
\begin{equation}%
\|f_n\|_{{\Bbb B}_{\infty,\infty}^{-2}} = O(\log n).
\end{equation}%\tag3.8
Here note that $\frac{p}{p-q} = -\frac1{\frac32-1}=-2$. 
Indeed, we have with (3.4)
\begin{eqnarray*}%$$\aligned
 \|f_n\|_{B_{\infty,\infty}^{-2}}
&=& \sup_{t>0}\, t \|P_t f_n\|_{\infty}
= \sup_{t>0}\, t\sup_x \int \frac{1}{(4\pi t)^{3/2}}e^{-\frac{|x-y|^2}{4t}}
           \,\rho_n(|y|)|e(y)|_{\ell^{\infty}}dy\\
&\leq& \frac2{(4\pi)^{3/2}}\sup_{t>0}\, 
  \sup_x \int \Big(\frac{|x-y|^2}{4t}\Big)^{1/2}e^{-\frac{|x-y|^2}{4t}}
             \frac{\rho_n(|y|)}{|x-y|(1+|y|^2)}dy\\
&\leq& \frac2{(4\pi)^{3/2}}(2e)^{-1/2} \sup_x 
 \int_{|y| \leq n+2} \frac{1}{|x-y|(1+|y|^2)}dy\,,
\end{eqnarray*}%\endaligned$$
where the last inequality is due to the fact that 
$s^{1/2}e^{-s} \leq (2e)^{-1/2}$ for all $s>0$.
Then we use  polar coordinates to get
\begin{eqnarray*}%$$\aligned
 \|f_n\|_{{\Bbb B}_{\infty,\infty}^{-2}}
&\leq& 
\frac1{(4\pi)^{3/2}}\big(\frac2{e}\big)^{1/2}
 \sup_x\, \int_0^{n+2} \frac{r^2}{1+r^2} dr \int_0^{\pi}
\frac{2\pi\, \sin\theta d\theta}{(|x|^2+r^2 -2|x|r \cos\theta)^{1/2}}\\
&=\,& \frac{2\pi}{(4\pi)^{3/2}}\big(\frac2{e}\big)^{1/2} \sup_x\, \int_0^{n+2}
 \frac{r^2dr}{1+r^2} \Big[\frac{(|x|^2+r^2-2|x|r\cos\theta)^{1/2}}{|x|r}
      \Big]_{\theta=0}^{\theta=\pi}\\
&=\,&\frac1{2(2\pi e)^{1/2}} \sup_x\, 
  \frac1{|x|} \int_0^{n+2}\frac{r[(|x|+r)-\big||x|-r\big|]}{1+r^2} dr\\
&=\,& [\sup_{|x|\geq n+2} \vee \sup_{|x|\leq n+2}]\,
   \frac1{2(2\pi e)^{1/2}}\, 
  \frac1{|x|} \int_0^{n+2}\frac{r[(|x|+r)-\big||x|-r\big|]}{1+r^2} dr\\
&=:& V_{\alpha,1} \vee V_{\alpha,2} \,.
\end{eqnarray*}%\endaligned$$
Then we can conclude (3.8) above, noting
\begin{eqnarray*}%$$\aligned
2(2\pi e)^{1/2} V_{\alpha,1}
&=& \sup_{|x|\geq n+2}\, 
   \frac1{|x|} \int_0^{n+2} \frac{2r^2}{1+r^2} dr \leq 2, \\
2(2\pi e)^{1/2}V_{\alpha,2}
&=& \sup_{|x|\leq n+2}\,\frac1{|x|} 
 \Big[\int_0^{|x|} \frac{2r^2}{1+r^2} dr
     +\int_{|x|}^{n+2} \frac{2|x|r}{1+r^2} dr\Big]\\
%&= \frac1{2(2\pi e)^{1/2}} \sup_{|x|\leq n+2}\,\frac1{|x|}
%   \Big(2(|x| -\tan^{-1}|x|) + |x|\Big[\log(1+r^2)\Big]_{|x|}^{n+2}\Big)\\
&\leq& 2+\log(1+(n+2)^2) = O(\log n).
\end{eqnarray*}%\endaligned$$

Thus, by (3.8) and since, as already seen above, the sequence
$\{\|(\alpha\cdot \operatorname{p}) f_n\|_1\}$ 
is uniformly bounded, we see the sequence 
$\{\|(\alpha\cdot \operatorname{p}) f_n\|_1^{2/3}
        \|f_n\|_{{\Bbb B}_{\infty,\infty}^{-2}}^{1/3}\}$ 
on the right-hand side of (3.6) is of order $O((\log n)^{1/3})$, 
while, for the left-hand side, we have by (3.5) with $q=\frac32$
\begin{eqnarray}%$$\aligned
\|f_n\|_{3/2} 
\!\!&\!\geq\!&\!\! 
   \Big(\int_{|x|\leq n} \frac1{(1+|x|^2)^{3/2}} dx\Big)^{2/3} 
= \Big(\int_0^n \frac{4\pi\, r^2}{(1+r^2)^{3/2}} dr \Big)^{2/3} 
                                                   \nonumber\\
\!\!&\!\geq\!&\!\!
   \Big(\int_1^n \frac{4\pi\, r^2}{(1+r^2)^{3/2}} dr \Big)^{2/3}
\geq  \Big(\int_1^n \frac{4\pi}{r}dr \Big)^{2/3} 
\geq (4\pi)^{2/3} (\log n)^{2/3}.
\end{eqnarray}%\endaligned\tag3.9$$ 
This means that inequality (3.6) or (1.16)/(1.30)  with 
$p=1,\, q=\frac32$ does not hold.

\bigskip\noindent
{\it An example for \hbox{\rm (1.59)} of Theorem \hbox{\rm 1.6}  
with $p=1$ to fail to hold.}

This case is with $x \in {\Bbb R}^4$ and  
$|x| = (x_1^2+x_2^2+x_3^2+x_4^2)^{1/2}$.
We can use the same arguments as above to construct 
a sequence 
$\{f_n\}$ in $C_0^{\infty}({\Bbb R}^4; {\Bbb C}^4)$ such that 
(1.59) fails to hold for any fixed constant $C$,
starting, instead of (3.2), from the following function
\begin{eqnarray}%$$\aligned
\hat{e}(x)
&:=& \frac1{(1+|x|^2)^{2}}(I_4+ i\beta\cdot x)
     \left(\begin{array}{cccc} 1\\ 0\\ 0\\ 0 \end{array}\right)
                                                  \nonumber\\
&\,=& \frac1{(1+|x|^2)^{2}}
    \left(\begin{array}{cccc}
    1&0&ix_3+x_4&ix_1+x_2 \\ 
    0&1&ix_1-x_2&-ix_3+x_4\\
    ix_3-x_4&ix_1+x_2&1&0\\
    ix_1-x_2&-ix_3-x_4&0&1\end{array}\right)
    \left(\begin{array}{cccc} 1\\ 0\\ 0\\ 0 \end{array}\right) 
                                                  \nonumber\\
&=& \frac1{(1+|x|^2)^{2}}
  \left(\begin{array}{cccc} 1\\0\\ix_3-x_4\\ix_1-x_2 \end{array}\right). 
\end{eqnarray}%\endaligned\tag3.10$$
It can be seen that $\hat{e}(x)$ satisfies the following equation 
\begin{equation}%
(\beta\cdot \operatorname{p})\hat{e}(x) 
 = \frac4{1+|x|^2} \hat{e}(x)\,,                              
\end{equation}%\tag3.11
and inequalities:
\begin{eqnarray}%$$\align
|\hat{e}(x)|_{\ell^{\infty}} 
  &=& \frac{1 \vee |ix_3-x_4| \vee |ix_1-x_2|}{(1+|x|^2)^{2}}
   \leq  \frac{1}{(1+|x|^2)^{3/2}}\,,                       \\%\tag3.12\\ 
|\hat{e}(x)|_{\ell^{q}}^q
  &=& \frac{1+|ix_3-x_4|^q + |ix_1-x_2|^q}{(1+|x|^2)^{2q}}
   = \frac{1+(x_3^2+x_4^2)^{q/2} + (x_1^2+x_2^2)^{q/2}}{(1+|x|^2)^{2q}} 
                   \nonumber\\
  &\geq& \frac{(1+|x|^2)^{q/2}}{(1+|x|^2)^{2q}}
   = \Big(\frac{1}{1+|x|^2}\Big)^{3q/2} \qquad (1\leq q \leq 2)\,.%\tag3.13
\end{eqnarray}%\endalign$$ 

For each positive integer $n$, put 
$f_n(x) = \rho_n(|x|)\hat{e}(x)$, where $\rho_n(r)$ is the same nonnegative
cutoff function in $ C_0^{\infty}({\Bbb R})$ as before such that 
 $\rho_n(r) = 1\, \, (r \leq n)\,\,; \, = 0 \,\,(r \geq n+2)$, and further 
$|\rho_n'(r)| \equiv |(d/dr) \rho_n(r)| \leq 1$  for all $r \geq 0$.
Then it is evident that $f_n$ belongs to $C_0^{\infty}({\Bbb R}^4;{\Bbb C}^4)$.

We are going to see that 
inequality (1.59), corresponding to (1.16) in the case for 
$\beta\cdot \operatorname{p}$, does not hold with any constant $C>0$ 
for $p=1$, $q=\frac43$ and hence $\frac{p}{q} =\frac34$. Indeed, 
there exists no constant $C$ such that, for all $n$,
\begin{equation}%
\|f_n\|_{4/3} \leq C \|(\alpha\cdot \operatorname{p})f_n\|_{1}^{3/4}
  \|f_n\|^{1/4}_{B^{-3}_{\infty,\infty}}\,.
\end{equation}%\tag3.14

First, we show that the sequence 
$\{(\beta\cdot \operatorname{p})f_n\}_{n=1}^{\infty}$
is uniformly bounded in $L^1$. Indeed, since 
\begin{eqnarray*}%$$\aligned
 (\beta\cdot \operatorname{p})f_n(x)
&=& \rho_n(|x|) (\beta\cdot \operatorname{p})\hat{e}(x)
  +\big((\beta\cdot \operatorname{p})\rho_n(|x|)\big)\hat{e}(x)\\
&=& \rho_n(|x|) \,\frac{4}{1+|x|^2}\hat{e}(x)
   -i\rho_n'(|x|)\,\frac{\beta\cdot x}{|x|}\hat{e}(x)\\
&=& \frac{4\rho_n(|x|)}{(1+|x|^2)^{3}}
    \left(\begin{array}{cccc} 1\\0\\ix_3-x_4\\ix_1-x_2\end{array}\right)
  +\frac{\rho_n'(|x|)}{|x|(1+|x|^2)^{2}}
    \left(\begin{array}{cccc} 
           |x|^2\\0\\-ix_3+x_4\\-ix_1+x_2\end{array}\right),
\end{eqnarray*}%\endaligned$$
we can estimate the $L^1$ norm of $(\beta\cdot \operatorname{p})f_n$, 
noting $\rho_n'(|x|) = 0$ for $|x| \leq n$ and $|x| \geq n+2$ and using 
 polar coordinates, to get
\begin{eqnarray}%$$\aligned
 \|(\beta\cdot \operatorname{p})f_n\|_1
&\leq& \int_{\{x\in {\Bbb R}^4; \,|x|\leq n+2\}}
      \frac{4(1+|ix_3-x_4|+|ix_1-x_2|)}{(1+|x|^2)^{3}}dx \nonumber\\
  &&\qquad+\int_{\{x\in {\Bbb R}^4; \,n \leq |x|\leq n+2\}}
   \frac{|x|^2+|-ix_3+x_4|+|-ix_1+x_2|}{|x|(1+|x|^2)^{2}} dx \nonumber\\
&\leq& \int_{|x|\leq n+2}\frac{4(1+\sqrt{2}|x|)}{(1+|x|^2)^{3}}dx
    +\int_{n \leq |x|\leq n+2}
      \frac{|x|^2+\sqrt{2}|x|}{|x|(1+|x|^2)^{2}} dx \nonumber\\
&=& \int_0^{n+2}\frac{4(1+\sqrt{2}r) 2\pi^2 r^3 dr}{(1+r^2)^{3}}
  +\int_n^{n+2}\frac{(r+\sqrt{2}) 2\pi^2 r^3 dr}{(1+r^2)^{2}} \nonumber\\
&\leq& 8\pi^2 \int_0^{n+2} \frac{2}{1+r^2} dr
     +2\pi^2 \int_n^{n+2}2 dr                                 \nonumber\\
&=&16\pi^2\tan^{-1}(n+2) + 8\pi^2
\leq  16\pi^2\cdot \frac{\pi}2 +8\pi^2,
\end{eqnarray}%\endaligned\tag3.15$$
where in the second inequality we have used that
$(r+\sqrt{2})r^3 \leq 2(1+r^2)^2$ and 
 $(1+\sqrt{2}r)r^3 \leq 2(1+r^2)^2$ for all $r\geq 0$. 
Thus we have shown 
the sequence $\{\|(\alpha\cdot \operatorname{p})f_n\|_1\}$ 
is uniformly bounded.

Next, we study how $\{f_n\}_{n=1}^{\infty}$ behaves in the norm of 
$B_{\infty,\infty}^{-3}({\Bbb R}^4; {\Bbb C}^4)$ for large $n$.
In fact, we show
\begin{equation}%
\|f_n\|_{{\Bbb B}_{\infty,\infty}^{-3}} = O(\log n).
\end{equation}%\tag3.16  
Here we note that $\frac{p}{p-q} = -\frac1{\frac43-1}=-3$.
Indeed, we have with (3.12)
\begin{eqnarray*}%$$\aligned
 \|f_n\|_{B_{\infty,\infty}^{-3}}
&=& \sup_{t>0}\, t^{3/2} \|P_t f_n\|_{\infty}
= \sup_{t>0}\, t^{3/2}\sup_x \int_{{\Bbb R}^4} 
   \frac{1}{(4\pi t)^2}e^{-\frac{|x-y|^2}{4t}}
           \,\rho_n(|y|)|\hat{e}(y)|_{\ell^{\infty}}dy\\
&\leq& \frac2{(4\pi)^{2}}\sup_{t>0}\, 
  \sup_x \int \Big(\frac{|x-y|^2}{4t}\Big)^{1/2}e^{-\frac{|x-y|^2}{4t}}
             \frac{\rho_n(|y|)}{|x-y|(1+|y|^2)^{3/2}} dy\\
&\leq& \frac2{(4\pi)^{2}}(2e)^{-1/2} \sup_x 
 \int_{|y| \leq n+2} \frac{1}{|x-y|(1+|y|^2)^{3/2}}dy\,,
\end{eqnarray*}%\endaligned$$
where the last inequality is due to the fact that 
$s^{1/2}e^{-s} \leq (2e)^{-1/2}$ for all $s>0$.
Then we use polar coordinates and 
$sin^2\theta \leq \sin\theta \, (0\leq \theta \leq\pi)$ to get 
\begin{eqnarray*}%$$\aligned
 \|f_n\|_{{\Bbb B}_{\infty,\infty}^{-3}}
&\leq& 
\frac2{(4\pi)^{2}}(2e)^{-1/2}
 \sup_x\, \int_0^{n+2}\frac{r^3}{(1+r^2)^{3/2}} dr 
   \int_0^{\pi} \frac{4\pi\,\sin^2\theta d\theta}
   {(|x|^2+r^2 -2|x|r \cos\theta)^{1/2}}\\
&\leq\,& \frac2{(4\pi)^{2}}(2e)^{-1/2}
 \sup_x\, \int_0^{n+2} \frac{r^3}{(1+r^2)^{3/2}} dr 
\int_0^{\pi}\frac{4\pi\,\sin\theta d\theta}
   {(|x|^2+r^2 -2|x|r \cos\theta)^{1/2}}\\
&=\,& \frac{4\pi}{(4\pi)^{2}} (2e)^{-1/2} 
 \sup_x\,\int_0^{n+2} \frac{r^3}{(1+r^2)^{3/2}} dr 
 \Big[\frac{(|x|^2+r^2-2|x|r\cos\theta)^{1/2}}{|x|r}
      \Big]_{\theta=0}^{\theta=\pi}\\
&=\,&\frac1{4\pi(2e)^{1/2}} \sup_x\,  \frac1{|x|} 
\int_0^{n+2}\frac{r^2((|x|+r)-\big||x|-r\big|)}{(1+r^2)^{3/2}} dr\\
&=\,& [\sup_{|x|\geq n+2}\vee \sup_{|x|\leq n+2}]\,
\frac1{4\pi(2e)^{1/2}}\,  \frac1{|x|} 
\int_0^{n+2}\frac{r^2((|x|+r)-\big||x|-r\big|)}{(1+r^2)^{3/2}} dr\\
&=:& V_{\beta,1} + V_{\beta,2}\,.
\end{eqnarray*}%\endaligned$$
Then we can conclude (3.16), noting
\begin{eqnarray*}%$$\aligned
4\pi(2e)^{1/2} V_{\beta,1}
&=& \sup_{|x|\geq n+2}\, 
   \frac1{|x|} \int_0^{n+2} \frac{2r^3}{(1+r^2)^{3/2}} dr \leq 2, \\
4\pi(2e)^{1/2} V_{\beta,2}
&=&\sup_{|x| \leq n+2}\,\frac1{|x|} \Big\{
  \int_0^{|x|} \frac{2r^3}{(1+r^2)^{3/2}} dr
     +\int_{|x|}^{n+2} \frac{2|x|r^2}{(1+r^2)^{3/2}} dr\Big\}\\
%&= \sup_{|x|\leq n+2}\, \frac1{|x|}\Big\{
%2\Big[(1+r^2)^{1/2}+(1+r^2)^{-1/2}\Big]_0^{|x|} \\
% &\qquad +
%   2|x|\Big(\Big[\log(r+(1+r^2)^{1/2})\Big]_{r=|x|}^{r=n+2}
% -\Big[\sin\theta \Big]_{\theta=\tan^{-1}|x|}^{\theta=\tan^{-1}(n+2)}\Big)
%  \Big\}  \\
&\leq& 2 + \log[(n+2)+(1+(n+2)^2)^{1/2}]= O(\log n).
\end{eqnarray*}%\endaligned$$

Thus, by (3.16) and since, as already seen above, the sequence
$\{\|(\beta\cdot \operatorname{p}) f_n\|_1\}$ 
is uniformly bounded, we see the sequence 
$\{\|(\beta\cdot \operatorname{p}) f_n\|_1^{3/4}
        \|f_n\|_{{\Bbb B}_{\infty,\infty}^{-3}}^{1/4}\}$ 
on the right-hand side of (3.14) is of order $O((\log n)^{1/4})$, 
while, for the left-hand side, we have by (3.13) with $q=\frac43$
\begin{eqnarray}%$$\aligned
\|f_n\|_{4/3} 
\!\!&\!\geq\!&\!\! 
 \Big(\int_{|x|\leq n} |\hat{e}(x)|_{\ell^{4/3}}^{4/3} dx \Big)^{3/4}
 \geq \Big(\int_{|x|\leq n} \Big(\frac{1}{1+|x|^2}\Big)^{(3/2)\cdot(4/3)}dx
\Big)^{3/4}                      \nonumber\\
\!\!&\!=\!&\!\! 
 \Big(\int_0^n \frac{2\pi^2 r^3 dr}{(1+r^2)^2}\Big)^{3/4}
 = O((\log n)^{3/4})
\end{eqnarray}%\endaligned\tag3.17$$ 
for large $n$. This means that inequality (3.14) or (1.59)  
with $p=1,\, q=\frac43$ does not hold.

%%%%%%%%%%%%%%%%%%%%%%%%%%%%%%%

\medskip
\section{Proof of Theorem 1.1}%sect.4

{\it Proof of Theorem \hbox{\rm 1.1 (i)}}. 
We follow the lucid arguments used in Ledoux [Le].
The proof is divided into three steps. 
In step I, we mention the weak-type inequality (1.14) given by [BES] 
with the idea of [Le] to sketch its proof,  
for the paper to be somehow self-contained. 
In step II we show the inequlity (1.28) in the special case
under the condition $f \in L^q({\bf R}^3;{\bf C}^4)$ and then the general case
in step III. 

\medskip
I. So we begin with  a sketch of proof of inequality (1.14). 

To do so, assume that $f$ satisfies 
$M_{\alpha\cdot \operatorname{p}; p}(f) < \infty$. 
Note that this implies with (1.22) that 
$\|(\alpha\cdot \operatorname{p})f\|_p < \infty$.
And further assume that our $f$ satisfies
$\|f\|_{B^{p/(p-q)}_{\infty,\infty}} < \infty$.
We may suppose by our convention (1.7) of notations and by homogeneity that 
\begin{equation}%
\|f\|_{B^{p/(p-q)}_{\infty,\infty}} 
= \sup_{t>0} t^{-p/2(p-q)}\|P_t f\|_{\infty} \leq 1.
\end{equation}%\tag4.1
Therefore $|P_t f|_{\ell^{\infty}} \leq t^{p/2(p-q)}$ pointwise. 
For $u>0$, put  $t=t_u \equiv u^{2(p-q)/p}$, so that
$|P_{t_u}f|_{\ell^{\infty}} \leq u$. Hence that $|f|_{\ell^{\infty}} \geq 2u$
pointwise implies that 
$|f-P_{t_u}f|_{\ell^{\infty}} \geq |f|_{\ell^{\infty}}- |P_t f|_{\infty}
\geq u$ pointwise.
Then
\begin{eqnarray*}%$$\aligned
u^q \big|\{|f|_{\ell^{\infty}}\geq 2u\}\big|
&\leq& u^q \big|\{|f-P_{t_u}f|_{\ell^{\infty}}\geq u\}\big|\\
&\leq& u^{q}\int\frac{|f-P_{t_u}f|_{\ell^{\infty}}^p}{u^p}dx
= u^{q}\int \vee_{k=1}^4 \frac{|f_k-e^{t_u\Delta}f_k|^p}{u^p}dx\\
&\leq& u^{q-p} \int \sum_{k=1}^4 |f_k- e^{t_u\Delta}f_k|^p dx\\
&=& u^{q-p} \int |f-P_{t_u}f|_{\ell^{p}}^p dx
= u^{q-p} \|f-P_{t_u}f\|_p^p.
\end{eqnarray*}%\endaligned$$

In [BES], it is shown that
\begin{equation}% 
\|f-P_{t_u}f\|_p 
\leq c_0 {t_u}^{1/2}\|(\alpha\cdot \operatorname{p})f\|_p.
\end{equation}%\tag4.2$$
with a positive constant $c_0$ depending only on $p$. 
Then by (4.2) and  since $q-p+p(p-q)/p =0$, we have
$$
u^q \big|\{|f|_{\ell^{\infty}} \geq 2u\}\big|
\leq c_0 u^{q-p} t_u^{p/2}\int |(\alpha\cdot \operatorname{p})f|_{\ell^p}^p dx
= c_0\int |(\alpha\cdot \operatorname{p})f|_{\ell^p}^p dx. 
$$
This yields the weak type inequality (1.14), taking account of 
definition of $\|f\|_{q,\infty}$ in (1.15). 

%%%%
\medskip
II. Next we want to replace the weak $L^q$ norm on the left-hand side of (1.14)
by the strong $L^q$ norm. Here we note with (1.22) that (1.14) holds also with 
$M_{\alpha\cdot \operatorname{p}; p}(f)$ in place of
$\|(\alpha\cdot \operatorname{p})f\|_p$. We show inequality (1.28) for 
$f$ which satisfies
$M_{\alpha\cdot \operatorname{p}; p}(f) < \infty$
and (4.1), i.e. $\|f\|_{B^{p/(p-q)}_{\infty,\infty}} \leq 1$, as in step I,
and the extra condition $f \in L^q({\Bbb R}^3;{\Bbb C}^4)$. 
In step III below, we shall remove this latter condition.  

Then what we need to show is that there exists a constant $C$ 
(depending only on $q$ and $p$) such that
\begin{equation}%
\int |f|_{\ell^q}^q dx 
\leq C\,M_{\alpha\cdot \operatorname{p}; p}(f)^p,
\end{equation}%\tag4.3$$
which amounts to our goal inequality (1.28), 
if only $f$ replaced by $f/\|f\|_{B^{p/(p-q)}_{\infty,\infty}}$ in (4.3).

\medskip
Now, for $u>0$, let $t=t_u =u^{2(p-q)/p}$ again.
Let $c\geq 5$ (depending on $q$ and $p$) to be specified later.

Note the `layer cake' representation [LLo, p.26, Theorem 1.13]
for any nonnegative measurable function $\psi(x)$:
\begin{equation}%
 \psi(x) = \int_0^{\infty} \chi_{\{\psi >s\}}(x) \,ds.
\end{equation}%\tag4.4$$
In particular, we have 
$$
|f(x)|_{\ell^q}^q = \int_0^{\infty} \chi_{\{|f|_{\ell^q}^q>s\}}(x) \,ds
            = \int_0^{\infty} \chi_{\{|f|_{\ell^q} >u\}}(x) \,d(u^q),
$$
with $d(u^q) = qu^{q-1}du$, so that by Fubini's theorem 
\begin{eqnarray}%$$\aligned
 \frac1{20^q}\|f\|_q^q 
&=& \frac1{20^q}\int |f(x)|_{\ell^q}^q dx
= \frac1{20^q}\int dx 
    \int_0^{\infty} \chi_{\{|f|_{\ell^q} > u\}}(x) \,d(u^q) \nonumber\\
&=& \frac1{20^q}\int_0^{\infty} d(u^q) 
  \int \chi_{\{|f|_{\ell^q} \geq u\}}(x) dx 
= \int_0^{\infty} \big|\{|f|_{\ell^q} \geq 20 u\}\big| d(u^q).\quad	
\end{eqnarray}%\endaligned\tag4.5$$
For every $u>0$ and for $f(x) = {}^t(f_1(x),f_2(x),f_3(x),f_4(x))$, let
\begin{eqnarray}%$$\aligned
&&f_u(x) = {}^t(f_{u,1}(x),f_{u,2}(x),f_{u,3}(x),f_{u,4}(x)),\nonumber\\
&&f_{u,k}(x) 
:= (f_k(x)-u)^+ \wedge ((c-1)u) + (f_k(x)+u)^- \vee (-(c-1)u),
\quad k=1,2,3,4,                                             \nonumber\\
\end{eqnarray}%\endaligned\tag4.6$$
for any $c>1$. 
Here, as in (1.7), $a\vee b$ denotes $\max\{a,b\}$, while
$a\wedge b$ denotes $\min\{a,b\}$.

Notice that  $f_u$ also satisfies the same condition as $f$.
Each $f_{u,k}(x)$ satisfies $0\leq |f_{u,k}(x)| \leq (c-1)u$. 
It vanishes when $|f_k(x)| \leq u$ and is equal to $(c-1)u$
when $f_k(x) \geq cu$, and to $-(c-1)u$ when $f_k(x) \leq -cu$.

%%%%%%%%
We see that, 
since on the set $\{|f_k| \geq 5u\}$, we have 
$|f_{u,k}| \geq 4u$ for each fixed $k$, and that 
on the set $\{|f|_{\ell^{\infty}} \geq 5u\}$, we have 
$|f_u|_{\ell^{\infty}} \geq 4u$.
We have
\begin{equation}%
  |f_{u,k}| \leq |f_{u,k} -e^{t_u\Delta}f_{u,k}|
            + e^{t_u\Delta}|f_{u,k}-f_k|+ |e^{t_u\Delta}f_k|,
\quad k=1,2,3,4.
\end{equation}%\tag4.7$$
By noting the notation (1.7) of the $\ell^p$/$\ell^{\infty}$ norm
of a four-vector we have
\begin{eqnarray}%$$\aligned
\int_0^{\infty} \big|\{|f|_{\ell^q} \geq 20u\}\big| d(u^q)
&\leq\,& \int_0^{\infty} \big|\{|f|_{\ell^{\infty}} \geq 5u\}\big| d(u^q)
    \nonumber\\
&\leq\,& \int_0^{\infty}\big|\{|f_u|_{\infty} \geq 4u\}\big| d(u^q)
= \int_0^{\infty}\big|\{\vee_{k=1}^4 |f_{u,k}| \geq 4u \}\big| d(u^q)
    \nonumber\\
&\leq\,& \int_0^{\infty}
\big|\{\vee_{k=1}^4|f_{u,k}- e^{t_u\Delta}f_{u,k}| \geq u \}
   \big| d(u^q) \nonumber\\
&& +\int_0^{\infty}\big|
\{\vee_{k=1}^4 e^{t_u\Delta}|f_{u,k}- f_k| \geq 2u \}\big|d(u^q)\nonumber\\
&=:& J_1+J_2,
\end{eqnarray}%\endaligned\tag4.8$$
where we have used the fact that $|P_{t_u}(f)|_{\ell^{\infty}}\leq u$,
which holds by our choice of $f$ in (4.1).

We shall estimate the last member $J_1 + J_2$ of (4.8).
First, to treat the second term $J_2$, 
we confirm that 
\begin{equation}%
|f_{u,k}-f_k|
= |f_{u,k}-f_k|\chi_{\{|f_k| \leq cu\}}
 +|f_{u,k}-f_k|\chi_{\{|f_k| > cu\}} 
\leq u +|f_k|\chi_{\{|f_k| > cu\}}.
\end{equation}%\tag4.9$$
This is checked with (4.6) as follows. Indeed, we see (4.6) imply that
\[ f_{u,k}(x)-f_k(x)= \left\{\begin{array}{rl}
       (-u) \wedge (-f_k(x)+(c-1)u), 
                &\quad \hbox{\rm if}\,\, f_k(x) \geq u\,,\\
        u \vee (-f_k(x)-(c-1)u), 
                &\quad \hbox{\rm if}\,\, f_k(x) \leq -u\,.
 \end{array}\right.\]
This further implies on the one hand that
\[ f_{u,k}(x)-f_k(x)= \left\{\begin{array}{rl}
         -u, &\quad \hbox{\rm if}\,\, u\leq f_k(x)\leq cu\,,\\
         u, &\quad \hbox{\rm if}\,\, -u \geq f_k(x) \geq -cu\,,
 \end{array}\right. \]
so that $|f_{u,k}(x)-f_k(x)|= u$, if $u\leq|f_k(x)|\leq cu$,
and on the other hand that 
\[
 f_{u,k}(x)-f_k(x)= \left\{\begin{array}{rl}
         -f_k(x)+(c-1)u \geq -f_k(x), 
               &\quad \hbox{\rm if}\,\, f_k(x)\geq cu\,,\\
         -f_k(x)-(c-1)u \leq -f_k(x), 
               &\quad \hbox{\rm if}\,\, f_k(x)\leq -cu\,,
 \end{array}\right. \]
so that $|f_{u,k}(x)-f_k(x)|\leq |f_k(x)|$, if $|f_k(x)|\geq cu$.
This yields (4.9).

\medskip 
Then, since $e^{t_u\Delta}$ is positivity-preserving, it follows that
\begin{eqnarray}%$$\aligned
J_2 &=& \int_0^{\infty} 
\big|\{\vee_{k=1}^4 e^{t_u\Delta}|f_{u,k}- f_k| \geq 2u \}\big|d(u^q)
     \nonumber\\
&\leq& \int_0^{\infty} 
  \big|\{\vee_{k=1}^4 e^{t_u\Delta}|f_k|
                  \chi_{\{|f_k|>cu\}} \geq u\}\big|d(u^q)\nonumber\\
&\leq& \int_0^{\infty} 
\Big(\int \vee_{k=1}^4 \frac{e^{t_u\Delta}|f_k|\chi_{\{|f_k|> cu\}}}
{u} dx\Big)d(u^q)\nonumber\\
&=&\int_0^{\infty} 
\frac1{u}\Big(\int \vee_{k=1}^4 |f_k|\chi_{\{|f_k|> cu\}}dx\Big)d(u^q)
\leq \int_0^{\infty} 
\frac1{u}\Big(\int |f(x)|_{\ell^{\infty}}
   \chi_{\{|f|_{\ell^{\infty}} > cu\}}dx\Big)d(u^q)\nonumber\\
&&\leq \frac{q}{q-1}\int |f|_{\ell^q}
\Big(\int_0^{\infty}\chi_{\{|f|_{\ell^q} > cu\}}d(u^{q-1})\Big) dx 
=\frac{q}{q-1}\frac1{c^{q-1}}\|f\|_q^q.
\end{eqnarray}%\endaligned\tag4.10$$
Here the last fourth equality is due to that
\begin{eqnarray*}%$$\aligned
\int e^{t_u\Delta}|f_k|\chi_{\{|f_k|>cu\}} dx
&=& \int \Big(\int(e^{t_u\Delta}(x-y)|f_k(y)|
    \chi_{\{|f_k(y)|>cu\}} dy\Big) dx\\ 
&=&\int|f_k(y)|\chi_{\{|f_k(y)|>cu\}} dy,
\end{eqnarray*}%\endaligned$$
because the heat kernel $e^{t_u\Delta}(x-y)$ satisfies
$\int e^{t_u\Delta}(x)dx =1$ for $t_u>0$, and 
the last second inequality is due to that 
$\vee_{k=1}^4 |f_k(x)| \leq |f(x)|_{\ell^{\infty}}
                               \leq |f(x)|_{\ell^q}$ by (1.7).

Next, as for the first term $J_1$ of the last member of (4.8), 
we have by (4.2)
\begin{eqnarray*}%$$\aligned
\big|\{\vee_{k=1}^4 |f_{u,k}- e^{t_u\Delta}f_{u,k}| \geq u\}\big|
&\leq& \int \vee_{k=1}^4 
\frac{|f_{u,k}- e^{t_u\Delta}f_{u,k}|^p}{u^p} dx\\
&\leq& u^{-p}\int \sum_{k=1}^4|f_{u,k}- e^{t_u\Delta}f_{u,k}|^p dx\\
&=& u^{-p}\int |f_{u}- P_{t_u}(f_{u})|_{\ell^p}^p dx\\
&\leq& c_0u^{-p}{t_u}^{p/2}
\int |(\alpha\cdot \operatorname{p})f_u|_{\ell^p}^p dx
= c_0u^{-q}\|(\alpha\cdot \operatorname{p})f_u\|_p^p\\ 
&\leq& C_0u^{-q}M_{\alpha\cdot \operatorname{p}; p}(f_u)^p\,,
\end{eqnarray*}%\endaligned$$
with $C_0 := 2^{1-(1/p)}c_0$, where the last inequality is due to 
(1.22), so that
\begin{equation}%$$\aligned
 J_1 \leq C_0 \int_0^{\infty} d(u^q) 
  u^{-q} M_{\alpha\cdot \operatorname{p}; p}(f_u)^p\,.
\end{equation}%\endaligned\tag4.11$$

\medskip
For (4.11), we want to show the following lemma. 

\begin{lemma}%{Lemma 4.1} 
Let $f = {}^t(f_1,f_2,f_3,f_4)$ satisfy 
$M_{\alpha\cdot \operatorname{p}; p}(f) <\infty$ and 
$\|f\|_{B^{p/(p-q)}_{\infty,\infty}} \leq 1$. Let
$f_u = {}^t(f_{u,1},f_{u,2},f_{u,3},f_{u,4})$ as in (4.6).
Then 
\begin{equation}%
 \int_0^{\infty} d(u^q) 
u^{-q} M_{\alpha\cdot \operatorname{p}; p}(f_u)^p 
= q (\log c) M_{\alpha\cdot \operatorname{p}; p}(f)^p.
\end{equation}%\tag4.12$$
\end{lemma}

{\it Proof}. For $f_u$ in (4.6) instead of $f$, 
we have by (1.21)
\begin{eqnarray}%$$\aligned
&&M_{\alpha\cdot \operatorname{p}; p}(f_u)^p \nonumber\\
&=\,& \int \big(|(\partial_1+i\partial_2) f_{u,1}|^p 
             +|\partial_3 f_{u,1}|^p\big) dx
 +\int \big(|(\partial_1-i\partial_2) f_{u,2}|^p 
             +|\partial_3 f_{u,2}|^p\big) dx \nonumber\\
&&+\int \big(|(\partial_1+i\partial_2) f_{u,3}|^p 
             +|\partial_3 f_{u,3}|^p\big) dx
 +\int \big(|(\partial_1-i\partial_2) f_{u,4}|^p  
             +|\partial_3 f_{u,4}|^p\big) dx \nonumber\\
& =:& F_1(u) +F_2(u) + F_3(u) + F_4(u).
\end{eqnarray}%\endaligned\tag4.13$$
Therefore
$$
 \int_0^{\infty} d(u^q) 
u^{-q} M_{\alpha\cdot \operatorname{p}; p}(f_u)^p
 = \sum_{k=1}^4 \int_0^{\infty} d(u^q) u^{-q}
   [F_1(u) +F_2(u) + F_3(u) + F_4(u)].
$$

We compute the integral of the first term on the right-hand side concerning  
$F_1(u)$. Before that, we note that
\begin{eqnarray}%$$\aligned
 F_1(u) 
&=&  \int_{u\leq |f_1(x)| \leq cu}
  (|(\partial_1+i\partial_2) f_{u,1}|^p +|\partial_3 f_{u,1}|^p) dx
            \nonumber\\
&=& \int_{u\leq |f_1(x)| \leq cu}
  (|(\partial_1+i\partial_2) f_{1}|^p +|\partial_3 f_{1}|^p) dx\,,
\end{eqnarray}%\endaligned\tag4.14$$
as the $x$-integration in the third member of (4.14)
may be done only on the set
$\{x;\,u\leq |f_1(x)| \leq cu\}$
 because $f_{u,1}(x)=0$ when $|f_1(x)|\leq u$, 
and $f_{u,1}(x)$ is constant (with $|f_{u,1}(x)|=(c-1)u$) 
when $|f_1(x)|\geq cu$.  
Further, the last equality in (4.14) is due to the fact that 
$\partial_j f_{u,1}(x)= \partial_j f_1(x),\,\, j=1,2,3$,
on the set $\{x;\, u\leq |f_1(x)| \leq cu\}$.

Thus, through (4.14) we have  
\begin{eqnarray}%$$\aligned
 \int_0^{\infty} d(u^q) u^{-q} F_1(u)
&=& \int_0^{\infty} d(u^q) u^{-q}  \int_{u\leq |f_1(x)| \leq cu}
  (|(\partial_1+i\partial_2) f_{1}|^p +|\partial_3 f_{1}|^p) dx
      \nonumber\\
&=&  q \int\, dx
  (|(\partial_1+i\partial_2) f_{1}|^p +|\partial_3 f_{1}|^p) 
  \int_{\frac{|f_1(x)|}{c}}^{|f_1(x)|} \frac{du}{u} \nonumber\\
&=& q (\log c) \int
  (|(\partial_1+i\partial_2) f_{1}|^p +|\partial_3 f_{1}|^p) dx.
\end{eqnarray}%\endaligned\tag4.15$$
In the same way for $F_2(u),\, F_3(u),\, F_4(u)$ in (4.13), we can get
\begin{eqnarray*}%$$\aligned
 \int_0^{\infty} d(u^q) u^{-q} F_2(u)
&=& q (\log c) \int
  (|(\partial_1-i\partial_2) f_{2}|^p +|\partial_3 f_{2}|^p) dx,\\
\int_0^{\infty} d(u^q) u^{-q} F_3(u)
&=& q (\log c) \int
  (|(\partial_1+i\partial_2) f_{3}|^p +|\partial_3 f_{3}|^p]) dx,\\
\int_0^{\infty} d(u^q) u^{-q} F_4(u)
&=& q (\log c) \int
  (|(\partial_1-i\partial_2) f_{4}|^p +|\partial_3 f_{4}|^p)dx.\\
\end{eqnarray*}%\endaligned$$
So we obtain
\begin{eqnarray*}%$$\aligned
 \int_0^{\infty} d(u^q) 
u^{-q} M_{\alpha\cdot \operatorname{p}; p}(f_u)^p
&=& q (\log c) \, 
  \big[\|(\alpha\cdot \operatorname{p})P_{13}f\|_p^p 
+ \|(\alpha\cdot \operatorname{p})P_{24}f\|_p^p\big] \nonumber\\
&=& q(\log c) M_{\alpha\cdot \operatorname{p}; p}(f)^p,
\end{eqnarray*}%\endaligned$$
establishing (4.12) of the lemma.    
\qed 

\medskip
Then, noting (4.5)/(4.8) to put together (4.10) and (4.11) 
with Lemma 4.1, we get 
\begin{equation}
\frac1{{20}^q}\|f\|_q^q 
\leq  C_0 q (\log c) M_{\alpha\cdot \operatorname{p}; p}(f)^p
  + \frac{q}{q-1}\frac1{c^{q-1}}\|f\|_q^q\,.
\end{equation}%4.16
Thus, since $\|f\|_q$ is finite by assumption, 
taking $c$ sufficiently large  in (4.16) and putting
$C= \frac{C_0 q(\log c)}{\frac1{20^q}-\frac{q}{q-1}\frac1{c^{q-1}}}$,
 we have shown the desired inequality (4.3) in step II. 
In the whole arguments in step II we need the condition 
$f \in L^q({\Bbb R}^3;{\Bbb C}^4)$, i.e. that $\|f\|_q < \infty$, 
only here in (4.16) so that we can obtain inequality (4.3) from (4.16).

%%%%%%
\medskip
III.  Finally we show that if  
$M_{\alpha\cdot \operatorname{p}; p}(f) < \infty$ and
$\|f\|_{B^{p/(p-q)}({\Bbb R}^3;{\Bbb C}^4)} \leq 1$, then
$f \in L^q({\Bbb R}^3;{\Bbb C}^4)$, and  that 
$\|f\|_q \leq C\,M_{\alpha\cdot \operatorname{p}; p}(f)$ 
with a constant $C$ independent of $f$.
 
We already know by the weak type inequality (1.14) that
$\|f\|_{q,\infty} <\infty$. Therefore, in view of the second member 
 of (4.8), we may consider, for every $0<\varepsilon <1$,
\begin{equation}%
N_{\varepsilon}(f) := \int_{u= \varepsilon}^{u= 1/\varepsilon}
\big|\{|f|_{\ell^{\infty}} \geq 5u\}\big| d(u^q) < \infty.
\end{equation}%\tag4.17$$
Note that
\begin{equation}%
 \frac1{20^q}\|f\|_q^q 
 \leq \lim_{\varepsilon\rightarrow 0} N_{\varepsilon}(f).
\end{equation}%\tag4.18$$

By modifying the arguments in (4.8)--(4.10) and (4.16), we obtain
\begin{eqnarray}%$$\aligned
N_{\varepsilon}(f) 
&\leq\,&  C_0 q (\log c) M_{\alpha\cdot \operatorname{p}; p}(f)^p
  + \int_{u= \varepsilon}^{u= 1/\varepsilon}\frac1{u}
  \Big(\int |f(x)|_{\ell^{\infty}}\, \chi_{\{|f|_{\ell^{\infty}} >cu\}}(x)
   \,dx\Big) d(u^q) \nonumber\\
 &=:& I_1 + I_2.
\end{eqnarray}%\endaligned\tag4.19$$
The layer cake representation (4.4) leads the second term $I_2$ 
on the right-hand side to
\begin{eqnarray*}%$$\aligned
I_2 &=&\int dx \int_{u= \varepsilon}^{u= \frac1{\varepsilon}}
\frac1{u}\, d(u^q)
  \int \chi_{\{|f|_{\ell^{\infty}} >s\}}(x)
      \chi_{\{|f|_{\ell^{\infty}}>cu\}}(x) ds\\
&=& \int dx \int_{u= \varepsilon}^{u= \frac1{\varepsilon}}\frac1{u}
  \,d(u^q)\Big[\int_0^{cu} \chi_{\{|f|_{\ell^{\infty}}>cu\}}(x)ds
    + \int_{cu}^{\infty}  \chi_{\{|f|_{\ell^{\infty}}>s\}}(x)ds\Big]\\
&=& c\int dx \int_{u= \varepsilon}^{u= \frac1{\varepsilon}}
   \chi_{\{|f|_{\ell^{\infty}}>cu\}}(x) d(u^q) 
 + \int dx \int_{u= \varepsilon}^{u= \frac1{\varepsilon}}qu^{q-2}du 
   \int_{cu}^{\infty}\chi_{\{|f|_{\ell^{\infty}}>s\}}(x)ds.
\end{eqnarray*}%\endaligned$$
Then by integration by parts we have
\begin{eqnarray}%$$\aligned
I_2&=\,& c\int dx \int_{u= \varepsilon}^{u= \frac1{\varepsilon}}
  \chi_{\{|f|_{\ell^{\infty}}>cu\}}(x) d(u^q) \nonumber\\
 &&+ \int dx \Big[\frac{q}{q-1}u^{q-1}\int_{cu}^{\infty} 
 \chi_{\{|f|_{\ell^{\infty}}>s\}}(x) ds 
    \Big]_{u=\varepsilon}^{u=\frac1{\varepsilon}} \nonumber\\
 &&+ \int dx \int_{\varepsilon}^{\frac1{\varepsilon}}\frac{q}{q-1}u^{q-1}
 c\chi_{\{|f|_{\ell^{\infty}}>cu\}}(x)du \nonumber\\
&=\,& c \int dx \int_{u= \varepsilon}^{u= \frac1{\varepsilon}}
   \chi_{\{|f|_{\ell^{\infty}}>cu\}}(x) d(u^q)
  + \frac{c}{q-1}\int dx \int_{u=\varepsilon}^{u=\frac1{\varepsilon}}
  \chi_{\{|f|_{\ell^{\infty}}>cu\}}(x)d(u^q) \nonumber\\
&&\qquad + \frac{q}{q-1}
\Big[\frac1{\varepsilon^{q-1}}\int_{c\frac1{\varepsilon}}^{\infty}
  \chi_{\{|f|_{\ell^{\infty}}>s\}}(x)ds
 -\varepsilon^{q-1}\int_{c\varepsilon}^{\infty}
   \chi_{\{|f|_{\ell^{\infty}}>s\}}(x)ds \Big] \nonumber\\
&\leq&\, \frac{cq}{q-1}\int dx \int_{u= \varepsilon}^{u= \frac1{\varepsilon}}
    \chi_{\{|f|_{\ell^{\infty}}\geq cu\}}(x)d(u^q) 
  + \frac{cq}{q-1}\frac1{\varepsilon^{q-1}} \int dx
     \int_{\frac1{\varepsilon}}^{\infty}
          \chi_{\{|f|_{\ell^{\infty}}\geq cu\}}(x)du \nonumber\\
&=\,& \frac{cq}{q-1}\int_{\varepsilon}^{\frac1{\varepsilon}}
    \big|\{|f|_{\ell^{\infty}} \geq cu\}\big|d(u^q)
  +  \frac{cq}{q-1}\frac1{\varepsilon^{q-1}}
    \int_{\frac1{\varepsilon}}^{\infty}
          \big|\{|f|_{\ell^{\infty}} \geq cu\}\big| du \nonumber\\
&=:& I_{21} + I_{22},
\end{eqnarray}%\endaligned\tag4.20$$
where the last equality is due to Fubini's theorem, so that
$I_2 \leq I_{21} + I_{22}$.
Changing, in $I_{21}$ and $I_{22}$,  the variable $cu=5s$ 
and writing $u$ for $s$ again, we see by (4.17) and by the definition
(1.15) of weak $L^q$ norm,
\begin{eqnarray}%$$\aligned
I_{21} &=& \frac{cq}{q-1}\int_{\varepsilon}^{\frac1{\varepsilon}}
     \big|\{|f|_{\ell^{\infty}} \geq cu\}\big|d(u^q)
= \frac{cq}{q-1}\big(\frac5{c}\big)^q 
    \int^{u= \frac{c}{5}\frac1{\varepsilon}}_{u=\frac{c}{5}\varepsilon}
    \big|\{|f|_{\ell^{\infty}}\geq 5u\}\big|d(u^q) \nonumber\\
&=& \frac{q}{q-1}\frac{5^q}{c^{q-1}} 
    \Big(\int^{u= \frac1{\varepsilon}}_{u=\varepsilon}
        +\int^{u= \frac{c}{5}\frac1{\varepsilon}}_{u=\frac{1}{\varepsilon}}
       - \int^{u= \frac{c}{5}\varepsilon}_{u=\varepsilon}\Big)
    \big|\{|f|_{\ell^{\infty}} \geq 5u\}\big|d(u^q) \nonumber\\
&\leq& \frac{q}{q-1}\frac{5^q}{c^{q-1}} \Big\{N_{\varepsilon}(f)
   + \int^{u= \frac{c}{5}\frac1{\varepsilon}}_{u=\frac{1}{\varepsilon}}
      (5u)^{-q}(5u)^q\big|\{|f|_{\ell^{\infty}} \geq 5u\}\big|d(u^q)\Big\}
             \nonumber\\
&\leq& \frac{q}{q-1}\frac{5^q}{c^{q-1}} \Big\{N_{\varepsilon}(f)
   + \|f\|_{q,\infty}^q
    \int^{u= \frac{c}{5}\frac1{\varepsilon}}_{u=\frac{1}{\varepsilon}}
       (5u)^{-q}d(u^q)\Big\} \nonumber\\
&=& \frac{q}{q-1}\frac{5^q}{c^{q-1}} \Big\{N_{\varepsilon}(f)
   + \|f\|_{q,\infty}^q
    \frac{q}{5^q} 
\Big[\log u\Big]^{\frac{c}{5}\frac1{\varepsilon}}_{\frac{1}{\varepsilon}} 
\Big\} \nonumber\\
&=& \frac{q}{q-1}\frac{5^q}{c^{q-1}} N_{\varepsilon}(f)
 +\frac{q}{q-1}\frac{\log\frac{c}{5}}{c^{q-1}}\|f\|_{q,\infty}^q.
\end{eqnarray}%\endaligned\tag4.21$$
For $I_{22}$ we have
\begin{eqnarray}%$$\aligned 
I_{22} &=& \frac{cq}{q-1}\frac1{\varepsilon^{q-1}}
    \int_{\frac1{\varepsilon}}^{\infty} (cu)^{-q}
   \big[(cu)^q \big|\{|f|_{\ell^{\infty}} \geq cu\}\big|\big]du\nonumber\\
&\leq& \frac{cq}{q-1}\frac1{\varepsilon^{q-1}}\|f\|_{q,\infty}^q
    \int_{\frac1{\varepsilon}}^{\infty} (cu)^{-q}du
= \frac{q}{(q-1)^2}\frac1{c^{q-1}}\|f\|_{q,\infty}^q.
\end{eqnarray}%\endaligned\tag4.22$$
Then
$$
 I_2 \leq I_{21} +I_{22}
\leq \frac{q}{q-1}\frac{5^q}{c^{q-1}}N_{\varepsilon}(f)
 +\frac{q}{q-1}\frac1{c^{q-1}}\|f\|_{q,\infty}^q
\Big(\frac1{q-1} + \log\frac{c}{5}\Big).
$$
Therefore from (4.19)
\begin{eqnarray*}%$$\aligned
N_{\varepsilon}(f)
&\leq& I_1  +I_2\\
&\leq& C_0 q (\log c) M_{\alpha\cdot \operatorname{p}; p}(f)^p
 +\frac{q}{q-1}\frac{5^q}{c^{q-1}}N_{\varepsilon}(f)\\
&&\qquad\qquad +\frac{q}{q-1}\frac1{c^{q-1}}\|f\|_{q,\infty}^q
   \Big(\frac1{q-1} + \log\frac{c}{5}\Big)\\
&\leq& \frac{q}{q-1}\frac{5^q}{c^{q-1}}N_{\varepsilon}(f)
 + \Big[C_0 q (\log c) +\frac{q}{q-1}\frac1{c^{q-1}}
    \Big(\frac1{q-1} + \log\frac{c}{5}\Big)
    \Big]M_{\alpha\cdot \operatorname{p}; p}(f)^p,
\end{eqnarray*}%\endaligned$$
where the last inequality is due to the fact that by (1.14) and (1.22)  
$\|f\|_{q,\infty} \leq \|(\alpha\cdot \operatorname{p})f\|_p 
\leq M_{\alpha\cdot \operatorname{p}; p}(f)$.
Then take $c$ large (if necessary, larger than
the $c$ chosen once already at the end of step II)  such that 
$1-\frac{q}{q-1}\frac{5^q}{c^{q-1}} < \frac12$, and we have with (4.18)
\begin{equation}%
\|f\|_q^q \leq 
  2\cdot 20^q \Big[C_0 q (\log c) +\frac{q}{q-1}\frac1{c^{q-1}}
    \Big(\frac1{q-1} + \log\frac{c}{5}\Big)
    \Big] M_{\alpha\cdot \operatorname{p}; p}(f)^p.
\end{equation}%\tag4.23$$
Thus, taking
$C := 2^{1/q} 20 
   \Big[C_0 q (\log c) +\frac{q}{q-1}\frac1{c^{q-1}}
    \Big(\frac1{q-1} + \log\frac{c}{5}\Big)\Big]^{1/q}
$ 
and noting homogeneity, we have shown the 
desired inequality (1.28), ending the proof of Theorem 1.1 (i). 
\qed 

%%%%%%%%%%%%%%%%%%%%%
{\it Proof of Theorem \hbox{\rm 1.1 (ii)}}. 
In case $p>1$, in our previous paper [IS] we have shown 
that $H_{\alpha\cdot \operatorname{p},0}^{1,p}({\Bbb R}^3;{\Bbb C}^4) 
    = H_0^{1,p}({\Bbb R}^3;{\Bbb C}^4)$,
so that the norms 
$\|f\|_{M_{\alpha\cdot \operatorname{p}},1,p} 
:= (\|f\|_p^p +M_{\alpha\cdot \operatorname{p}; p}(f)^p)^{1/p}$ and  
$\|f\|_{\alpha\cdot \operatorname{p},1,p} 
:= (\|f\|_p^p +\|(\alpha\cdot \operatorname{p})f\|_p^p)^{1/p}$
are equivalent to the norm 
$\|f\|_{1,p} := (\|f\|_p^p +\|\nabla f\|_p^p)^{1/p}$. But this
may not be sufficient to derive (1.30).

To show the assertion, we need show that for $p>1$ the two semi-norms 
$\|(\alpha\cdot \operatorname{p})f\|_p$ and $\|\nabla f\|_p$ are equivalent. 
However, noting the two inequalities (1.22), we have only to show 
the following lemma.

%\bigskip\noindent
\begin{lemma}%{Lemma 4.2}  
For $1<p<\infty$, there exists a positive constant
$C$ such that  
\begin{equation}%
\|\nabla f\|_p \leq C \|(\alpha\cdot \operatorname{p})f\|_p
\end{equation}%\tag4.24$$
for every $f \in C_0^{\infty}({\Bbb R}^3;{\Bbb C}^4)$.
\end{lemma}

%\bigskip
{\it Proof}.  We give two proofs.

(i) (A first proof with functional analysis) 
In the proof of [IS, Proposition 3.1], we had already seen this fact
of the lemma.  Here let us briefly sketch the argument.

Let $f= {}^t (f_1,f_2,f_3,f_4) \in C_0^{\infty}({\Bbb R}^3;{\Bbb C}^4)$ 
so that
 $(\alpha\cdot \operatorname{p})f \in L^p({\Bbb R}^3;{\Bbb C}^4)$, and
$$g= {}^t (g_1,g_2,g_3,g_4) := (\alpha\cdot \operatorname{p})f 
= -i[\alpha_1\partial_1 f +\alpha_2\partial_2 f +\alpha_3\partial_3 f],
$$
belongs to $L^p({\Bbb R}^3;{\Bbb C}^4)$.

Since 
$ -\Delta f = (\alpha\cdot \operatorname{p})^2f
= (\alpha\cdot \operatorname{p})g$,
we have
$
 \Delta (\partial_j f) 
= i[\alpha_1\partial_1  +\alpha_2\partial_2  +\alpha_3\partial_3]
\partial_jg, \,\, (j=1,2,3), 
$
where the derivatives are taken in distribution sense.
Then we can show for each $j=1,2,3$, $k=1,2,3,4$, that 
there exist constants $C_{j,kl},\,\, k,l =1,2,3,4$, such that
\begin{eqnarray*}%$$\aligned
|\langle \partial_j f_k, \Delta \phi\rangle|
&\leq& \big[(C_{j,k1}\|g_1\|_p +C_{j,k2}\|g_2\|_p 
    +C_{j,k3}\|g_3\|_p +C_{j,k4}\|g_4\|_p\big] \|\Delta\phi\|_{p'}\\
&\leq& C(\sum_{l=1}^4\|g_l\|_p^p)^{1/p}\|\Delta\phi\|_{p'} 
= C\|g\|_p\|\psi\|_{p'}
\end{eqnarray*}%\endaligned$$
for all $\phi \in C_0^{\infty}({\Bbb R}^3)$
with $C := (\sum_{l=1}^4 {C_{j,kl}}^{p'})^{1/p'}$, where the last second
inequality is due to H\"older's inequality with $\frac1{p}+\frac1{p'}=1$. 
Hence
$
|\langle \partial_j f_k, \psi\rangle|
\leq  C\|g\|_p\|\psi\|_{p'}
$ 
 for all $\psi \in L^{p'}({\Bbb R}^3)$, since 
for $p>1$ the space $\Delta (C_0^{\infty}({\Bbb R}^3))$
is dense in $L^{p'}({\Bbb R}^3)$, 
so that $\partial_j f_k$ belongs to $L^p({\Bbb R}^3)$
for $j=1,2,3$, $k=1,2,3,4$, and
$$%\aligned
\|\partial_j f_k\|_p
\leq C\|g\|_p = C\|(\alpha\cdot \operatorname{p})f\|_p.
%\endaligned
$$
This proves the desired inequality (4.24).

(ii) (A second proof with pseudodifferential calculus)

To show the assertion, we have only to show that
for $j=1,2,3$, $-i\partial_j/(\alpha\cdot \operatorname{p})$ is a bounded 
operator on $L^p({\Bbb R}^3; {\Bbb C}^4)$. To see it, since 
$(\alpha\cdot \operatorname{p})^2 = -\Delta$, we note that
$$
\frac{-i\partial_j}{\alpha\cdot \operatorname{p}}
 = \frac{-i\partial_j}{-\Delta}(\alpha\cdot \operatorname{p})
= \frac{-i\partial_j}{(-\Delta)^{1/2}}
   \sum_{k=1}^3 \frac{-i\alpha_k \partial_k}{(-\Delta)^{1/2}}
= - \sum_{k=1}^3 \alpha_k\cdot  R_jR_k,
$$ 
where $R_k =\frac{-i\partial_k}{(-\Delta)^{1/2}},\, k=1,2,3,$ 
is the Riesz transform which is a pseudo-differntial operator having symbol
$i\xi_k/|\xi|$, and if $1<p<\infty$, we have $\|R_k g\|_p \leq C \|g\|_p$ 
with a constant $C>0$, e.g. 
by the Calderon--Zygmund theorem [e.g. S, 4.2, Theorem 3, p.29] or by 
Fefferman's theorem [Fe, Theorem, a, p.414].
Therefore we obtain for each $j=1,2,3$,
$$
\|[-i\partial_j/(\alpha\cdot \operatorname{p})]f\|_p \leq 3C^2\|f\|_p.
$$
This proves (4.24), again showing the lemma.  
\qed 

\medskip
Thus we have proved Theorem 1.1 (ii), completing the proof of Theorem 1.1.
\qed 

%%%%%%%%%%%%%%%%%%%%%%%%%%%%
\medskip
\section{Proof of Corollaries 1.2, 1.3, Theorem 1.4 , Corollary 1.5 
and Theorem 1.6}%\sect.5 

$\,\,$\indent
{\it Proof of Corollary \hbox{\rm 1.2}}. 
Let $h := {}^t(h_1,h_2)$ be a ${\Bbb C}^2$-valued function and put 
$f= {}^t(f_1,f_2,f_3,f_4)$ with $f_1=h_1,\,\, f_2=h_2, f_3=f_4=0$.
Then (1.33) is nothing but (1.28). This proves Corollary 1.2 (i).
(ii) can be seen as in the proof of Theorem 1.1 (ii).   
\qed 

\bigskip
{\it Proof of Corollary \hbox{\rm 1.3}}.
Let $\psi$ be a ${\Bbb C}$-valued function and put 
$f= {}^t(f_1,f_2,f_3,f_4)$ with $f_2=\psi,\,\, f_1=f_3=f_4=0$.
Then (1.37) is nothing but (1.28). This proves Corollary 1.3 (i).
(ii) can be seen as in the proof of Theorem 1.1 (ii).  
\qed 

\bigskip
{\it Proof of Theorem \hbox{\rm 1.4}}. 
The proof is divided into two parts (a) and (b). First in (a),
we show (1.41) for the operator $(\sigma\cdot\operatorname{p})^{(a)}$ 
in (1.40a), and then in (b) for the other two 
$(\sigma\cdot\operatorname{p})^{(b)}$, $(\sigma\cdot\operatorname{p})^{(c)}$
in (1.40bc).

\medskip
(a) The case for $(\sigma\cdot\operatorname{p})^{(a)}$ in (1.40a):
First we are going to show (4.16) with 
$\alpha\cdot\operatorname{p}$ replaced by 
$(\sigma\cdot\operatorname{p})^{(a)}$ in (1.40a), and then
the proof proceeds to use almost 
the same arguments as in steps I, II, III of
the proof of Theorem 1.1 (i). In step II we shall not
 need to introduce some other semi-norm like 
$M_{(\alpha\cdot \operatorname{p})^{(a)}; p}(f)$
than $\|(\alpha\cdot\operatorname{p})^{(a)}f\|_p$, and 
have only to go with the semi-norm  
$\|(\sigma\cdot\operatorname{p})^{(a)}f\|_p$
for ${\Bbb C}^2$-valued functions $f(x)= {}^t(f_1(x),f_2(x))$ 
on ${\Bbb R}^2$.

\medskip
I. In the same way as before, 
we can show an inequality corresponding to (1.14), 
i.e. that there exists a constant $C$ such that
$
 \|f\|_{q,\infty} 
\leq C\|(\sigma\cdot\operatorname{p})^{(a)}f\|_p^{p/q}
               \|f\|^{1-(p/q)}_{B_{\infty,\infty}^{p/(p-q)}}
$
for every $f= {}^t(f_1,f_2)$ which satisfies 
$(\sigma\cdot\operatorname{p})^{(a)}f \in L^p({\Bbb R}^2; {\Bbb C}^2)$
and belongs to $B_{\infty,\infty}^{p/(p-q)}({\Bbb R}^2; {\Bbb C}^2)$.

\medskip
II. This step contains a slight improvement in its own. 
We want to replace the weak $L^q$ norm by the strong $L^q$ norm.
Under the same hypothesis as in step I above but 
with $\|f\|_{B_{\infty,\infty}^{p/(p-q)}} \leq 1$,
we are going to show the following inequality: 
\begin{equation}%
\int |f|_{\ell^q}^qdx \leq 
C\|(\sigma\cdot\operatorname{p})^{(a)}f\|_p^p\,
\end{equation}%\tag5.1$$
with a constant $C$ independent $f$, 
a sharper inequality than the previous (4.3),
assuming the extra condition $f \in L^q({\Bbb R}^2; {\Bbb C}^2)$,
which will turn out to be unnecessary in step III below.

To this end, we can proceed as in II of the proof of 
Theorem 1.1 (i), 
Section 4, to obtain an anlogous version of (4.8) : 
\begin{eqnarray*}%$$\aligned
\frac1{20^q}\|f\|_q^q 
&=\,& \int_0^{\infty} |\{|f|_{\ell^q} \geq 20u\}|d(u^q)\\
&\leq\,& \int_0^{\infty}|\{\vee_{k=1}^2 
      |f_{u,k}-e^{t_u\Delta} f_{u,k}|\geq u\}|d(u^q) \\
\qquad &&\qquad+ \int_0^{\infty}|
  \{\vee_{k=1}^2 \{e^{t_u\Delta} |f_{u,k}-f_k|\geq 2u\}|d(u^q)\\
&=:& J'_1 + J'_2,
\end{eqnarray*}%\endaligned$$
where  $\Delta$ is the Laplacian in ${\Bbb R}^2$,
$f(x) := {}^t (f_{1}(x),f_{2}(x)) \in L^q({\Bbb R}^2;\,{\Bbb C}^2)$
with $\|f\|_{B_{\infty,\infty}^{p/(p-q)}}\leq 1$ 
and $f_u(x) := {}^t (f_{u,1}(x),f_{u,2}(x))$ is given by (4.6)
with the subscription moving over $\{1,2\}$, not $\{1,2,3,4\}$.
By the same arguments used before to get (4.10) and (4.11), 
respectively, we have 
$
 J'_2 \leq \frac{q}{q-1}\frac1{c^{q-1}} \|f\|_q^q\, 
$ 
and
\begin{eqnarray*}%$$\aligned
 J'_1 
&\leq& C_0\int_0^{\infty}d(u^q)u^{-q}
        \|(\sigma\cdot\operatorname{p})^{(a)}f_u\|_p^p\\ 
&=& C_0 \int_0^{\infty}d(u^q)u^{-q}\big[
    \|(\partial_1+i\partial_2)f_{u,1}\|_p^p 
      +\|(\partial_1-i\partial_2)f_{u,2}\|_p^p \big]\,.
\end{eqnarray*}%\endaligned$$
Noting that
\begin{eqnarray*}%$$\aligned
\int_0^{\infty}d(u^q)u^{-q}\|(\partial_1+i\partial_2)f_{u,1}\|_p^p
&=& \int d(u^q)u^{-q}\int_{u \leq |f_1(x)| \leq cu}
      |(\partial_1+i\partial_2)f_{u,1}(x)|^p dx\\
&=& \int_0^{\infty}d(u^q)u^{-q}\int_{u\leq |f_1(x)| \leq cu} 
      |(\partial_1+i\partial_2)f_{1}(x)|^pdx\\
&=& q (\log c) \int |(\partial_1+i\partial_2)f_{1}(x)|^pdx\\
&=& q (\log c) \|(\partial_1+i\partial_2)f_{1}\|_p^p\,,
\end{eqnarray*}%\endaligned$$
and in the same way
$$
\int_0^{\infty}d(u^q)u^{-q}\|(\partial_1-i\partial_2)f_{u,2}\|_p^p
= q (\log c) \|(\partial_1-i\partial_2)f_{2}\|_p^p\,, 
$$
we have
$
 J'_1 \leq C_0 \,q \,(\log c)\,
      \|(\sigma\cdot\operatorname{p})^{(a)}f\|_p^p\,.
$ 
Thus 
$$
 \frac1{20^q}\|f\|_q^q\leq J'_1+J'_2
\leq C_0 \,q \,(\log c) \|(\sigma\cdot\operatorname{p})^{(a)}f\|_p^p
 + \frac{q}{q-1}\frac1{c^{q-1}} \|f\|_q^q\,,
$$
whence we get the desired inequality (4.1), 
taking 
$C= \frac{C_0 q (\log c)}{\frac1{20^q}-\frac{q}{q-1}\frac1{c^{q-1}}}$
for $c$ sufficiently large.

\medskip
III. Finally we remove the condition that 
$f \in L^q({\Bbb R}^2; {\Bbb C}^2)$ assumed in step II. In fact, we show
that if $\|(\sigma\cdot\operatorname{p})^{(a)}f\| <\infty$ and 
$\|f\|_{B_{\infty,\infty}^{p/(p-q)}} \leq 1$, then
$f \in L^q({\Bbb R}^2; {\Bbb C}^2)$. 

The proof proceeds in the same way as in III of the proof of Theorem 1.1 (i), 
Section 4. Indeed, with the corresponding $N_{\varepsilon}(f)$ as in (4.17) 
and (4.18), we can show, instead of (4.19), 
$$
N_{\varepsilon}(f) 
\leq  C_0\, q\, (\log c)  \|(\sigma\cdot \operatorname{p})^{(a)}f\|_p^p
  + \int_{u= \varepsilon}^{u= 1/\varepsilon}\frac1{u}
  \Big(\int |f(x)|_{\ell^{\infty}}\, \chi_{\{|f|_{\ell^{\infty}} >cu\}}(x)
   \,dx\Big) d(u^q)\,.
$$
Estimating, in the same way as before, the two terms on the right-hand side, 
we can obtain the desired inequality
$
 \|f\|_q^q \leq C \|(\sigma\cdot\operatorname{p}^{(a)})f\|_p^p.
$ 
This shows (1.41) in Theorem 1.4 for $(\sigma\cdot\operatorname{p})^{(a)}$
in (1.40a).

%%%%%
\medskip
(b) The other cases for $(\sigma\cdot\operatorname{p})^{(b)}$ and 
$(\sigma\cdot\operatorname{p})^{(c)}$ in (1.40bc): 
Each of these two cases is reduced to the case (a) for 
$(\sigma\cdot\operatorname{p})^{(a)}$ by a linear transformation. 
The idea is based on the following lemma.

%\bigskip\noindent
\begin{lemma}%{Lemma 5.1} 
The three 2-dimensional Weyl--Dirac (or Pauli) operators
$(\sigma\cdot\operatorname{p})^{(a)},\, (\sigma\cdot\operatorname{p})^{(b)},\,
(\sigma\cdot\operatorname{p})^{(c)}$ in \hbox{\rm (1.40abc)} 
are unitarily equivalent. 
In fact, there exist unitary $2\times 2$-matrices $N$, $N'$ such that
for $f= {}^t(f_1,f_2)$ and $h ={}^t(h_1,h_2) =: Nf$, 
$h ={}^t(h_1,h_2) =: N'f$, 
\begin{eqnarray}%$$\align
 (\sigma\cdot\operatorname{p})^{(a)}h 
&=& (\sigma\cdot\operatorname{p})^{(a)}Nf
 = N(\sigma\cdot\operatorname{p})^{(b)}f,\, 
  \,\,\, \hbox{\rm with}\,\, h ={}^t(h_1,h_2)= Nf\,,  \\%\tag5.2
%%%%
(\sigma\cdot\operatorname{p})^{(a)}h 
&=& (\sigma\cdot\operatorname{p})^{(a)}N'f
 = N'(\sigma\cdot\operatorname{p})^{(c)}f\,,
  \,\,\, \hbox{\rm with}\,\, h ={}^t(h_1,h_2)= N'f\,.  %\tag5.3
\end{eqnarray}%\endalign$$
\end{lemma}

%\bigskip
{\it Proof}. Take matrices 
$N  := \frac1{\sqrt{2}}\left(\begin{array}{cc} 1 &-i \\ 1& i \end{array}\right)$,
$N' := \frac1{\sqrt{2}}\left(\begin{array}{cc} 1 & -1 \\ 1& 1 \end{array}\right)$, 
which are unitary.
We have $N^{-1} = \frac1{\sqrt{2}}\left(\begin{array}{cc} 
                                  1&1 \\ i& -i \end{array}\right)$,
$(N')^{-1} = \frac1{\sqrt{2}}\left(\begin{array}{cc} 
                                  1 & 1 \\ -1&1 \end{array}\right)$,
and
\begin{eqnarray*}%$$\aligned
\frac{1}{\sqrt{2}}\left(\begin{array}{cc} 1&-i \\ 1&i \end{array}\right)
\left(\begin{array}{cc} 
      \partial_1 & \partial_2 \\\partial_2 & -\partial_1 \end{array}\right)
\frac1{\sqrt{2}}\left(\begin{array}{cc} 1&1 \\ i&-i \end{array}\right)
&=&\left(\begin{array}{cc} 
      0&\partial_1-i\partial_2\\ \partial_1+i\partial_2&0\end{array}\right)
\,,\\
%%%%
\frac{1}{\sqrt{2}}\left(\begin{array}{cc} 1 & -1 \\ 1 &1 \end{array}\right)
\left(\begin{array}{cc} 
     \partial_1 & -i\partial_2 \\i\partial_2 &-\partial_1\end{array}\right)
\frac1{\sqrt{2}}\left(\begin{array}{cc} 1&1 \\ -1&1 \end{array}\right)
&=&\left(\begin{array}{cc} 
     0&\partial_1-i\partial_2\\\partial_1+i\partial_2&0\end{array}\right)\,.
\end{eqnarray*}%\endaligned$$
Taking into account the definition (1.40abc) of 
$(\sigma\cdot \operatorname{p})^{(a)}$,
$(\sigma\cdot \operatorname{p})^{(b)}$,
$(\sigma\cdot \operatorname{p})^{(c)}$
yields (5.2) and (5.3), showing Lemma 5.1.    
\qed 

%%%%%%%%%%%%%%%
\bigskip
Now we continue the proof (b) of Theorem 1.4. 
Take the same matrices $N$ and $N'$ as in Lemma 5.1, which we see 
reduce the cases
$(\sigma\cdot\operatorname{p})^{(b)}$ and $(\sigma\cdot\operatorname{p})^{(c)}$
to the case $(\sigma\cdot\operatorname{p})^{(a)}$.

Note the bounds of  the matrix norms of them and their inverses satisfy 
 that for $1\leq r \leq \infty$, 
\begin{eqnarray}%$$\aligned
&& \|N\|_{\ell^r\rightarrow \ell^r} \leq \sqrt{2},
\quad 
\|N^{-1}\|_{\ell^r\rightarrow \ell^r} \leq \sqrt{2}\,; \nonumber\\
&& \|N'\|_{\ell^r\rightarrow \ell^r} \leq \sqrt{2},
\quad 
\|(N')^{-1}\|_{\ell^r\rightarrow \ell^r} \leq \sqrt{2}\,.
\end{eqnarray}%\endaligned\tag5.4$$
It follows that if  $h = Nf$ or $h=N'f$, then
\begin{equation}%
 \|f\|_r \leq \sqrt{2}\|h\|_r, \qquad  \|h\|_r \leq \sqrt{2}\|f\|_r.
\end{equation}%\tag5.5$$

First, we treat the case $(\sigma\cdot\operatorname{p})^{(b)}$
with $N$. We have by (5.2) in Lemma 5.1 and (5.4)
\begin{equation}%
\|(\sigma\cdot \operatorname{p})^{(a)}h\|_p 
=  \|N(\sigma\cdot \operatorname{p})^{(b)}f\|_p
\leq \|N\|_{\ell^p\rightarrow \ell^p}
  \|(\sigma\cdot \operatorname{p})^{(b)}f\|_p
\leq \sqrt{2} \|(\sigma\cdot \operatorname{p})^{(b)}f\|_p\,.
\end{equation}%\tag5.6$$
We note that $P_t$ commutes with $N$ to get
\begin{eqnarray*}%$$\aligned
|(P_t h)(x)|_{\ell^{\infty}} 
&=& |(P_t Nf)(x)|_{\ell^{\infty}} = |(NP_t f)(x)|_{\ell^{\infty}}\\ 
&\leq& \|N\|_{\ell^{\infty}\rightarrow \ell^{\infty}}
                           |(P_t f)(x)|_{\ell^{\infty}} 
\leq \sqrt{2}|(P_t f)(x)|_{\ell^{\infty}}\,,
\end{eqnarray*}%\endaligned$$
whence
\begin{eqnarray}%$$\aligned
 \|h\|_{B_{\infty,\infty}^{p/(p-q)}}
&=& \sup_{t>0} \|P_t h\|_{\infty}
= \sup_{t>0}\sup_x |(P_t h)(x)|_{\ell^{\infty}} \nonumber\\ 
&\leq& \sqrt{2}\sup_{t>0}\sup_x |(P_t f)(x)|_{\ell^{\infty}} 
= \sqrt{2}\sup_{t>0} \|P_t f\|_{\infty}
= \sqrt{2}\|f\|_{B_{\infty,\infty}^{p/(p-q)}}\,.\quad
\end{eqnarray}%\endaligned\tag5.7$$
Then, since we already know (1.41) holds for $(\sigma\cdot \operatorname{p})^{(a)}$
with $h$ in place of $f$, we combine it with (5.5), (5.6), (5.7)  
to get
$$
\|f\|_q \leq \sqrt{2}\|h\|_q 
 \leq  \sqrt{2}C\|(\sigma\cdot \operatorname{p})^{(a)} h\|_p^{p/q}
    \|h\|_{B_{\infty,\infty}^{p/(p-q)}}^{1-(p/q)}
\leq 2C \|(\sigma\cdot \operatorname{p})^{(b)}f\|_p^{p/q}
\|f\|_{B_{\infty,\infty}^{p/(p-q)}}^{1-(p/q)}\,,
$$
which yields the desired inequality (1.41) for 
$(\sigma\cdot\operatorname{p})^{(b)}$.

\par
Next, as for the other last case $(\sigma\cdot \operatorname{p})^{(c)}$,
exactly the same arguments apply to it as those just made in the case 
$(\sigma\cdot \operatorname{p})^{(b)}$ above, 
with the matrix $N$, relation (5.2) replaced by the matrix $N'$, 
relation (5.3).

This completes the proof of Theorem 1.4.  
\qed 

\bigskip
{\it Proof of Corollary \hbox{\rm 1.5}}.  
(1.42) follows from Corollary 1.3 (1.39) because our function
$\psi(x)=\psi(x_1,x_2)$ here is independent of $x_3$, or from
Theorem 1.4 (1.41) for $h= {}^t(h_1,h_2)$ with $h_1=0,\, h_2=\psi$. 
\qed 

\bigskip
{\it Proof of Theorem \hbox{\rm 1.6}}. 
The proof is done by analogous arguments used to prove Theorem 1.1.  
We only note that Lemma 4.1 is replaced by the following lemma, which 
can be shown in the same way as before.

\begin{lemma}%{Lemma 5.2} 
For $f = {}^t(f_1,f_2,f_3,f_4)$, one has 
\begin{equation}%
 \int_0^{\infty} d(u^q) 
u^{-q} M_{\beta\cdot \operatorname{p}; p}(f_u)^p 
= q (\log c) M_{\beta\cdot \operatorname{p}; p}(f)^p.
\end{equation}%\tag5.8$$
\end{lemma}

Here we only note with (1.50) that the proof turns out to deal, 
instead of (4.13), with
\begin{eqnarray*}%$$\aligned
&&M_{\beta\cdot \operatorname{p}; p}(f_u)^p\\
&=& \!\int\! \big(|(\partial_1+i\partial_2) f_{u,1}|^p 
             +|(\partial_3 +i\partial_4) f_{u,1}|^p\big) dx
 +\!\int\! \big(|(\partial_1-i\partial_2) f_{u,2}|^p 
             +|(\partial_3 -i\partial_4)f_{u,2}|^p\big) dx\\
&&\!\!\!\!+\!\int\! \big(|(\partial_1+i\partial_2) f_{u,3}|^p 
             +|(\partial_3 -i\partial_4)f_{u,3}|^p\big) dx
 +\!\int\! \big(|(\partial_1-i\partial_2) f_{u,4}|^p  
             +|(\partial_3 +i\partial_4)f_{u,4}|^p\big) dx.
\end{eqnarray*}%\endaligned$$
\qed 

%%%%%%%%%%%%%%%%%%%%%%%%%%%%%%%%%%%%%%%%%%%%%%
\section{Concluding Comments}%\sect.7

We have originated a version of improved Sobolev embedding theorem 
for vector-valued functions involved with the 
three-dimensional Dirac operator $D=\alpha\cdot \operatorname{p}$, 
the three-dimensional Weyl--Dirac 
(or Pauli) operator $D=\sigma\cdot \operatorname{p}$,
and the four-dimensional Euclidian Dirac operator 
$D=\beta\cdot \operatorname{p}$. To this end we have introduced 
in Section 1 the corresponding 
first-order-derivative semi-norms $M_{\alpha\cdot \operatorname{p}; p}(f)$,
$M_{\sigma\cdot \operatorname{p}; p}(h)$ 
and $M_{\beta\cdot \operatorname{p}; p}(f)$ by decomposing them into 
two parts: $D= D_1+D_2$. Although the used decomposition 
looked to be artificial, it turns out there are other meaningful 
decompositions which give the same semi-norms as thus defined.
In fact, we have characterized, in Proposition 1.0
for $\alpha\cdot \operatorname{p}$ and its counterpart for 
$\sigma\cdot \operatorname{p}$ and $\beta\cdot \operatorname{p}$, 
which kind of decompositions are fit for our semi-norms at all. 
It turns out that they should be those 
which satisfy the condition that {\it each row of the matrices of both 
the parts $D_1$ and $D_2$ contains only one nonzero entry}. 
Why one needs this condition is simply because our proof given 
in Section 4  needs it. 

In this section 
we will make some further comments and observe that after all 
this semi-norm is of reasonably good and optimal choice,
having intrinsic and universal character and being 
an intermediate one in strength lying between both the semi-norm
$\|(\alpha\cdot \operatorname{p})f\|_p$,  
$\|(\sigma\cdot \operatorname{p})h\|_p$ or  
$\|(\beta\cdot \operatorname{p})f\|_p$
and the seminorm 
$\|\nabla f\|_p$, $\|\nabla h\|_p$ or $\|\nabla f\|_p$, respectively. 
We describe only with the 4-dimensional Euclidian Dirac operator, 
as we can deal with the other two operators just in the same way. 

So consider the 4-dimensional Euclidian Dirac operator
$D:= \beta\cdot \operatorname{p}$ in (1.46) 
and its decomposition into the sum of its two parts : $D= D_1 +D_2$. 
Ignoring the order of the pair $(D_1, D_2)$, we regard 
the two decomposition $(D_1, D_2)$ and $(D_2, D_1)$ as the same.
Then there are totally $\frac12\cdot 2^7=64$ 
decompositions including the trivial decomposition with
$(D_1, D_2) = (D, 0)$ or $(D_1, D_2) = (0, D)$. The set of all 
decompositions of $D= \beta\cdot \operatorname{p}$ is denoted by 
$\hbox{\rm Decom}(D)$. Let $\hbox{\rm Decom}_1(D)$ be the subset of
all $(D_1,D_2)$ in $\hbox{\rm Decom}(D)$ which satisfy the condition 
that {\it each row of $D_1$ and $D_2$ contains only one nonzero entry}. 
It is seen that $\hbox{\rm Decom}_1(D)$  consists of $\frac12\cdot 2^4=8$ 
decompositions of $D$.
The decompositions (1.47), (1.53), (1.54) and (1.55)
are examples of elements of $\hbox{\rm Decom}_1(D)$.
With the decomposition (1.47), i.e. 
$((\beta\cdot \operatorname{p})P_{13}, 
      (\beta\cdot \operatorname{p})P_{24})$,
we have defined the semi-norm $M_{\beta\cdot \operatorname{p}; 1}(f)$ 
by (1.48), that is, 
\begin{equation}
M_{\beta\cdot \operatorname{p};\, p} (f)
:= \big[\,\|(\beta\cdot \operatorname{p})P_{13}f\|_p
   +\|(\beta\cdot \operatorname{p})P_{24}f\|_p\,\big]^{1/p}.
\end{equation}%(6.1) 
We have shown Theorem 1.6, a version of improved Sobolev embedding theorem for 
vector-valued functions, that inequality (1.57) holds with this semi-norm 
$M_{\beta\cdot \operatorname{p}; 1}(f)$ for $1\leq p <\infty$, 
and also seen in Section 3 that in case of $p=1$ 
one cannot replace the semi-norm
$M_{\beta\cdot \operatorname{p};\, p} (f)$ on the right by a weaker one
$\|(\beta\cdot \operatorname{p})f\|_p$, though one can 
for $1<p<\infty$. 
Actually we have
\begin{equation}%
 (D_1,D_2)\, \in\, \hbox{\rm Decom}_1(D) 
\,\,\, \Rightarrow \,\,\,
M_{\beta\cdot \operatorname{p};\, p} (f) :=
 M_{D_1 \vee D_2;p}(f) =\big[\,\|D_1 f\|_p +\|D_2 f\|_p\,\big]^{1/p}.
\end{equation}%\tag6.2$$
Thus our semi-norm $M_{\beta\cdot \operatorname{p};\, p} (f)$
is characterized as the one associated with $\hbox{\rm Decom}_1(D)$.
At this point also notice that this semi-norm has the {\it very} 
expression (1.50) with symmetric arrangement of eight terms in its last member.
Inequality (1.51) shows that $M_{\beta\cdot \operatorname{p};\, p} (f)$
is lying in strength between the semi-norms 
$\|(\beta\cdot \operatorname{p})f\|_p$ and $\|\nabla f\|_p$.
Notice that the condition that 
{\it each row of $D_1$ and $D_2$ contains only one nonzero entry} 
is satisfied by neither the 3-dimensional Dirac operator 
(1.17),  3-dimensional Weyl--Dirac (or Pauli) operator (1.31) nor 
4-dimensional Euclidian Dirac operator (1.46) {\it themselves}. 
Otherwise, our proof could establish for $p=1$  inequality (1.28) 
of Theorem 1.1, (1.35) of Corollary 1.2 and (1.57) of Theorem 1.6
with the semi-norm 
$\|(\alpha\cdot \operatorname{p})f\|_1$,
$\|(\sigma\cdot \operatorname{p})h\|_1$ and
$\|(\beta\cdot \operatorname{p})f\|_1$ in place of 
$M_{\alpha\cdot \operatorname{p}; 1}(f)$,
$M_{\sigma\cdot \operatorname{p}; 1}(h)$ and 
$M_{\beta\cdot \operatorname{p}; 1}(f)$ on the right-hand side. 
But this is not in general possible because we have counterexamples 
as given in Section 3.

In K\"ahler Geometry and/or Spin Geometry (e.g. [Fr], [LawM]), 
the four-dimensional Euclidian Dirac operator $D$ appears as an operator
acting on the Clifford algebra $CL({\Bbb R}^4)$, 
which is canonically isomorphic to the exterior algebra 
$\Lambda^*({\Bbb R}^4) \equiv \Lambda^*(T^*({\Bbb R}^4))$. On this 
$\Lambda^*({\Bbb R}^4)$, in turn, there act two canonical first-order 
differential operators, namely, the exterior derivative 
$d: \Lambda^*({\Bbb R}^4) \rightarrow \Lambda^*({\Bbb R}^4)$ and 
its formal adjoint 
$d^*: \Lambda^*({\Bbb R}^4) \rightarrow \Lambda^*({\Bbb R}^4)$,
which satisfy $d^2= {d^*}^2=0$. Then the fact is that 
the Dirac operator $D$ is considered to decompose into their sum: 
$D \cong d+d^*$. In passing, it is conversely along with such 
a decomposition that the Dirac operator of {\it even} infinite dimension 
is defined on a Fock space in [A1, A2]. 

In this connection, notice that $\hbox{\rm Decom}_1(D)$ contains two pairs 
$(D_1,D_2)$, (1.53) and (1.54), which satisfy the one condition 
$D_2 = D_1^*$, but neither of the elements of $\hbox{\rm Decom}_1(D)$ 
satisfy the other condition $D_1^2 = D_2^2 = 0$.
We ask: how about the inequality 
\begin{equation}%
  \|f\|_q \leq C M_{D_1\vee D_2;p}(f)^{p/q}
                 \|f\|_{B^{p/(p-q)}_{\infty,\infty}}^{1-(p/q)}\,
\end{equation}%\tag6.3$$ 
like (1.57) for the decompositions {\it not} belonging to 
$\hbox{\rm Decom}_1(D)$, to hold with a fixed constant $C>0$ for 
all functions $f(x) = {}^t(f_1(x), f_2(x),f_3(x),f_4(x))$ on ${\Bbb R}^4$ ? 
To answer it, consider the following three decompositions 
$D \equiv \beta\cdot \operatorname{p} =D_1+D_2$ in 
$\hbox{\rm Decom}(D) \setminus \hbox{\rm Decom}_1(D)$ which are typical 
in some sense :

\noindent
$\underline{M_{\beta}^{(4)}}$
\begin{subequations}
\begin{align}
&\beta\cdot \operatorname{p}
= (\beta\cdot \operatorname{p})P_{12}
   + (\beta\cdot \operatorname{p})P_{34}\nonumber\\
&\qquad=  \left(\begin{array}{cccc}
0&0&0&0\\ 
0&0&0&0\\
\operatorname{p}_3+i\operatorname{p}_4
&\operatorname{p}_1-i\operatorname{p}_2&0&0\\ 
\operatorname{p}_1+i\operatorname{p}_2 
&-(\operatorname{p}_3-i\operatorname{p}_4)&0&0
\end{array}\right)
+  \left(\begin{array}{cccc}
0&0&\operatorname{p}_3-i\operatorname{p}_4
&\operatorname{p}_1-i\operatorname{p}_2 \\ 
0&0&\operatorname{p}_1+i\operatorname{p}_2 
&-(\operatorname{p}_3+i\operatorname{p}_4)\\
0&0&0&0\\ 
0&0&0&0\end{array}\right)\,,\\% \tag6.4a
&M^{(4)}_{\beta\cdot \operatorname{p}; 1}(f)
 := \|(\beta\cdot \operatorname{p})P_{12} f\|_1 
             + \|(\beta\cdot \operatorname{p})P_{34}f\|_1\nonumber\\
&\,\,\,\qquad\qquad 
  = \|(\partial_1+i\partial_2)f_{1}-(\partial_3-i\partial_4)f_{2}\|_1
  + \|(\partial_1-i\partial_2)f_{2}+(\partial_3+i\partial_4)f_{1}\|_1
  \nonumber\\
 &\,\,\,\qquad\qquad
  \quad+\|(\partial_1+i\partial_2)f_{3}-(\partial_3+i\partial_4)f_{4}\|_1
  +\|(\partial_1-i\partial_2)f_{4}+(\partial_3-i\partial_4)f_{3}\|_1\,;
  %\tag6.34\\
\end{align}
\end{subequations}
%%%

\noindent
$\underline{M_{\beta}^{(5)}}$
\begin{subequations}
\begin{align}
&\beta\cdot \operatorname{p}
= \left(\begin{array}{cccc}
0&0&\operatorname{p}_3-i\operatorname{p}_4
&\operatorname{p}_1-i\operatorname{p}_2 \\ 
0&0&0&0\\
0&\operatorname{p}_1-i\operatorname{p}_2&0&0\\ 
0&-(\operatorname{p}_3-i\operatorname{p}_4)& 0 &0
\end{array}\right)
+ \left(\begin{array}{cccc}
0&0&0&0\\ 
0&0&\operatorname{p}_1+i\operatorname{p}_2 
&-(\operatorname{p}_3+i\operatorname{p}_4)\\
\operatorname{p}_3+i\operatorname{p}_4&0&0&0\\ 
\operatorname{p}_1+i\operatorname{p}_2&0& 0 &0
\end{array}\right)\nonumber\\
&\qquad
=: (\beta\cdot \operatorname{p})_5 +(\beta\cdot \operatorname{p})_6\,,
\\% \tag6.5a
&M^{(5)}_{\beta\cdot \operatorname{p}; 1}(f)
:= \|(\beta\cdot \operatorname{p})_5 f\|_1 
             + \|(\beta\cdot \operatorname{p})_6f\|_1\nonumber\\
&\,\,\,\qquad\qquad
 = \|(\partial_1+i\partial_2)f_{1}\|_1+\|(\partial_3+i\partial_4)f_{1}\|_1
  +\|(\partial_1-i\partial_2)f_{2}\|_1 +\|(\partial_3-i\partial_4)f_{2}\|_1
  \nonumber\\
&\,\,\,\qquad\qquad
 \quad+\|(\partial_1+i\partial_2)f_{3}-(\partial_3+i\partial_4)f_{4}\|_1
  +\|(\partial_1-i\partial_2)f_{4}+(\partial_3-i\partial_4)f_{3}\|_1\,;
\end{align}%\tag6.5b
\end{subequations}
%%%

\noindent
$\underline{M_{\beta}^{(6)}}$
\begin{subequations}
\begin{align}
&\beta\cdot \operatorname{p}
= \left(\begin{array}{cccc}
0&0&\operatorname{p}_3-i\operatorname{p}_4
&\operatorname{p}_1-i\operatorname{p}_2 \\ 
0&0&0&-(\operatorname{p}_3+i\operatorname{p}_4)\\
0&\operatorname{p}_1-i\operatorname{p}_2&0&0\\ 
0&-(\operatorname{p}_3-i\operatorname{p}_4)& 0 &0
\end{array}\right)
+ \left(\begin{array}{cccc}
0&0&0&0\\ 
0&0&\operatorname{p}_1+i\operatorname{p}_2 
&0\\
\operatorname{p}_3+i\operatorname{p}_4&0&0&0\\ 
\operatorname{p}_1+i\operatorname{p}_2&0& 0 &0
\end{array}\right)\nonumber\\
&\qquad
=: (\beta\cdot \operatorname{p})_7 +(\beta\cdot \operatorname{p})_8\,,
\\%\tag6.6a
&M^{(6)}_{\beta\cdot \operatorname{p}; 1}(f)
 := \|(\beta\cdot \operatorname{p})_7 f\|_1 
             + \|(\beta\cdot \operatorname{p})_8f\|_1\nonumber\\
&\,\,\,\qquad\qquad
 = \|(\partial_1+i\partial_2)f_{1}\|_1+\|(\partial_3+i\partial_4)f_{1}\|_1
  +\|(\partial_1-i\partial_2)f_{2}\|_1 +\|(\partial_3-i\partial_4)f_{2}\|_1
  \nonumber\\
&\,\,\,\qquad\qquad
 \quad+\|(\partial_1+i\partial_2)f_{3}\| + \|(\partial_3+i\partial_4)f_{4}\|_1
  +\|(\partial_1-i\partial_2)f_{4} + (\partial_3-i\partial_4)f_{3}\|_1\,.
\end{align}%\tag6.6b
\end{subequations}
Here the first decomposition (6.4a) and the second (6.5a) enjoy 
the same property as the Dirac operator $D$ mentioned above in connection
with  K\"ahler Geometry and/or Spin Geometry.
Further, the former (6.4a), which we have already 
referred to in Section 1 below Proof of Proposition 1.0 and also below 
equations (1.56a, b, c),
has a beauty of symmetry. The latter (6.5a) has another beauty that 
each nonzero entry of $(\beta\cdot \operatorname{p})_6$ is either of the two 
Cauchy--Riemann operators in the variables $(x_1,x_2)$ and  
$(x_3,x_4)$, while that of $(\beta\cdot \operatorname{p})_5$ 
either of their adjoints. 
The third decomposition (6.6a), which is a slight modification of (6.4a),
looks artificial, lacking in beauty of symmetry and satisfying neither
$(\beta\cdot \operatorname{p})_8 = {(\beta\cdot \operatorname{p})_7}^*$ nor 
${(\beta\cdot \operatorname{p})_7}^2 = {(\beta\cdot \operatorname{p})_8}^2=0$. 

Our answer from the present paper is affimative for 
$1 < p <\infty$, as already shown in Theorem 1.6 (ii), 
because, for any decomposition $(D_1,D_2) \in \hbox{\rm Decom}(D)$,
the semi-norm $M_{D_1\vee D_2;p}(f)$ is equivalent
to the semi-norms $\|(\beta\cdot \operatorname{p})f\|_p$ and
$\|\nabla f\|_p$ as seen in (1.51).
However, as for $p=1$, it will be negative,
so long as one requires that ${D_1}^2={D_2}^2=0$.
Thus the problem is when $p=1$.

Comparing with the semi-norm 
$M_{\beta\cdot \operatorname{p}; 1}(f)$ in (1.50) for $p=1$, 
we note (cf. (1.51))
\begin{equation}%
\|(\beta\cdot \operatorname{p})f\|_1
= M^{(4)}_{\beta\cdot \operatorname{p}; 1}(f) 
\leq M^{(5)}_{\beta\cdot \operatorname{p}; 1}(f) 
\leq M^{(6)}_{\beta\cdot \operatorname{p}; 1}(f) 
\leq M_{\beta\cdot \operatorname{p}; 1}(f) 
\leq \|\nabla f\|_1\,,
\end{equation}%\tag6.7$$
where these three semi-norms are not equivalent to one another.
Hence we also realize that $M^{(6)}_{\beta\cdot \operatorname{p}; 1}(f)$ is 
{\it next} weaker than $M_{\beta\cdot \operatorname{p}; 1}(f)$,
and $M^{(5)}_{\beta\cdot \operatorname{p}; 1}(f)$ is {\it next} weaker 
than $M^{(6)}_{\beta\cdot \operatorname{p}; 1}(f)$.

%%%%
\begin{prp}%{Proposition 6.1} 
For $p=1$, inequality \hbox{\rm (6.3)} 
does not hold with the semi-norm $M_{D_1 \vee D_2;1}(f)$ 
replaced by 
$M^{(4)}_{\beta\cdot \operatorname{p}; 1}(f)$ in \hbox{\rm (6.4b)} and 
$M^{(5)}_{\beta\cdot \operatorname{p}; 1}(f)$ in \hbox{\rm (6.5b)}
corresponding to the decompositions  \hbox{\rm (6.4a)} and  \hbox{\rm (6.5a)},
respectively.
\end{prp}

The proof of Proposition 6.1 is omitted. We give only some notes here.
As to $M^{(4)}_{\beta\cdot \operatorname{p}; 1}(f)$ in 
 \hbox{\rm (6.4b)}, the asertion is clear, because
the last member of this semi-norm  is the same 
as (1.49), namely, $M^{(4)}_{\beta\cdot \operatorname{p}; 1}(f) 
=\|(\beta\cdot \operatorname{p})f\|_1$. 
As to $M^{(5)}_{\beta\cdot \operatorname{p}; 1}(f)$ in  
\hbox{\rm (6.5b)}, we can show the same sequence $\{f_n\}_{n=1}^{\infty}$ 
used to construct the counterexample in Section 3 violates inequality (6.3)
for $p=1, \, q= \frac43$, so that $\frac{p}{p-q} = -3$.

\medskip
It should be probably approriate to mention here 
whether the present work has any connection with those of [BoBr] and [LanSt]. 
They proved an inequality of the form
$$ 
\|u\|_{n/(n-1)} \leq C\,(\|du\|_1+\|d^*u\|_1)
$$
holds with a constant $C>0$ for all smooth $m$-forms $u$ on ${\Bbb R}^n$,
when $m$ is neither $1$ nor $n-1$. 
For $m=1$, it holds with $\|d^*u\|_1$ replaced by $\|d^*u\|_{H^1}$, 
and for $m=n-1$, with $\|du\|_1$ replaced by $\|du\|_{H^1}$, 
where $H^1$ is the real Hardy space. 
This looks a little similar since (1.57) implies that
$\|f\|_q \leq C_1 M_{\beta\cdot \operatorname{p}; 1}(f) 
    + C_2 \|f\|_{B^{1/(1-q)}_{\infty,\infty}}$ with constants $C_1, \, C_2 >0$.
But we don't know whether it is related to our results, partly because, 
though it will be the case $n=4$, $m=1$ and $q= \frac43$, 
so that if our paper should have a relation, 
as Proposition 6.1 above says, inequality (6.3) fails to hold 
for the semi-norms 
$M^{(4)}_{\beta\cdot \operatorname{p}; 1}(f)$ in \hbox{\rm (6.4b)} and 
$M^{(5)}_{\beta\cdot \operatorname{p}; 1}(f)$ in \hbox{\rm (6.5b)}  
in place of $M_{D_1,D_2; 1}(f)$.

\medskip
Finally, as for the third semi-norm 
$M^{(6)}_{\beta\cdot \operatorname{p}; 1}(f)$ in (6.6b) associated with the 
decomposition (6.6a), it is not clear whether or not (6.3) holds, 
although we learn in Theorem 1.6 that it holds for its {\it next} 
stronger semi-norm $M_{\beta\cdot \operatorname{p}; 1}(f)$, but in
Proposition 6.1 above that it does not for its {\it next weaker} 
semi-norm $M^{(5)}_{\beta\cdot \operatorname{p}; 1}(f)$. 
However, it should be probably noted here that the sequence 
$\{f_n\}$ used to construct the counterexample in Section 3 
{\it does not violate but keeps} inequality (1.57) with semi-norm 
$M^{(6)}_{\beta\cdot \operatorname{p}; 1}(f)$
in place of $M_{\beta\cdot \operatorname{p}; 1}(f)$.
Needless to say, this sequence $\{f_n\}$ of course keeps inequality (1.57) safe, though.

%%%%%%%%%%%%%%%%%%%%%%
\section{Summary}%\sect.7

In this work we have extended the improved Sobolev embedding theorem (1.1), 
which originally is for single-valued functions, to a 
vector-valued version, 
(1.28) and (1.30), which are connected with the three-dimensional
massless Dirac operator $\alpha\cdot \operatorname{p}$ in (1.4)/(1.17):
\begin{eqnarray*}%$$\align
1\leq p<q<\infty: \quad
 \|f\|_{q} &\leq& C M_{\alpha\cdot \operatorname{p}; p}(f)^{p/q}
  \|f\|_{B^{p/(p-q)}_{\infty,\infty}}^{1-(p/q)}, 
\qquad\qquad\qquad\qquad\qquad(1.28)\\
1< p<q<\infty: \quad
\|f\|_{q} &\leq& C\|(\alpha\cdot \operatorname{p})f\|_p^{p/q}
  \|f\|_{B^{p/(p-q)}_{\infty,\infty}}^{1-(p/q)}, 
\qquad\qquad\qquad\qquad\qquad(1.30) 
\end{eqnarray*}%\endalign$$
where  $f(x)= {}^t(f_1(x),f_2(x),f_3(x),f_4(x))$ are 
${\Bbb C}^4$-valued functions on ${\Bbb R}^3$.
The  first-order-derivative semi-norm 
$M_{\alpha\cdot \operatorname{p}; p}(f)$ on the right 
of (1.28) is at first defined by (1.19) with the rather artificial
decomposition (1.18) of $\alpha\cdot \operatorname{p}$ 
into the sum of its two parts, but then can be seen, through its explicit 
expression (1.21), to coincide with the ones to be defined 
with the other decompositions like (1.24), (1.25) and (1.26), just 
as clarified in Proposition 1.0. This will reveal the semi-norm  
$M_{\alpha\cdot \operatorname{p}; p}(f)$ to have an intrinsic meaning.

When $1< p<q<\infty$, the semi-norm 
$M_{\alpha\cdot \operatorname{p}; p}(f)$
is equivalent to the semi-norm $\|(\alpha\cdot \operatorname{p})f\|_p$
as well as $\|\nabla f\|_p$. Therefore, in this case it is no wonder that 
inequality (1.30) holds, because (1.28) is reduced to (1.30) which is 
also equivalent to (1.13). It also is an improvement of the (1.14) 
that has the weak $L^q$ norm on the left-hand side.

But when $p=1$, these three first-order-derivative semi-norms 
are not equivalent to one another, cf. (1.22).
In this case, (1.16)/(1.30) does not hold in general. A counterexample 
is given in Section 3. 
Further, for $p=1$ two inequalities (1.28) and (1.14) cannot be compared 
so as to say which of them is sharper.

Analogous improved Sobolev embedding theorems are also given 
for the three-dimensional Weyl--Dirac (or Pauli) operator 
$\sigma\cdot \operatorname{p}$ in (1.31),  
the Cauchy--Riemann operator $\frac12(\partial_1 + i\partial_2)$
and the four-dimensional Euclidian Dirac operator
$\beta\cdot \operatorname{p}$ in (1.46). Here, for the last one 
$\beta\cdot \operatorname{p}$, in the same way as for 
$\alpha\cdot \operatorname{p}$, the semi-norm  
$M_{\beta\cdot \operatorname{p}; p}(f)$, which is defined at first by (1.48) 
with the rather artificial decomposition (1.47), 
turns out to coincide with the ones to be defined with the other 
decompositions like (1.53), (1.54) and (1.55), and so to be meaningful.
Noted is in Section 2, 5$^o$ that all the results are also vaild for the 
other represntations of the three-dimensional massless and the four-dimensional 
Euclidian Dirac operators.

However, exceptionally for the {\it two}-dimensional Weyl--Dirac (or Pauli) 
operator 
$(\sigma\cdot \operatorname{p})^{(2)}$ in (1.40abc), 
we have proved an inequality 
which is just expected as (1.16) for all $1\leq p<q<\infty$: 
$$%\begin{equation}%
\quad \|f \|_{q} \leq C \|(\sigma\cdot\operatorname{p})^{(2)}f\|_p^{p/q}
  \|f \|_{B^{p/(p-q)}_{\infty,\infty}}^{1-(p/q)},
   \qquad\qquad\qquad\qquad\qquad\qquad\qquad\qquad\qquad(1.41)
$$%\end{equation}%\tag1.41$$
for ${\Bbb C}^2$-valued functions $f(x)= {}^t(f_1(x),f_2(x))$ on ${\Bbb R}^2$,
which might be said to be a {\it true} extension of the single-valued (1.1) 
to the vector-valued version.

%%%%%%%%%%%%%%%%%%%%%%%%%
%%%%%%%%%%%%%%%%%%%%%%
\bigskip
\section{Acknowledgements} 

It is a pleasure to thank Hideo Kozono for a
kind instructive lecture around the Besov spaces concerned at the 
early stage of this work and  useful discussions on some related problem 
about elliptic and div-curl systems.
We also wish to thank Yasuhiro Nakagawa and Asao Arai for beneficial 
discussions together with bibliographical information
about the decomposition of the Dirac operator $D=d+d^*$ 
in K\"ahler Geometry and Spin Geometry. 

The research of the authors is supported in part by Grant-in-Aid for 
Sientific Research No. 20540161 and No. 23540191, 
Japan Society for the Promotion of Science.

%%%%%%%%%%%%%%%%%%%%%%%%%%%%%%%%%%%%%%%%%%%%%%%%%%%

%%%%%%%%%%%%%%%%%%%%%%%%%%%%%%%%%%%%%%%%%%%%%%%%%%%


\begin{thebibliography}{99}%\section*{References} 
%%%%%%%%%%%%%%%%%%%%%%%%%%%%%%%%%%%%%%%%%%%%%%%%%%%
%%%%%%%%%%%%%%%%%%%%%%%%%%%%%%%%%%%%%%%%%%%%%%%%%%%

\bibitem[A1]{A1}
Arai,~A. : Path integral representation of the index of K\"ahler-Dirac 
operators on an infinite-dimensional manifold, 
{\it J. Functional Analysis} {\bf 82} (1989), 330--369.

\bibitem[A2]{A2}
Arai,~A. : A general class of infinite-dimensional Dirac operators 
and path integral representation of their index,
{\it J. Functional Analysis} {\bf 105} (1982), 342--408.

\bibitem[BES]{BES}
Balinsky,~A., Evans,~W. D., Sait{\= o},~Y. :
Dirac-Sobolev inequalities and estimates for the zero 
modes of massless Dirac operators,
{\it J. Math. Phys.} {\bf 49} (2008), 043524-1 -- 043524-10. 


\bibitem[BEU]{BEU}
Balinsky,~A., Evans,~W. D., Umeda,~T.: 
The Dirac-Hardy and Dirac-Sobolev inequalities in $L^1$, 
{\it Publ. RIMS, Kyoto Univ.} {\bf 47} (2011), 791--801


\bibitem[BeSa]{BeSa}
Bethe,~H.~A., Salpeter,~E.~E. :
{\it Quantum mechanics of one- and two-electron atoms},
Reprint of the 1957 original, A Plenum/Rosetta Edition,
 Plenum Publishing Corp., New York 1977


\bibitem[BoBr]{BoBr}
Bourgain,~J., Brezis,~H.:  
New estimates for the Laplacian, the div-curl, and related Hodge systems,
{\it C. R. Math. Acad.Sci. Paris} {\bf 338} (2004), 539--543.

\bibitem[CDPX]{CDPX}
Cohen,~A., DeVore,~R., Meyer,~Y., Petrushev,~P., Xu,~H.:
{Nonlinear approximation and the space $BV({\Bbb R}^2)$}, 
{\it Amer. J. Math.} {\bf 11} (1999), 587--628. 


\bibitem[CMO]{CMO}
Cohen,~A., Meyer,~Y., ~Oru,~F. :
Improved Sobolev inequalities, {\it S\'eminaire X-EDP}, 
\'Ecole Polytechnique, France  1998

\bibitem[Fe]{Fe}
Fefferman,~C.: 
$L^p$ bounds for pseudo-differential operators,
{\it Israel J. Math.} {\bf 14} (1973), 413--417. 

\bibitem[Fr]{Fr}
Friedrich,~T.: 
{\it Dirac Operators in Riemannian Geometry}, 
Translated from the 1997 German original by Andreas Nestke, 
Amer. Math. Soc., Providence, Rhode Island 2000. 

\bibitem[G]{G}
Good,~R.~H.: 
Properties of the Dirac matrices,
{\it Rev. Mod. Phys.} {\bf 27} (1955), No. 3, 187--211. 

\bibitem[IS]{IS}
Ichinose,~T., Sait{\= o},~Y.: 
Dirac--Sobolev spaces and Sobolev spaces,
{\it Funkcialaj Ekvacioj} {\bf 53} (2010), 291--310. 

\bibitem[ItZ]{ItZ}
Itzykson,~C., Zuber,~J.~B.: 
{\it Quantum Field Theory}, McGraw-Hill, New York 1980.

\bibitem[LanSt]{LanSt}
Lanzani,~L., Stein, E.M.: 
A note on div curl inequalities, {\it Math. Res. Lett.}
{\bf 12} (2005), 57--61. 


\bibitem[LawM]{LawM}
Lawson,~H.~B.,  Michelsohn,~M-L.: 
{\it Spin Geometry}, 
Princeton University Press, Princeton, NJ 1989.

\bibitem[Le]{Le}
Ledoux,~M.: 
On improved Sobolev embedding theorems,
{\it Math. Res. Lett.} {\bf 10} (2003), 659--669. 

\bibitem[LLo]{LLo}
Lieb,~E.~H., Loss,~M.: 
{\it Analysis, 2nd edition},
Amer. Math. Soc., Providence, Rhode Island  2001.

\bibitem[LoY]{LoY}
Loss,~M., Yau,~H.~T.: 
Stability of Coulomb systems with 
magnetic fields. III. Zero energy bound states of the Pauli operators,
{\it Commun. Math. Phys.} 
{\bf 104} (1986),  283--290. 

\bibitem[P]{P}
Pauli,~W.: 
Contributions math\'ematiques \`a la th\'eorie des matrices 
de Dirac,
{\it Ann. Inst. H. Poncar\'e}
{\bf 6} (1936), 109--136. 

\bibitem[St]{St}
Stein,~E.~M.: 
{\it Singular Integrals and Differentiability Properties
    of Functions}, 
Princeton University Press  1970.

\bibitem[T]{T}
Triebel,~H.: 
{\it Interpolation Theory, Function Spaces, Differential 
Operators,  2nd edition},
Pohann Ambrosius Barth, Heidelberg 1995.


\bibitem[W]{W}
Weisstein,~Eric~W. :  ``Dirac Matrices." From MathWorld--A Wolfram Web Resource.
 http://mathworld.wolfram.com/DiracMatrices.html 


\end{thebibliography}
\end{document}